\documentclass[3p, 10pt, twocolumn]{elsarticle}
\usepackage{multicol}
\usepackage{color}
\usepackage{etoolbox}
\usepackage{float}
\usepackage{ifthen}

\usepackage{amssymb, amsmath}

\usepackage[nodots]{numcompress}

\usepackage{placeins}

\journal{High-Energy-Density-Physics}

\allowdisplaybreaks

\begin{document}

\begin{frontmatter}



\title{Nonrelativistic grey S$_\textrm{n}$-transport radiative-shock solutions}


\author[lanl_xcp]{J.M.~Ferguson}

\author[tamu_nuen]{J.E.~Morel}

\author[lanl_ccs2]{R.B.~Lowrie}


\address[lanl_xcp]{XCP-Division, Los Alamos National Laboratory, Los Alamos, NM 87545, USA}
\address[tamu_nuen]{Department of Nuclear Engineering, Texas A \& M University, College Station, TX 77843, USA}
\address[lanl_ccs2]{CCS-Division, Los Alamos National Laboratory, Los Alamos, NM 87545, USA}

\begin{abstract}
We present semi-analytic radiative-shock solutions in which grey S$_{\textrm{n}}$-transport is used to model the radiation, and we include both constant cross sections and cross sections that depend on temperature and density.
These new solutions solve for a variable Eddington factor (VEF) across the shock domain, which allows for interesting physics not seen before in radiative-shock solutions.
Comparisons are made with the grey nonequilibrium-diffusion radiative-shock solutions of Lowrie and Edwards \cite{LE2008}, which assumed that the Eddington factor is constant across the shock domain.
{\color{black}
It is our experience that the local Mach number is monotonic when producing nonequilibrium-diffusion solutions, but that this monotonicity may disappear while integrating the precursor region to produce S$_{\text{n}}$-transport solutions.}
For temperature- and density-dependent cross sections we show evidence of a spike in the VEF in the far upstream portion of the radiative-shock precursor.
We show evidence of an adaptation zone in the precursor region, adjacent to the embedded hydrodynamic shock, as conjectured by Drake \cite{Drake2007a, Drake2007b}, and also confirm his expectation that the precursor temperatures adjacent to the Zel'dovich spike take values that are greater than the downstream post-shock equilibrium temperature.
We also show evidence that the radiation energy density can be nonmonotonic under the Zel'dovich spike, which is indicative of anti-diffusive radiation flow as predicted by McClarren and Drake \cite{McD2010}.
We compare the angle dependence of the radiation flow for the S$_{\textrm{n}}$-transport and nonequilibrium-diffusion radiation solutions, and show that there are considerable differences in the radiation flow between these models across the shock structure.
Finally, we analyze the radiation flow to understand the cause of the adaptation zone, as well as the structure of the S$_{\textrm{n}}$-transport radiation-intensity solutions across the shock structure.
\end{abstract}


\end{frontmatter}



\section{Introduction}
In this paper we present semi-analytic, time-independent, 1D, planar, nonrelativistic, frequency-independent (``grey''), nonequilibrium, radiative-shock solutions using a variable Eddington factor (VEF) computed from angularly-discretized (``S$_\textrm{n}$'') radiation transport (RT).
Previous work by Sen and Guess \cite{SG1957}, and Lowrie and Rauenzahn \cite{LR2007}, presented semi-analytic nonrelativistic equilibrium-diffusion radiative-shock solutions, for which the material-radiation system is assumed to be in thermal equilibrium and the radiation field is linearly anisotropic.
A linearly anisotropic radiation field implies that the radiation pressure is one-third of the radiation energy density which is associated with using a constant Eddington factor.
Their work confirmed that continuous shock wave solutions existed and that radiation heat-conduction affected the material over a considerable distance ahead of the material shock into the radiation precursor.
Semi-analytic, nonrelativistic, nonequilibrium-diffusion radiative-shock solutions were originally presented by Heaslet and Baldwin \cite{HB1963}, and more recently by Lowrie and Edwards \cite{LE2008}, which allow for the material and radiation temperatures to have separate values, but still assume that the radiation field is linearly anisotropic.
However, the work by Heaslet and Baldwin neglected the radiation energy density and radiation pressure terms which were retained by Lowrie and Edwards.
By separating the material and radiation temperatures those solutions provided a clearer understanding of the material response to the radiation, specifically, that the embedded hydrodynamic shock and the Zel'dovich temperature spike may exist independently of one another.

The work presented herein extends the work by Lowrie and Edwards \cite{LE2008} by modeling the radiation with S$_{\textrm{n}}$-transport in order to describe its angular dependence.
This provides a better understanding of how the radiation flows through optically thick radiative shocks by allowing the Eddington factor to vary spatially across the shock, instead of using a constant Eddington factor.
The idea of using S$_{\textrm{n}}$-transport to describe the radiation, for radiative-shock solutions, appears to have been first recommended in the conclusion of the paper by Sen and Guess \cite{SG1957}, where they say, ``It seems best to treat the radiation as a series of flux streams, in the manner of Chandrasekhar'', who was the first to discretize the angular variable and integrate over it using quadrature methods \cite{Chandrasekhar1960}.
Further, we verify conjectures made by Drake \cite{Drake2007a, Drake2007b} that there should exist an adaptation zone adjacent to the embedded hydrodynamic shock, and subsequently that the temperatures very near the embedded hydrodynamic shock can take values that are greater than the value of the downstream equilibrium temperature.
We also verify the prediction made by McClarren and Drake \cite{McD2010} that it is possible for the radiation energy density to be nonmonotonic under the Zel'dovich spike while the radiation flux is not near its equilibrium value.
This set of ideas goes against the canonical literature for radiation hydrodynamics \cite{ZR2002, MM1999}.
As a matter of practical utility, the solutions described here have already been used as a code-verification tool for a radiation-hydrodynamic (RH) code \cite{JSD2014}.

{\color{black}
The solution method developed by Lowrie and Edwards \cite{LE2008} relies on the local Mach number being monotonic across the shock structure.
Whether this is strictly true mathematically remains an open problem.
We found no evidence that producing nonequilibrium-diffusion solutions violated this requirement for monotonicity.
We did find that producing grey S$_{\text{n}}$-transport solutions may violate this monotonicity requirement in the precursor region.
When the local Mach number becomes nonmonotonic it can cause our solution method to fail, although it is not a guarantee of failure.
}

{\color{black}
The physical model used in this paper assumes that the system is optically-thick.
As such, radiation cannot escape the material through either equilibrium boundary.
Other authors have investigated other RH environments and solution methods.
An analytic model of radiative shocks in a mixed, optically thick-thin environment was investigated by McClarren and co-authors \cite{McDMH2010}.
Self-similar solutions via asymptotic analysis have been presented as an extension of the original Marshak solution by Lane and McClarren \cite{LMc2013}.
Other self-similar solutions have been produced by considering the method of Lie groups \cite{CA1986, FMB2011}.
Asymptotic solutions of the second-kind, based on Barenblatt's work \cite{Barenblatt1996}, were analyzed by Liang and Keilty \cite{LK2000}.
Ion-electron shocks were recently studied by Masser, Wohlbier and Lowrie \cite{MWL2011}.
An initial attempt was made to study the effect of the radiation's frequency-dependence on the shock structure in the work by Holgado, Ferguson and McClarren \cite{HFM2015}.
In short, our particular solution method applies to a specific, theoretical RH environment, and care must be used when applying its analysis.
}

\begin{figure}[!t]
  \vspace{-20pt}
  \hspace{-10pt}
  \includegraphics[width = 1.1 \columnwidth]{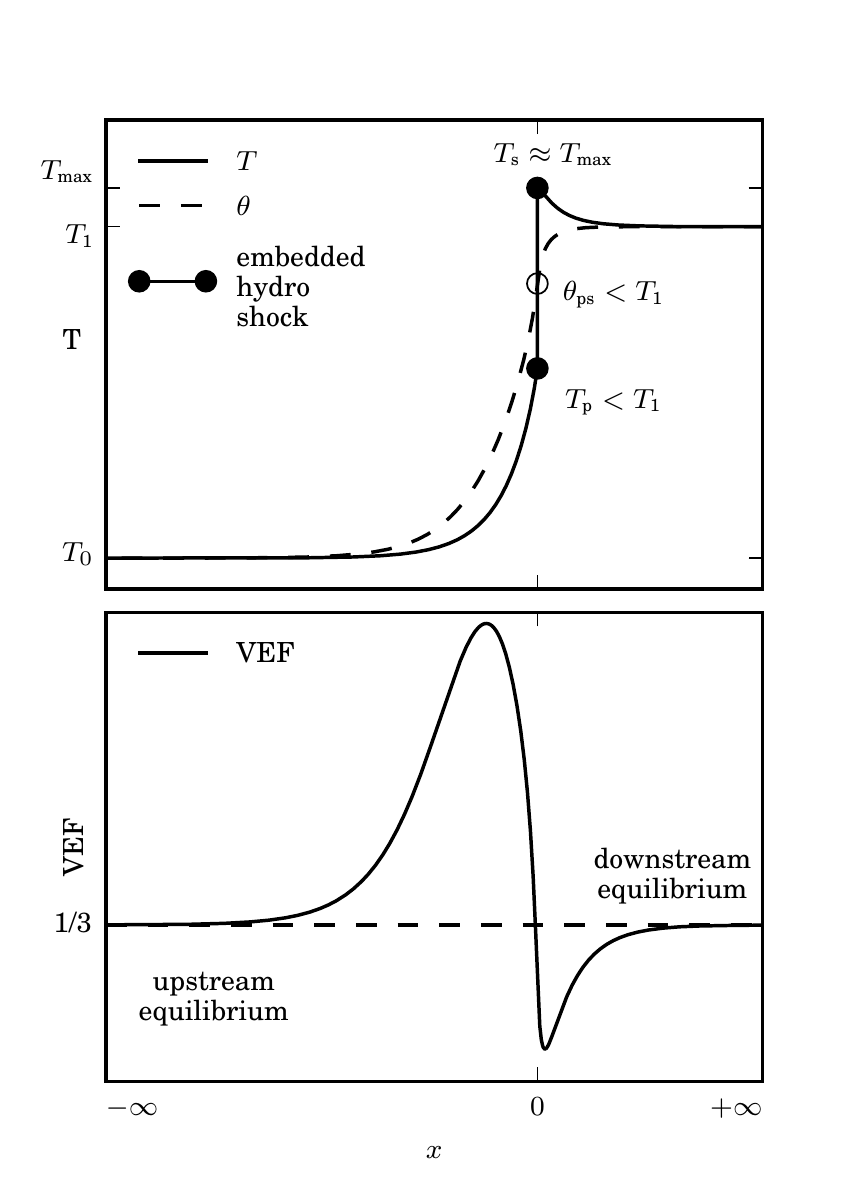}
\caption{
A shock solution showing the material and radiation temperatures, $T$ and $\theta$, respectively, in the top plot, and the variable Eddington factor (VEF) in the bottom plot.
The upstream ($x \approx - \infty$) and downstream ($x \approx + \infty$) equilibria regions of the shock are labeled, and the unshocked material is traveling rightward into the shock, and the shock is traveling leftward.
An embedded hydrodynamic shock exists between state-``p'' and state-``s''.
The material temperature is discontinuous across the embedded hydrodynamic shock and labeled $T_{\textrm{\tiny p}}$ and $T_{\textrm{\tiny s}}$, whereas the radiation temperature is continuous and labeled $\theta_{\textrm{\tiny ps}}$.
\label{fig:generic_M2_shock}}
\end{figure}
\begin{figure}[!t]
  \vspace{-20pt}
  \hspace{-10pt}
  \includegraphics[width = 1.1 \columnwidth]{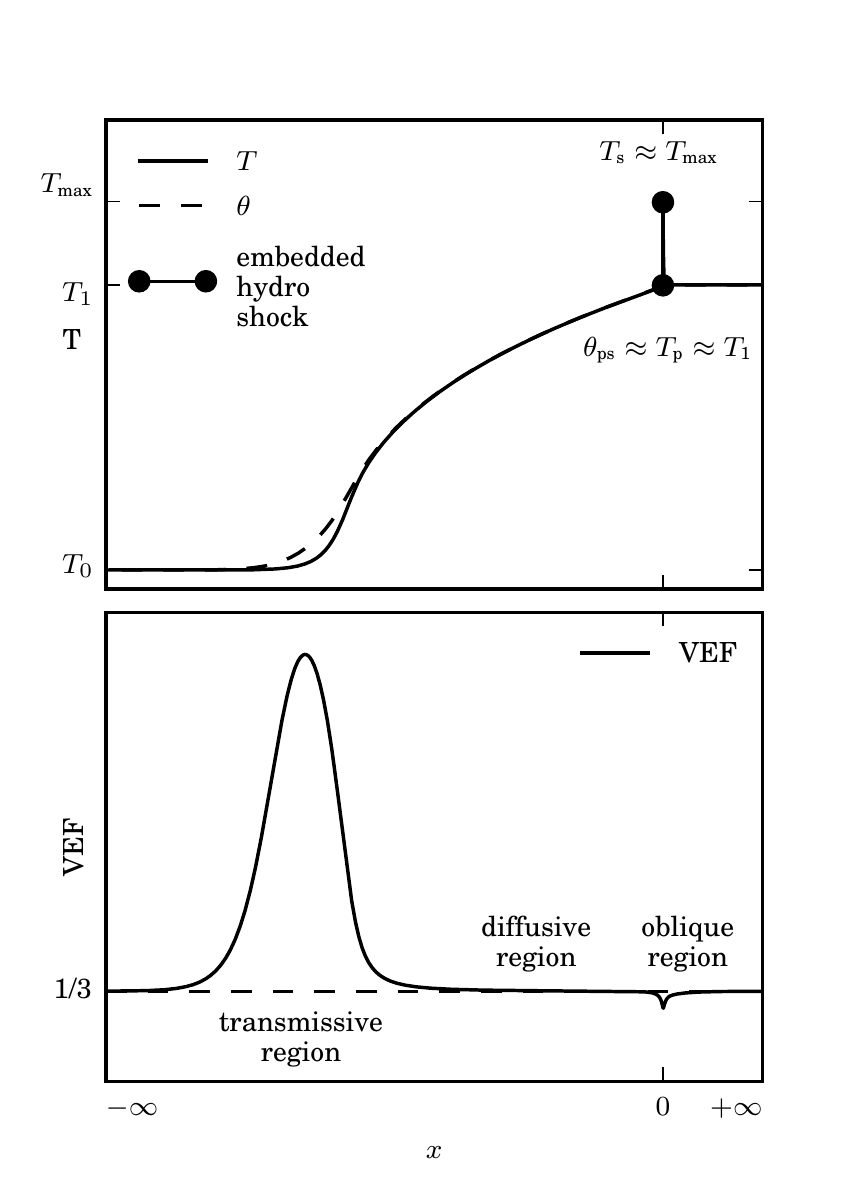}
\caption{
A different shock solution than shown in Figure \ref{fig:generic_M2_shock}.
Here, the transmissive, diffusive, and oblique regions of the radiation flow are labeled.
The transmissive region corresponds to the region where $f > 1/3$, and the radiation intensity reaches its maximum value at $\mu = -1$.
The diffusive region corresponds to the region where $f \approx 1/3$ over a considerable distance, and the radiation is dominantly isotropic.
The oblique region corresponds to the region where $f < 1/3$, and the radiation intensity reaches its maximum value near $\mu = 0$.
\label{fig:generic_M5_shock}}
\end{figure}
The rest of this paper is devoted to describing the solution method, presenting results obtained from it and comparing them with previous results, and analyzing specific features of the results.
In Section \ref{sec:governing_equations}, the necessary RH equations are collected, nondimensionalized, and reduced to steady-state ordinary differential equations (ODEs).
In Section \ref{sec:problem_statement}, the problem to be solved is defined.
In Section \ref{sec:reduced_equations}, the nondimensionalized steady-state ODEs are manipulated into 1) two steady-state ODEs for the ``RH solve'', which is a two-point boundary value problem, and 2) ``n'' separate ODEs for the ``RT solve'', which represent ``n'' separate initial-value problems, and where there are ``n'' ODEs for S$_{\textrm{n}}$-transport.
In Section \ref{sec:results}, the results are presented, compared with previous results, and analyzed.
In Section \ref{sec:summary}, we summarize our work and make recommendations for future work.
In \ref{app:app5B}, we nondimensionalize the ideal-gas $\gamma$-law equation-of-state used in this paper to show that the values of our nondimensional constants are consistent with the physics that we analyze.
In \ref{app:the_solution_procedure}, we present the solution procedure.
\section{The governing equations}
\label{sec:governing_equations}
We seek time-independent, 1D, planar, frequency-independent (``grey''), radiative-shock solutions of the nonrelativistic RH equations in which the angular dependence of the radiation is modeled using S$_{\textrm{n}}$-transport.
We use the Euler equations of hydrodynamics coupled to the radiation-momentum and radiation-energy sources along with the RT equation to describe the RH system.
The length scale of the radiative shock is set by the radiation mean-free-path \cite{SG1957, T1965}, which is greater than the mean-free-path of any set of interactions between material particles, so that viscosity and material heat-conduction do not effect the radiative-shock structure.
The nonrelativistic Euler equations are accurate through ${\cal O}(\beta)$, where $\beta = u/c$ is the ratio of the fluid velocity to the speed-of-light, and we use the mixed-frame RT equation, accurate through ${\cal O}(\beta)$, with lab-frame radiation variables and comoving-frame cross sections.
We add an ${\cal O}(\beta^2)$ term to the RT equation in order to force it, and the radiation-energy and momentum sources to be exactly zero in equilibrium.
Therefore, the overall approximation is exact through ${\cal O}(\beta)$ with errors of ${\cal O}(\beta^2)$.

In Subsection \ref{subsec:the_equations_of_radiation_hydrodynamics} the RH equations are collected, and the radiation sources and radiation variables are defined in terms of angular moments.
In Subsection \ref{subsec:nondimensional_steady_state_ODEs}, the radiation sources and radiation variables are nondimensionalized, along with the RH equations which are rewritten as steady-state ODEs.
\subsection{The equations of radiation hydrodynamics}
\label{subsec:the_equations_of_radiation_hydrodynamics}
The 1D RH equations are the Euler equations of hydrodynamics coupled to the radiation-momentum and energy sources, along with the direction-dependent grey RT equation, which are all correct through ${\cal O}(\beta)$:
\begin{subequations}
\label{eq:RH_equations}
  \begin{gather}
    \partial_t \rho + \partial_x \left( \rho \, u \right) = 0 \, , \\
    \partial_t \left( \rho \, u \right) + \partial_x \left( \rho \, u^2 + p \right) = - S_{\textrm{\tiny rp}} \, , \\
    \partial_t E + \partial_x \left[ u \left( E + p \right) \right] = - S_{\textrm{\tiny re}} \, ,\\
    \frac{1}{c} \, \partial_t I + \mu \, \partial_x I = Q \, . \label{eq:radiation_transport_streaming}
  \end{gather}
The directionally-dependent radiation source term, $Q = Q(\mu)$, is
  \begin{multline}
  \label{eq:radiation_transport_source}
    Q = - \sigma_{\textrm{t}} \, I + \frac{\sigma_{\textrm{s}}}{4 \, \pi} \, c \, {\cal E} + \frac{\sigma_{\textrm{a}}}{4 \, \pi} \, a_{\textrm{\tiny R}} \, c \, T^4 \\
      - 2 \, \frac{\sigma_{\textrm{s}}}{4 \, \pi} \, \beta \, {\cal F} + \beta \, \mu \left( \sigma_{\textrm{t}} \, I + 3 \, \frac{\sigma_{\textrm{s}}}{4 \, \pi} \, c \, {\cal E} + 3 \, \frac{\sigma_{\textrm{a}}}{4 \, \pi} \, a_{\textrm{\tiny R}} \, c \, T^4 \right) \\
      + Q_{\textrm{\tiny eq}} \, ,
  \end{multline}
\end{subequations}
where $\partial_t$ and $\partial_x$ are the time and space derivatives, $\rho$ is the mass density, $u$ is the material velocity, $p$ is the material pressure, $S_{\textrm{\tiny rp}}$ is the radiation-momentum source and $- S_{\textrm{\tiny rp}}$ is a material-momentum source, $E = \tfrac{1}{2} \, \rho \, u^2 + \rho \, e$ is the material energy density, $S_{\textrm{\tiny re}}$ is the radiation-energy source and $- S_{\textrm{\tiny re}}$ is a material-energy source, $c$ is the speed of light, $I = I( \mu )$ is the direction-dependent radiation intensity, and $\mu$ is a direction-cosine.
The total cross section is the sum of the absorption and scattering cross sections, $\sigma_{\textrm{t}} = \sigma_{\textrm{a}} + \sigma_{\textrm{s}}$, all three of which are assumed to take comoving-frame values and to be independent of angle and frequency, ${\cal E}$ is the radiation energy density, ${\cal F}$ is the radiation flux, $a_{\textrm{\tiny R}}$ is the radiation constant, and $T$ is the material temperature; additionally, ${\cal P}$ is the radiation pressure, defined below.
The ${\cal O}(\beta^2)$ correction which ensures that the RT equation and the radiation sources go to zero in equilibrium, is labeled $Q_{\textrm{\tiny eq}}$.
The radiation energy density, radiation flux, and radiation pressure are the zeroth, first, and second angular moments of the radiation intensity, respectively:
\begin{subequations}
\label{eqs:EFP_as_angle_integrated_angular_moments}
  \begin{gather}
    {\cal E} \equiv \frac{2 \, \pi}{c} \int_{-1}^1 I( \mu ) \, d\mu \, , \\
    {\cal F} \equiv 2 \, \pi \int_{-1}^1 \mu \, I( \mu ) \, d\mu \, , \\
    {\cal P} \equiv \frac{2 \, \pi}{c} \int_{-1}^1 \mu^2 \, I( \mu ) \, d\mu \, . \label{eq:P_angular_moment}
  \end{gather}
\end{subequations}
The radiation-energy and radiation-momentum sources are the zeroth and first angular moments of the radiation source, $Q(\mu)$, respectively:
\begin{subequations}
\label{eq:radiation_sources}
  \begin{align}
  \nonumber
    S_{\textrm{\tiny re}}
    & = 2 \, \pi \int_{-1}^1 Q( \mu ) \, d\mu \\
  \nonumber
    & = \sigma_{\textrm{a}} \, c \, \left( a_{\textrm{\tiny R}} \, T^4 - {\cal E} \right) + \beta \left( \sigma_{\textrm{a}} - \sigma_{\textrm{s}} \right) {\cal F} + Q_{\textrm{\tiny eq}}^{\textrm{\tiny re}} \, , \label{eq:radiation_energy_source} \\
    & = \partial_t {\cal E} + \partial_x {\cal F} \\
  \nonumber
    S_{\textrm{\tiny rp}}
    & = \frac{2 \, \pi}{c} \int_{-1}^1 \mu \, Q( \mu ) \, d\mu \\
  \nonumber
    & = - \frac{\sigma_{\textrm{t}}}{c} \, {\cal F} + \beta \left( \sigma_{\textrm{t}} \, {\cal P} + \sigma_{\textrm{s}} \, {\cal E} + \sigma_{\textrm{a}} \, a_{\textrm{\tiny R}} \, T^4 \right) \label{eq:radiation_momentum_source} \\
    & = \frac{1}{c^2} \, \partial_t {\cal F} + \partial_x {\cal P} \, ,
  \end{align}
\end{subequations}
in which equations (\ref{eqs:EFP_as_angle_integrated_angular_moments}) have been used.
The expression for the equilibrium radiation source, $Q_{\textrm{\tiny eq}} = Q_{\textrm{\tiny eq}}(\mu)$, is
\begin{gather}
  Q_{\textrm{\tiny eq}}( \mu ) = \frac{c}{\pi} \left[ \beta^2 \left( 2 \, \sigma_{\textrm{s}} - 3 \, \mu^2 \, \sigma_{\textrm{t}} \right) {\cal P} \right]_{\textrm{\tiny eq}} \, , \label{eq:Q_eq}
\end{gather}
which is a directionally-dependent, ${\cal O}(\beta^2)$ correction to the RT equation, (\ref{eq:radiation_transport_streaming}) and (\ref{eq:radiation_transport_source}), which forces the equation to be zero in equilibrium.
By extension, the radiation-energy (\ref{eq:radiation_energy_source}) and momentum source (\ref{eq:radiation_momentum_source}) are also zero in equilibrium.
The derivation of $Q_{\textrm{\tiny eq}}$ uses the equilibrium expressions for the radiation intensity, radiation energy density, and radiation flux,
\begin{subequations}
  \begin{gather}
    I_{\textrm{\tiny eq}} = \frac{a_{\textrm{\tiny R}} \, c \, T_{\textrm{\tiny eq}}^4}{4 \, \pi} \left( 1 + 4 \, \beta_{\textrm{\tiny eq}} \, \mu \right) \, , \label{eq:equilibrium_radiation_intensity} \\
    {\cal E}_{\textrm{\tiny eq}} = a_{\textrm{\tiny R}} \, T_{\textrm{\tiny eq}}^4 \, , \label{eq:equilibrium_radiation_energy_density} \\
    {\cal F}_{\textrm{\tiny eq}} = \frac{4}{3} \, u_{\textrm{\tiny eq}} \, a_{\textrm{\tiny R}} \, T_\textrm{\tiny eq}^4 \label{eq:equilibrium_radiation_flux} \, ,
  \end{gather}
where all expressions are correct through ${\cal O}(\beta)$.
Additionally, the equilibrium expressions for the radiation pressure and the VEF can be derived from (\ref{eq:equilibrium_radiation_intensity}) using (\ref{eq:P_angular_moment}):
  \begin{gather}
    {\cal P}_{\textrm{\tiny eq}} = \frac{1}{3} \, a_{\textrm{\tiny R}} \, T_{\textrm{\tiny eq}}^4 \, , \\
    f_{\textrm{\tiny eq}} \equiv \frac{{\cal P}_{\textrm{\tiny eq}}}{{\cal E}_{\textrm{\tiny eq}}} = \frac{1}{3} \, .
  \end{gather}
\end{subequations}
The expression for $Q_{\textrm{\tiny eq}}$ applies equally to the upstream and downstream equilibrium states.
The expression for the equilibrium radiation-energy source, $Q_{\textrm{\tiny eq}}^{\textrm{\tiny re}}$, is
\begin{gather}
  Q_{\textrm{\tiny eq}}^{\textrm{\tiny re}} = 4 \, c \left[ \beta^2 \left( \sigma_{\textrm{s}} - \sigma_{\textrm{a}} \right) {\cal P} \right]_{\textrm{\tiny eq}} \, ,
\end{gather}
which is the zeroth angular moment of $Q_{\textrm{\tiny eq}}$ (\ref{eq:Q_eq}), and ensures that the radiation-energy source (\ref{eq:radiation_energy_source}) goes to zero in equilibrium.
Since the expression for $Q_{\textrm{\tiny eq}}$ (\ref{eq:Q_eq}) is an even function of $\mu$ it does not contribute to the radiation-momentum source (\ref{eq:radiation_momentum_source}) since the integrand is an odd function of $\mu$ and the integration range is even.
The radiation internal-energy source is $S_{\textrm{\tiny rie}} = S_{\textrm{\tiny re}} - u S_{\textrm{\tiny rp}}$, and since we consider the fluid to be in thermodynamic equilibrium we write the internal-energy source as:
\begin{gather}
  \rho \, \frac{D e}{D t} + \rho \, p \, \frac{D}{D t} \left( \frac{1}{\rho} \right) = - S_{\textrm{\tiny rie}} \, , \label{eq:radiation_internal_energy_conservation}
\end{gather}
where $\left( D / D t \right) \left( \cdot \right) = \partial_t \left( \cdot \right) + u \, \partial_x \left( \cdot \right)$ is the material or substantive derivative.
The internal-energy source is implicitly contained within the RH equations since it can be derived directly from them.
\begin{figure}
  \vspace{-15pt}
  \hspace{-10pt}
  \includegraphics[width = 1.1 \columnwidth]{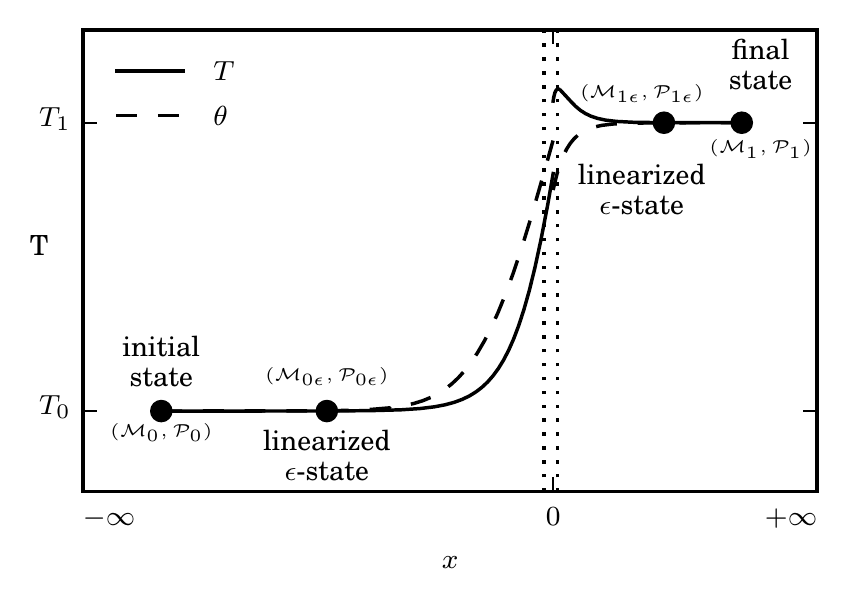}
\caption{
Given an initial state in the upstream equilibrium region, $({\cal M}_0, {\cal P}_0)$, the final state in the downstream equilibrium region, $({\cal M}_1, {\cal P}_1)$, is determined via the Rankine-Hugoniot conditions, as described in Subsection \ref{subsec:Rankine_Hugoniot_jump_conditions}.
A linearization procedure is used to move the solution away from the upstream and downstream equilibrium states, $({\cal M}_{0 \epsilon}, {\cal P}_{0 \epsilon})$ and $({\cal M}_{1 \epsilon}, {\cal P}_{1 \epsilon})$, respectively, as described in Subsection \ref{subsec:linearization_away_from_equilibrium}.
The ODEs are then integrated to determine the precursor and relaxation regions, which are derived in Subsection \ref{subsec:RH_ODEs}.
The two vertical dashed lines represent the location at which continuity of the radiation flux and the radiation pressure (see Figure \ref{fig:solution_procedure_PFM}) are achieved in the precursor and relaxation regions, and thus represent state-``p'' and state-``s'', as described in Subsection \ref{subsec:continuity_conditions}.
\label{fig:solution_procedure_Tx}}
\end{figure}
\begin{figure}[!t]
  \vspace{-41pt}
  \hspace{-10pt}
  \includegraphics[width = 1.1 \columnwidth]{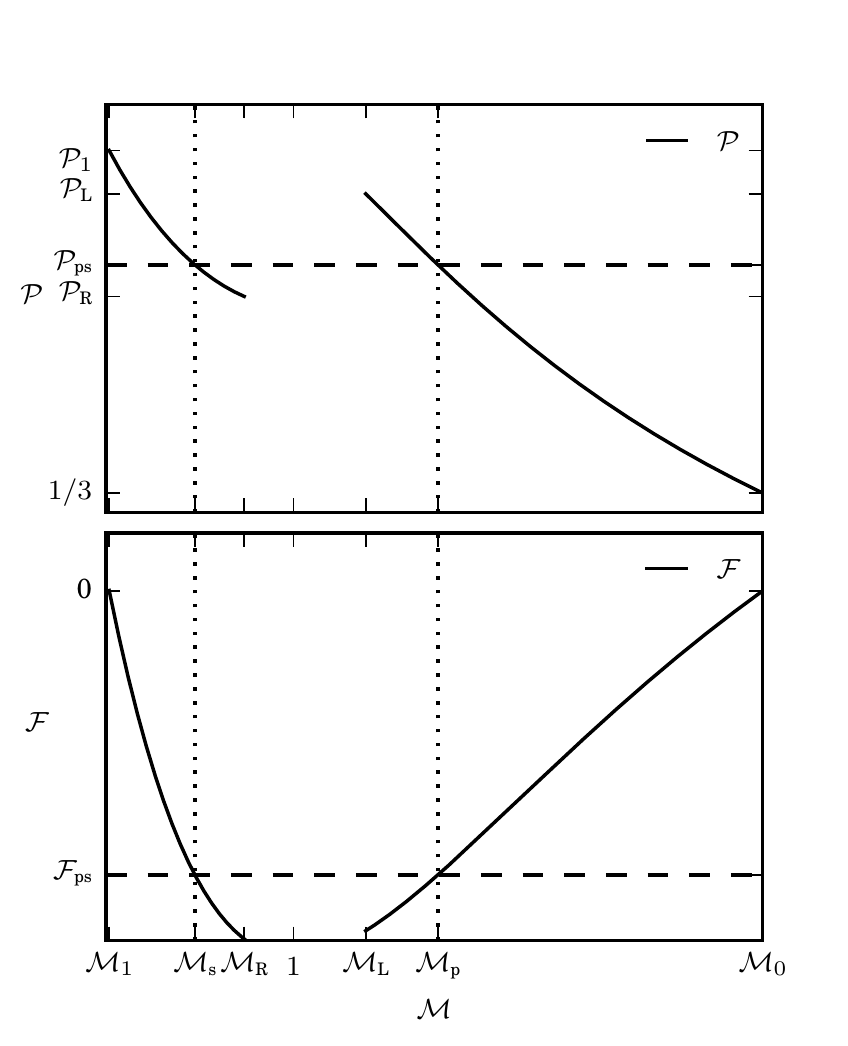}
\caption{
For the shock shown in Figure \ref{fig:solution_procedure_Tx}, the radiation pressure (top) and radiation flux (bottom) are shown as functions of the local Mach number.
{\color{black} The values of ${\cal M}_{\text{\tiny R}}$ and ${\cal M}_{\text{\tiny L}}$ are the integration endpoints in Mach-space for the relaxation and precursor regions, respectively.}
The values of ${\cal M}_{\text{\tiny s}}$ and ${\cal M}_{\text{\tiny p}}$ are defined where the horizontal lines cross the curves for ${\cal P}$ and ${\cal F}$ such that ${\cal P}({\cal M}_{\text{\tiny s}}) = {\cal P}({\cal M}_{\text{\tiny p}})$ and ${\cal F}({\cal M}_{\text{\tiny s}}) = {\cal F}({\cal M}_{\text{\tiny p}})$, simultaneously.
{\color{black} See Subsections \ref{subsec:integrating_the_RH_ODEs} and \ref{subsec:continuity_conditions}.}
\label{fig:solution_procedure_PFM}}
\end{figure}
\subsection{The nondimensional steady-state ODEs}
\label{subsec:nondimensional_steady_state_ODEs}
In this subsection we nondimensionalize the RH equations (\ref{eq:RH_equations}) and group the dimensional quantities.
Each dimensional variable is decomposed into a variable containing the value, and a separate symbol with a tilde over it which contains the dimension; e.g., $\tilde{x} = x \, \tilde{L}$ represents the dimensional variable $\tilde {x}$ as having the value $x$ and the dimension associated with $\tilde{L}$.
The following dimensional reference quantities are used in the nondimensionalization:
\begin{alignat*}{2}
  & \tilde{L} &\quad& \text{(reference length)} \, , \\
  & \tilde{\rho}_0 &\quad& \text{(reference material mass density)} \, , \\
  & \tilde{T}_0 &\quad& \text{(reference material temperature)} \, , \\
  & \tilde{a}_0 &\quad& \text{(reference material sound speed)} \, , \\
  & \tilde{c} &\quad& \text{(speed of light)} \, , \\
  & \tilde{a}_{\textrm{\tiny R}} &\quad& \text{(radiation constant)} \, .
\end{alignat*}
\begin{figure}[t!]
  \vspace{-20pt}
  \hspace{-25pt}
  \includegraphics[width = 1.2 \columnwidth]{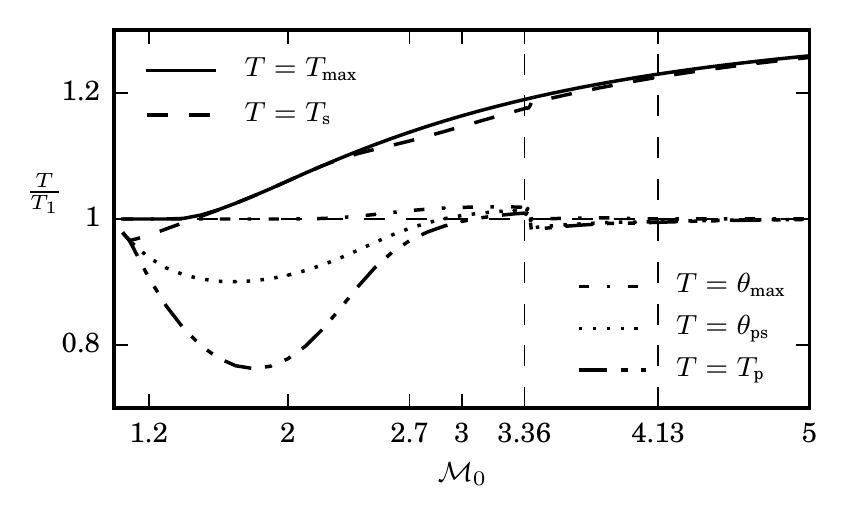}
\caption{
Comparison of characteristic temperature values across the shock structure, normalized by
$T_1$, as functions of
${\cal M}_0$.
There is a horizontal dashed line at $T / T_1 = 1$.
When $T_{\textrm{\tiny max}} / T_1 > 1$ a Zel'dovich temperature spike exists in the relaxation region of the solution.
When $T_{\textrm{\tiny max}} \neq T_{\textrm{\tiny s}}$ then the maximum temperature in the solution is not at state-``s''.
The normalized curves for $\theta_{\textrm{\tiny max}}$, $\theta_{\textrm{\tiny ps}}$, and $T_{\textrm{\tiny p}}$, all rise above a value of one for some value of ${\cal M}_0$, as discussed in Subsections \ref{subsec:describe_radiation_flow} and \ref{subsec:analyze_radiation_flow}.
According to the nonequilibrium-diffusion radiation model described in Subsection \ref{subsec:simplified_radiation_diffusion_models}, these values should never exceed one.
When $\theta_{\textrm{\tiny max}} / T_1 > 1$ the radiation temperature is nonmonotonic and the radiation field is anti-diffusive.
{\color{black} Values of ${\cal M}_0$ inside the vertical dashed lines represent the open range for which our solution method fails to converge, as described in Susbection \ref{subsec:solution_procedure}.
The plotted values in this region are taken from the solution's last iteration.}
\label{fig:TvM}}
\end{figure}
\noindent
The dimensional reference quantities with a subscript-``0'' are evaluated at the pre-shock, upstream, equilibrium state.
The reference length, $\tilde{L}$, is assumed to have units of cm, the reference density to have units of g/cm$^3$, the reference temperature to have units of eV, and the reference sound speed to have units of cm/s.
The nondimensional variables are then defined in terms of their dimensional counterparts as follows:
\begin{alignat*}{2}
  & x = \frac{\tilde{x}}{\tilde{L}} &\quad& \text{(spatial coordinate)} \, , \\
  & \rho = \frac{\tilde{\rho}}{\tilde{\rho}_0} &\quad& \text{(material mass density)} \, , \\
  & u = \frac{\tilde{u}}{\tilde{a}_0} &\quad& \text{(material velocity)} \, , \\
  & \beta = \frac{\tilde{u}}{\tilde{c}} &\quad& \text{(relativistic measure)} \, , \\
  & e = \frac{\tilde{e}}{\tilde{a}_0^2} &\quad& \text{(material specific internal-energy)} \, , \\
  & p = \frac{\tilde{p}}{\tilde{\rho} \, \tilde{a}_0^2} &\quad& \text{(material pressure)} \, , \\
  & T = \frac{\tilde{T}}{\tilde{T}_0} &\quad& \text{(material temperature)} \, , \\
  & \theta = \frac{\tilde{\theta}}{\tilde{T}_0} &\quad& \text{(radiation temperature)} \, , \\
  & {\cal E} = \frac{\tilde{{\cal E}}}{\tilde{a}_{\textrm{\tiny R}} \, \tilde{T}_0^4} &\quad& \text{(radiation energy density)} \, , \\
  & {\cal F} = \frac{\tilde{{\cal F}}}{\tilde{a}_{\textrm{\tiny R}} \, \tilde{c} \, \tilde{T}_0^4} &\quad& \text{(radiation flux)} \, , \\
  & {\cal P} = \frac{\tilde{{\cal P}}}{\tilde{a}_{\textrm{\tiny R}} \, \tilde{T}_0^4} &\quad& \text{(radiation pressure)} \, , \\
  & I = \frac{\tilde{I}}{\tilde{a}_{\textrm{\tiny R}} \, \tilde{c} \, \tilde{T}_0^4} &\quad& \text{(radiation intensity)} \, , \\
  & \sigma_{\textrm{a}} = \tilde{\sigma}_{\textrm{a}} \tilde{L} &\quad& \text{(absorption cross section)} \, , \\
  & \sigma_{\textrm{s}} = \tilde{\sigma}_{\textrm{s}} \tilde{L} &\quad& \text{(scattering cross section)} \, , \\
  & \sigma_{\textrm{t}} = \tilde{\sigma}_{\textrm{t}} \tilde{L} &\quad& \text{(total cross section)} \, .
\end{alignat*}
Since we neglect the time dependence of the problem we are not concerned with ensuring that the time variable is appropriately nondimensionalized.
The nondimensional radiation energy density, radiation flux, and radiation pressure are the first three angular moments of the nondimensional radiation intensity,
\begin{subequations}
\label{eq:nondimensional_radiation_variable_definitions}
  \begin{gather}
    {\cal E} = 2 \, \pi \int_{-1}^1 I(\mu) \, d\mu \, , \label{eq:nondimensional_E} \\
    {\cal F} = 2 \, \pi \int_{-1}^1 \mu \, I(\mu) \, d\mu \, , \label{eq:nondimensional_F} \\
    {\cal P} = 2 \, \pi \int_{-1}^1 \mu^2 \, I(\mu) \, d\mu \, , \label{eq:nondimensional_P} 
  \end{gather}
and the VEF is,
  \begin{gather}
    f = \frac{{\cal P}}{{\cal E}} = \frac{\int_{-1}^1 \mu^2 \, I(\mu) \, d\mu}{\int_{-1}^1 I(\mu) \, d\mu} \, .
  \end{gather}
\end{subequations}
The nondimensional steady-state RH equations (\ref{eq:RH_equations}), along with the internal-energy source (\ref{eq:radiation_internal_energy_conservation}), are:
\begin{subequations}
\label{eq:nondimensional_RH_equations_to_solve}
  \begin{gather}
    \partial_x \left( \rho \, u \right) = 0 \, , \label{eq:nondimensional_mass_conservation} \\
    \partial_x \left( \rho \, u^2 + p \right) = - P_0 \, S_{\textrm{\tiny rp}} \, , \label{eq:nondimensional_momentum_conservation} \\
    \partial_x \left[ u \left( E + p \right) \right] = - P_0 \, {\cal C} \, S_{\textrm{\tiny re}} \, , \label{eq:nondimensional_energy_conservation} \\
    \rho \, u \, \partial_x e + p \, \partial_x u = - P_0 \, {\cal C} \, S_{\textrm{\tiny rie}} \, , \label{eq:nondimensional_internal_energy_source}
  \end{gather}
\vspace{-20pt}
  \begin{multline}
    \mu \, \partial_x I = - \sigma_{\textrm{t}} \, I + \frac{\sigma_{\textrm{s}}}{4 \, \pi} \, {\cal E} + \frac{\sigma_{\textrm{a}}}{4 \, \pi} \, T^4 - 2 \, \frac{\sigma_{\textrm{s}}}{4 \, \pi} \, \beta \, {\cal F} \\
    + \beta \, \mu \left( \sigma_{\textrm{t}} \, I + \frac{3 \sigma_{\textrm{s}}}{4 \, \pi} \, {\cal E} + \frac{3 \sigma_{\textrm{a}}}{4 \, \pi} \, T^4 \right) \\
    + \frac{1}{\pi} \left[ \beta^2 \left( 2 \, \sigma_{\textrm{s}} - 3 \, \sigma_{\textrm{t}} \, \mu^2 \right) {\cal P} \right]_{\textrm{\tiny eq}} \, , \label{eq:transport_equation}
  \end{multline}
\end{subequations}
and the nondimensional radiation sources (\ref{eq:radiation_sources}) are:
\begin{subequations}
\label{eq:nondimensional_radiation_sources}
  \begin{align}
\nonumber
    S_{\textrm{\tiny re}}
    & = \sigma_{\textrm{a}} \left( T^4 - {\cal E} \right) + \beta \left( \sigma_{\textrm{a}} - \sigma_{\textrm{s}} \right) {\cal F} \\
\nonumber
    & + 4 \left[ \beta^2 \left( \sigma_{\textrm{s}} - \sigma_{\textrm{a}} \right) {\cal P} \right]_{\textrm{\tiny eq}} \\
    & = \partial_x {\cal F} \, , \label{eq:nondimensional_radiation_energy_source} \\
\nonumber
    S_{\textrm{\tiny rp}}
    & = - \sigma_{\textrm{t}} \, {\cal F} + \beta \left( \sigma_{\textrm{t}} \, {\cal P} + \sigma_{\textrm{s}} \, {\cal E} + \sigma_{\textrm{a}} \, T^4 \right) \\
    & = \partial_x {\cal P} \, , \label{eq:nondimensional_radiation_momentum_source} \\
    S_{\textrm{\tiny rie}}
    & = S_{\textrm{\tiny re}} - \beta \, S_{\textrm{\tiny rp}} \, .
  \end{align}
\end{subequations}
The nondimensional constant $P_0 \equiv \tilde{a}_{\textrm{\tiny R}} \, \tilde{T}_0^4 / \tilde{\rho}_0 \, \tilde{a}_0^2$ is a measure of the influence of radiation on the material flow dynamics.
As shown in \ref{app:app5B}, for the equation of state (EOS) used in this paper, $\tilde{a}_0$ is only a function of $\tilde{T}_0$, and so $P_0$ depends only on $\tilde{\rho}_0$ and $\tilde{T}_0$.
The nondimensional constant ${\cal C} \equiv \tilde{c} / \tilde{a}_0$ is the ratio of the speed of light to the reference material sound speed.
The nondimensional equilibrium radiation source, $Q_{\textrm{\tiny eq}}$, is the equilibrium term in the nondimensional RT equation (\ref{eq:transport_equation}), and the nondimensional equilibrium radiation-energy source, $Q_{\textrm{\tiny eq}}^{\textrm{\tiny re}}$, is the equilibrium term in the nondimensional radiation-energy source (\ref{eq:nondimensional_radiation_energy_source}).
These equilibrium sources apply equally well to both the upstream and downstream equilibrium states.
In equilibrium, the nondimensional radiation intensity, radiation energy density, radiation flux, radiation pressure, and VEF, are:
\begin{figure}[t!]
  \vspace{-20pt}
  \hspace{-20pt}
  \includegraphics[width = 1.1 \columnwidth]{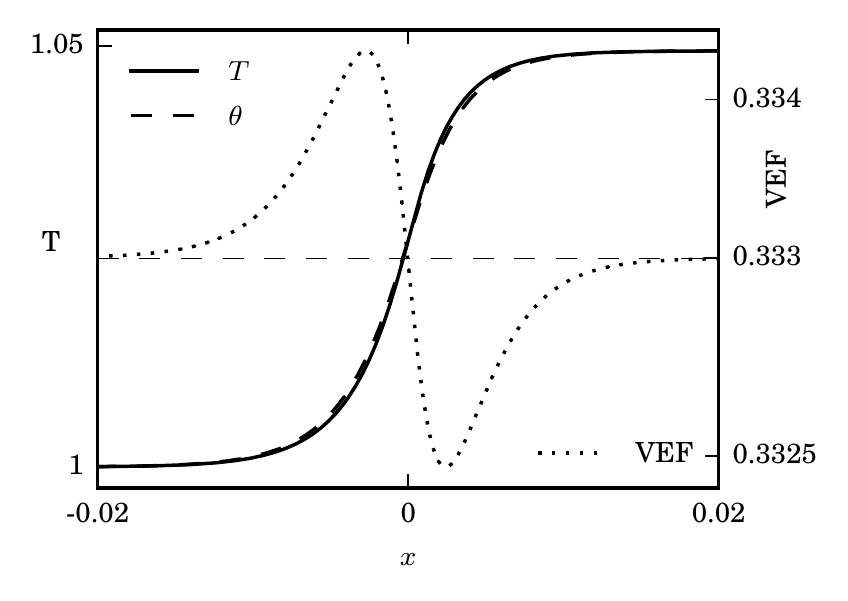}
\caption{
The radiative-shock solutions for the material temperature, $T$, radiation temperature, $\theta$, and the variable Eddington factor (VEF), for ${\cal M}_0 = 1.05$, $\sigma_{\textrm{t}} = 577.35 = \sigma_{\textrm{a}}$, $\tilde{T}_0 = 100$ eV and $\tilde{\rho}_0 = 1$ g/cm$^3$, such that $P_0 \approx 8.5 \times 10^{-5}$; these values were chosen to aide comparison with figures presented in \cite{LE2008}.
This radiative shock is continuous in all variables, and the VEF does not deviate significantly from one-third.
\label{fig:M1p05_Sn}}
\end{figure}
\begin{figure}[t!]
  \vspace{-20pt}
  \hspace{-20pt}
  \includegraphics[width = 1.1 \columnwidth]{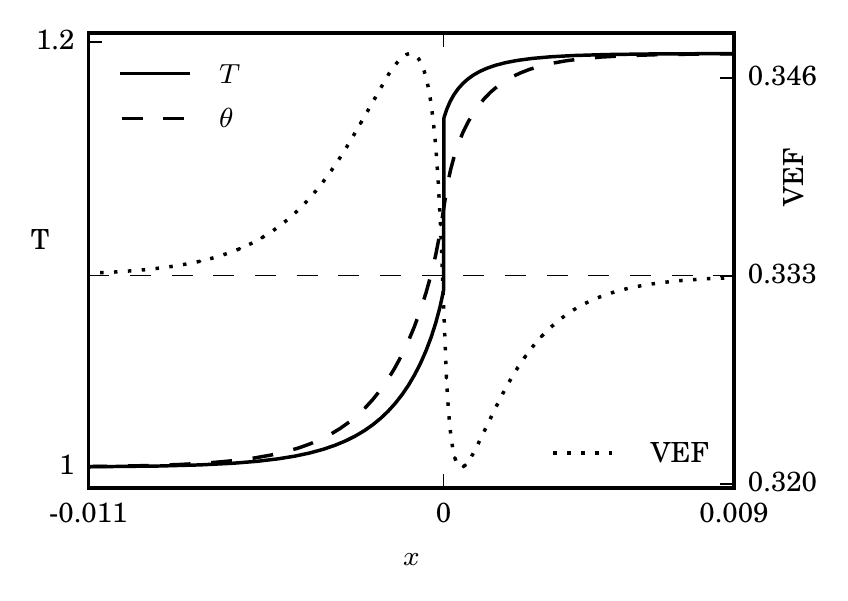}
\caption{
The radiative shock solution for ${\cal M}_0 = 1.2$, but with all other values being the same as in Figure \ref{fig:M1p05_Sn}.
This radiative shock contains an embedded hydrodynamic shock located at $x=0$, but does not contain a Zel'dovich temperature spike.
The VEF deviates slightly from one-third.
\label{fig:M1p2_Sn}}
\end{figure}
\begin{subequations}
\label{eqs:equilibrium_EFPf}
  \begin{gather}
    I_{\textrm{\tiny eq}}( \mu ) = \frac{{\cal E}_{\textrm{\tiny eq}}}{4 \, \pi} \left( 1 + 4 \, \beta_{\textrm{\tiny eq}} \, \mu \right) \, , \label{eq:equilibrium_EFPf_I} \\
    {\cal E}_{\textrm{\tiny eq}} = T_{\textrm{\tiny eq}}^4 \, , \\
    {\cal F}_{\textrm{\tiny eq}} = \frac{4}{3} \, \beta_{\textrm{\tiny eq}} \, T_{\textrm{\tiny eq}}^4 \, , \\
    {\cal P}_{\textrm{\tiny eq}} = \frac{1}{3} \, T_{\textrm{\tiny eq}}^4 \, , \\
    f_{\textrm{\tiny eq}} = \frac{1}{3} \, .
  \end{gather}
\end{subequations}
\section{Problem statement}
\label{sec:problem_statement}
In this section the problem to be solved is defined.
The fluid is assumed to flow in the $+x-$direction, while the shock is assumed to move in the $-x-$direction.
The reference state with subscript-``0'' refers to the pre-shock, upstream, equilibrium boundary condition, satisfied as $x \rightarrow - \infty$, while the subscript-``1'' refers to the post-shock, downstream, equilibrium boundary condition, satisfied as $x \rightarrow + \infty$.
The nondimensional pre-shock equilibrium state is assumed to be known since reference dimensional values, e.g., $\tilde{\rho}_0$ and $\tilde{T}_0$, are taken from this region, and we therefore choose to set $\rho_0 = 1$, $T_0 = 1$, ${\cal E}_0 = 1$, and ${\cal P}_0 = 1/3$.
The problem statement is:
\begin{itemize}
  \item \emph{Assume:} An ideal-gas $\gamma$-law EOS such that $p = \rho \, e \left( \gamma - 1 \right)$, for a fluid obeying Eulerian hydrodynamics and interacting with radiation described by grey S$_{\textrm{n}}$-transport, such that the material-radiation system is described by equations (\ref{eq:nondimensional_RH_equations_to_solve}) and (\ref{eq:nondimensional_radiation_sources}), with equilibrium boundary conditions given by equations (\ref{eqs:equilibrium_EFPf}).
  \item \emph{Given:} The values for $\gamma$, ${\cal M}_0$, $\tilde{\rho}_0$ and $\tilde{T}_0$, the functions $\sigma_{\textrm{a}}(\rho, T)$, $\sigma_{\textrm{s}}(\rho, T)$, and $\sigma_{\textrm{t}}( \rho, \, T ) = \sigma_{\textrm{a}}( \rho, \, T ) + \sigma_{\textrm{s}}( \rho, \, T )$, along with an initial guess for the VEF, $f( x )$, which we typically assume is a constant value of one-third across the shock domain, and the number ``n'' of angular directions for the S$_{\textrm{n}}$-transport radiation model.
The reason for specifying $\tilde{\rho}_0$ and $\tilde{T}_0$ is to obtain a value of $P_0$ that is consistent with the ideal-gas $\gamma$-law EOS.
  \item \emph{Calculate:} Values for the functions $p(x)$, $\rho(x)$, $u(x)$, $T(x)$, ${\cal M}(x)$, $I(x, \, \mu_{\textrm{\tiny m}})$, ${\cal E}(x)$, ${\cal F}(x)$, ${\cal P}(x)$, and $f(x)$, where $\mu_{\textrm{\tiny m}}$ is the discrete direction variable described in Subsection \ref{subsec:SN_ODE}.
\end{itemize}
\section{Reduced equations, the solution procedure, and two radiation models}
\label{sec:reduced_equations}
In this section the necessary equations for the RH solve and the RT solve are derived, and the general solution procedure is outlined, and two radiation models are described.
In Subsection \ref{subsec:nondimensional_EOS} the nondimensional ideal-gas $\gamma$-law EOS is presented.
This EOS is used with equations (\ref{eq:nondimensional_RH_equations_to_solve})-(\ref{eqs:equilibrium_EFPf}) to derive two simplified ODEs for the RH solve in Subsection \ref{subsec:RH_ODEs}.
These two RH ODEs represent a two-point boundary-value problem.
One boundary is specified by the upstream equilibrium state, while the downstream equilibrium boundary is determined from the Rankine-Hugoniot jump conditions, whose solution is discussed in Subsection \ref{subsec:Rankine_Hugoniot_jump_conditions}.
Since the two RH ODEs evaluate to zero at the equilibrium states, their integration cannot begin at the equilibrium states, and the solution must be moved to a nearby state, which is performed by a linearization procedure as described in Subsection \ref{subsec:linearization_away_from_equilibrium}.
{\color{black}
Integrating these ODEs is discussed in Subsection \ref{subsec:integrating_the_RH_ODEs}.
Their successful integration relies on the local Mach number being monotonic across the shock structure, which our experience shows to be a reasonable expectation when the VEF is constant, or when the local Mach number is sufficiently far from the adiabatic sonic point (ASP).}
After the two RH ODEs are integrated, continuity conditions are used to connect the precursor and relaxation regions of the radiative shock, as described in Subsection \ref{subsec:continuity_conditions}.
Figure \ref{fig:solution_procedure_Tx} illustrates the discussion contained in Subsections \ref{subsec:Rankine_Hugoniot_jump_conditions} - \ref{subsec:continuity_conditions}, and Figure \ref{fig:solution_procedure_PFM} further illustrates the discussion in Subsection \ref{subsec:continuity_conditions}.
Then, the RT equation (\ref{eq:transport_equation}) is directionally discretized in Subsection \ref{subsec:SN_ODE}, which produces n RT ODEs for S$_{\textrm{\tiny n}}$-transport.
Each of these n RT ODEs represents initial-value problem, so only the boundary at which the integration is to begin must be specified, as discussed in Subsection \ref{subsec:RT_initial_values_and_linearization}.
The RT ODEs evaluate to zero in equilibrium by construction so their integration cannot begin at the equilibrium state.
Therefore, the solution must be moved to a nearby state.
This is performed by a linearization procedure and is also discussed in Subsection \ref{subsec:RT_initial_values_and_linearization}.
Upon integration, the RT ODEs naturally arrive at the far equilibrium state, which is appropriate to an initial-value problem.
The quadrature definitions of the radiation energy density, radiation flux, and radiation pressure are provided in Subsection \ref{subsec:construct_RT_solutions}, so that their solutions can be reconstructed along with the solution for the VEF.
A brief overview of the solution procedure is given afterward, in Subsection \ref{subsec:solution_procedure}, and a more detailed description is given in \ref{app:the_solution_procedure}.
We close this section by discussing the analytic solution for the directionally-dependent radiation model in Subsection \ref{subsec:analytic_solution_for_I}, and discussing the nonequilibrium-diffusion radiation model in Subsection \ref{subsec:simplified_radiation_diffusion_models}.

In order to simplify the analysis several assumptions are made.
Generally, we assume that the material is homogeneous and in thermodynamic equilibrium across the shock domain.
The ion and electron temperatures are assumed to be equal throughout the material, and material heat-conduction and viscous effects are assumed to be negligible.
The S$_{\textrm{n}}$-transport model assumes that the angular distribution of the radiation intensity is well represented by a finite number of discrete directions.
The cross sections are restricted to depend only on the material density and temperature.
The grey assumption represents cross sections and radiation intensities as being frequency independent, and is particularly invalid when line or edge structures of the cross sections play a significant role.
Although the solutions presented here have been used to verify RH codes \cite{JSD2014}, and offer additional physical insight, the assumptions made must be kept in mind.
In short, these assumptions may not be consistent with material and radiation models in physics codes.
\subsection{The nondimensional ideal-gas $\gamma$-law EOS}
\label{subsec:nondimensional_EOS}
A discussion of the nondimensionalized ideal-gas $\gamma$-law EOS, including the sound speed, is provided in \ref{app:app5B}.
The nondimensional ideal-gas $\gamma$-law EOS, $p = \rho \, e \left( \gamma - 1 \right)$, produces the following expressions for the material internal-energy and pressure,
\begin{subequations}
\label{eq:EOS}
  \begin{gather}
    e = \frac{T}{\gamma \left( \gamma - 1 \right) } \, , \label{eq:EOS_e} \\
    p = \frac{\rho \, T}{\gamma} \, , \label{eq:EOS_p}
  \end{gather}
where the adiabatic index, $\gamma$, is assumed to be constant, and we further assume $\gamma = 5/3$, which is consistent with a monatomic gas.
This EOS is quite restrictive, but greatly simplifies the analysis.
The expression for the local Mach number, in terms of nondimensional variables, is included here for convenience:
  \begin{gather}
    {\cal M} = \frac{u}{\sqrt{T}} \, . \label{eq:local_Mach}
  \end{gather}
\end{subequations}
The steady-state ODEs for the RH solve (\ref{eq:nondimensional_mass_conservation}) - (\ref{eq:nondimensional_internal_energy_source}) using the ideal-gas $\gamma$-law EOS (\ref{eq:EOS}) are:
\begin{subequations}
\label{eqs:nondimensional_RH_ODEs}
  \begin{gather}
    \frac{d}{dx} \left( \rho \, u \right) = 0 \label{eq:mass_ODE} \, , \\
    \frac{d}{dx} \left( \rho \, u^2 + \frac{\rho \, T}{\gamma} + P_0 \, {\cal P} \right) = 0 \label{eq:momentum_ODE} \, , \\
    \frac{d}{dx} \left[ u \left( \frac{1}{2} \, \rho \, u^2 + \frac{\rho \, T}{\gamma - 1} \right) + P_0 \, {\cal C} \, {\cal F} \right] = 0 \label{eq:energy_ODE} \, , \\
    \frac{\rho \, u}{\gamma \left( \gamma - 1 \right)} \frac{dT}{dx} + \frac{\rho \, T}{\gamma} \frac{du}{dx} = - P_0 \, {\cal C} \, S_{\textrm{\tiny rie}} \, . \label{eq:internal_energy_ODE}
  \end{gather}
\end{subequations}
Again, these equations represent the steady-state expressions of mass conservation, total-momentum conservation, total-energy conservation, and the material internal-energy source coupled to the radiation internal-energy source, respectively.
The nondimensional radiation flux is derived from the radiation-momentum source (\ref{eq:nondimensional_radiation_momentum_source}):
\begin{gather}
\label{eq:nondimensional_radiation_flux}
  {\cal F} = - \frac{1}{\sigma_{\textrm{t}}} \frac{d{\cal P}}{dx} + \frac{1}{\sigma_{\textrm{t}}} \beta \left( \sigma_{\textrm{t}} \, {\cal P} + \sigma_{\textrm{s}} \, {\cal E} + \sigma_{\textrm{a}} \, T^4 \right) \, .
\end{gather}
Equations (\ref{eqs:nondimensional_RH_ODEs}) and (\ref{eq:nondimensional_radiation_flux}) hold across the spatial domain, specifically across an embedded hydrodynamic shock, and they are used in Subsections \ref{subsec:RH_ODEs}-\ref{subsec:continuity_conditions} to describe how the shock structure is determined.
\begin{figure}[t!]
  \vspace{-20pt}
  \hspace{-20pt}
  \includegraphics[width = 1.1 \columnwidth]{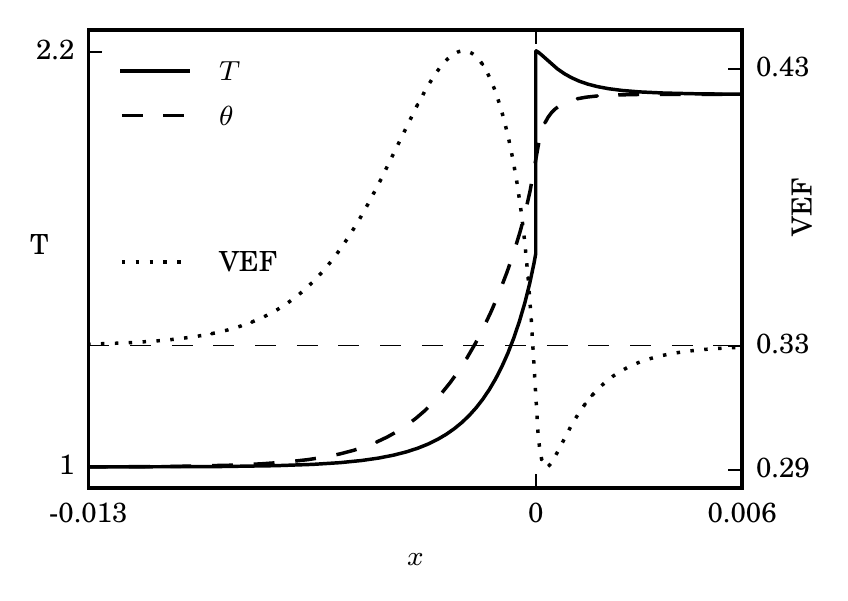}
\caption{
The radiative shock solution for ${\cal M}_0 = 2$, but with all other parameters being the same as in Figure \ref{fig:M1p05_Sn}.
This radiative shock solution contains an embedded hydrodynamic shock located at $x=0$, and a Zel'dovich spike, but the radiation temperature is still monotonic, and it is clear that $\theta_{\textrm{\tiny ps}} < T_1$ and $T_{\textrm{\tiny p}} < T_1$; see Figure \ref{fig:TvM}.
The VEF deviates from one-third over the whole shock domain, except at the equilibria end-states.
\label{fig:M2_Sn}}
\end{figure}
\begin{figure}[t!]
  \vspace{-20pt}
  \hspace{-20pt}
  \includegraphics[width = 1.1 \columnwidth]{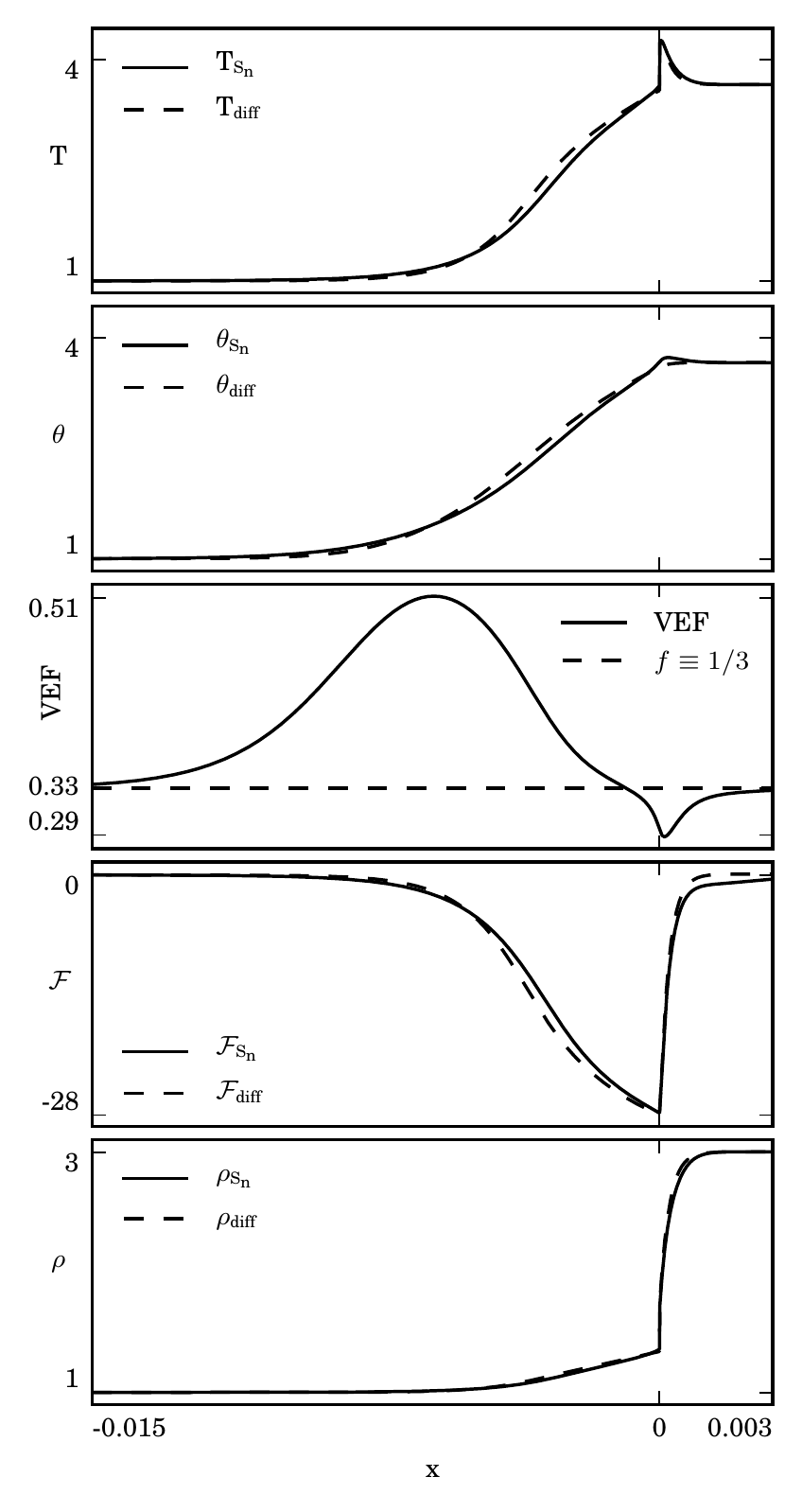}
\caption{
Radiative shock structures for ${\cal M}_0 = 3$ and $\sigma_{\textrm{t}} = 577.35 = \sigma_{\textrm{a}}$, comparing the nonequilibrium-diffusion and S$_{\textrm{n}}$-transport solutions.
In descending order, the material temperatures are compared, followed by the radiation temperatures, the Eddington factors, the radiation flux, and finally, the material density.
The general structure of $T$, $\theta$ and ${\cal F}$ are similar between the two solution methods, and $\rho$ appears to be unchanged, but the VEF is different, as expected.
\label{fig:M3_const}}
\end{figure}
\begin{figure}[t!]
  \vspace{-20pt}
  \hspace{-20pt}
  \includegraphics[width = 1.1 \columnwidth]{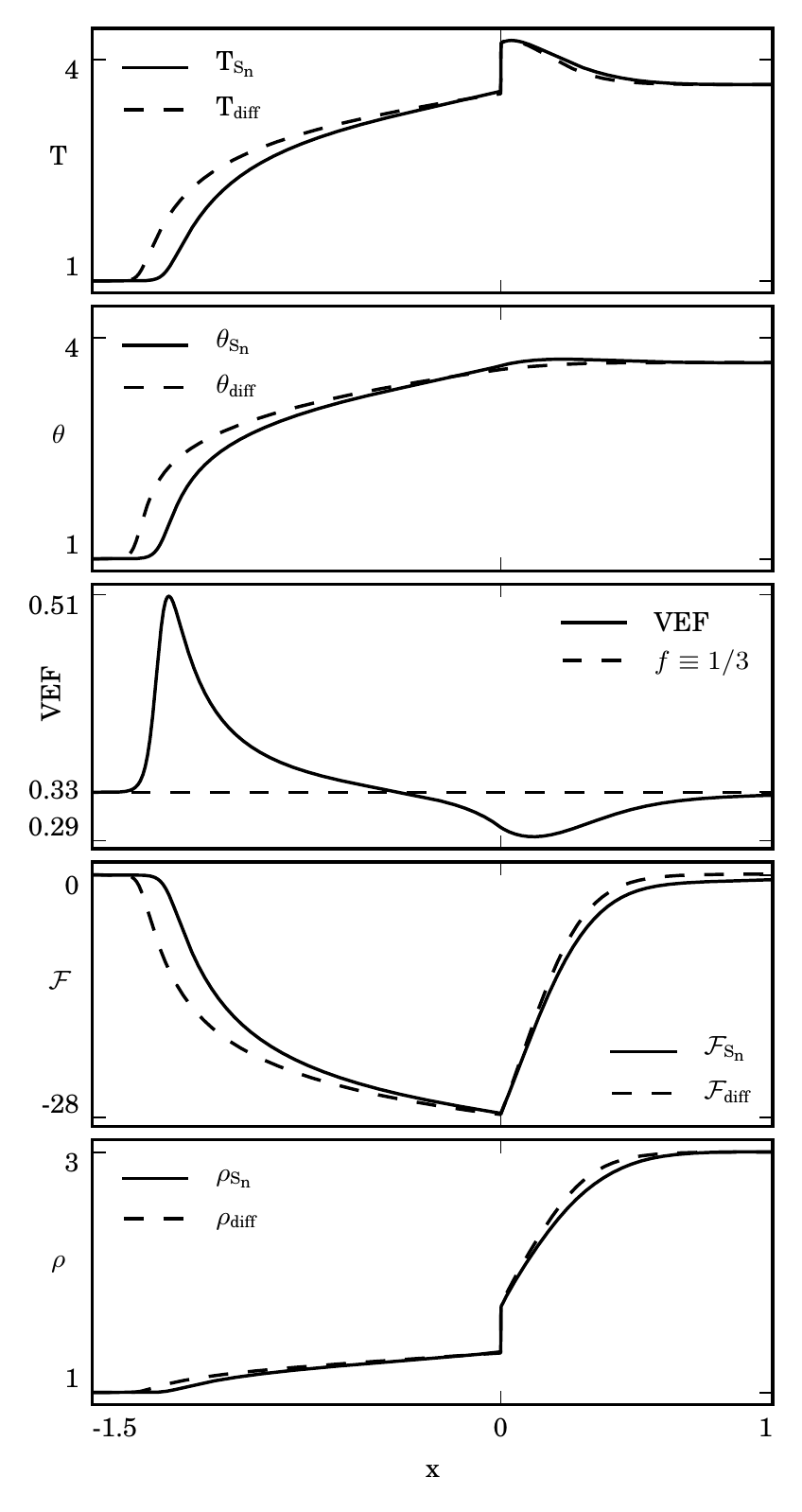}
\caption{
The same as Figure \ref{fig:M3_const}, but with Thomson scattering, $\sigma_{\textrm{s}} = 0.4006 \, \rho$, and Bremsstrahlung emission, $\sigma_{\textrm{a}} = 44.88 \, \rho^2 / \, T^{7/2}$.
Differences between all variables are noticeable, and there is a spatial shift between the nonequilibrium-diffusion solutions and the S$_{\textrm{n}}$-transport solutions in the far upstream precursor region due to the rapid increase in the VEF there, which causes the widths of the upstream precursors to be different.
In the relaxation region, all S$_{\textrm{n}}$-transport variables appear to be relaxing to their equilibrium values more slowly than the nonequilibrium-diffusion variables.
\label{fig:M3_Trho}}
\end{figure}
\subsection{The ODEs for the RH solve}
\label{subsec:RH_ODEs}
Integrating equation (\ref{eq:mass_ODE}) expresses the initial Mach number as an invariant of the problem,
\begin{gather}
\label{eq:integrated_mass}
  \rho \, u = \rho_0 \, u_0 = u_0 = {\cal M}_0 \, ,
\end{gather}
where the initial values $\rho_0 = 1$ and $T_0 = 1$ have been used.
The local Mach number (\ref{eq:local_Mach}) may now be expressed as a function of the material density and temperature:
\begin{gather}
\label{eq:local_Mach_number}
  {\cal M} = \frac{u}{\sqrt{T}} = \frac{{\cal M}_0}{\rho \, \sqrt{T}} \, .
\end{gather}
Integrating the equation of total-momentum conservation (\ref{eq:momentum_ODE}), and using (\ref{eq:integrated_mass}) and (\ref{eq:local_Mach_number}), produces expressions for the material density and temperature as functions of the local Mach number and the radiation pressure:
\begin{subequations}
\label{eqs:material_density_and_temperature}
  \begin{align}
  \nonumber
    \rho \left( {\cal M}, {\cal P} \right)
    & = \frac{{\cal M}_0^2 \left( \gamma \, {\cal M}^2 + 1 \right)}{{\cal M}^2 \left( \gamma \, {\cal M}_0^2 + 1 + \gamma \, P_0 \left( f_{\textrm{\tiny eq}} - {\cal P} \right) \right)} \\
    & = {\cal M}_0^2 \, \frac{g \left( {\cal M} \right)}{h \left( {\cal P} \right)} \, , \label{eq:material_density} \\
    T \left( {\cal M}, {\cal P} \right)
    & = \frac{{\cal M}_0^2}{\rho^2 {\cal M}^2} = \left( \frac{h \left( {\cal P} \right)}{{\cal M}_0 \, g \left( {\cal M} \right)} \right)^2 \, . \label{eq:material_temperature}
  \end{align}
\end{subequations}
As shown in equation (\ref{eq:material_density}), the material density is a separable function of the pair $\left( {\cal M}, {\cal P} \right)$, as indicated by the functions $g \left( {\cal M} \right)$ and $h \left( {\cal P} \right)$, as is the material temperature (\ref{eq:material_temperature}), and therefore the material pressure (\ref{eq:EOS_p}) as well.
Now, every variable in equations (\ref{eqs:nondimensional_RH_ODEs}) can be written as a function of the pair $({\cal M}, {\cal P})$.
We choose to use the pair $\left( {\cal M}, {\cal P} \right)$ as a coordinate basis because integrating in Mach-space is straight-forward in the sense that the integration begins near an equilibrium state and ends before the state at ${\cal M} = 1$ is reached, and because the radiation pressure is a continuous variable across the shock domain.
Equations (\ref{eq:momentum_ODE}) and (\ref{eq:internal_energy_ODE}) can be solved simultaneously for $d\rho / dx$ and $dT / dx$:
\begin{subequations}
  \begin{multline}
    \frac{d \rho}{dx} = \frac{P_0}{T \left( {\cal M}^2 - 1 \right)} \\
                      \times \left[ \frac{d {\cal P}}{dx} - \frac{\left( \gamma - 1 \right)}{{\cal M}_0} \, {\cal C} \, \rho \, S_{\textrm{\tiny rie}} \right] \, , \label{eq:drho_dx}
  \end{multline}
\vspace{-15pt}
  \begin{multline}
    \frac{dT}{dx} = \frac{P_0 \left( \gamma - 1 \right)}{\rho \, \left( {\cal M}^2 - 1 \right)} \\
                   \times \left[\frac{d{\cal P}}{dx} - \frac{\left( \gamma \, {\cal M}^2 - 1 \right)}{{\cal M}_0} \, {\cal C} \, \rho \, S_{\textrm{\tiny rie}} \right] \, , \label{eq:dT_dx}
  \end{multline}
\end{subequations}
where the $\left( \gamma \, {\cal M}^2 - 1 \right)$ term in $dT / dx$ is the source of the isothermal-sonic point discussed by Lowrie and Edwards \cite{LE2008}.
Integrating the total-energy conservation equation (\ref{eq:energy_ODE}), and using (\ref{eq:integrated_mass}), produces:
\begin{multline}
  \frac{d {\cal P}}{dx} = \frac{\sigma_{\textrm{t}} \, {\cal M}_0}{{\cal C} \, P_0} \left[ \frac{T - 1}{\gamma - 1} + \frac{{\cal M}_0^2}{2 \, \rho^2} \left( 1 - \rho^2 \right) \right. \\
  \left. + P_0 \left( \frac{\sigma_{\textrm{t}} \, {\cal P} + \sigma_{\textrm{s}} \, {\cal E} + \sigma_{\textrm{a}} \, T^4}{\rho \, \sigma_{\textrm{t}}} - \frac{4}{3} \right) \right] \, . \label{eq:dP_dx}
\end{multline}
As a reminder, the cross sections, $\sigma_{\textrm{t}} = \sigma_{\textrm{a}} + \sigma_{\textrm{s}}$, are only functions of $\rho$ and $T$, which are both strictly functions of the pair $({\cal M}, {\cal P})$.
Now, the derivative of the local Mach number (\ref{eq:local_Mach_number}), 
\begin{align}
\nonumber
  \frac{d {\cal M}}{dx}
  & = - {\cal M} \left( \frac{1}{\rho} \frac{d \rho}{dx} + \frac{1}{2 \, T} \frac{dT}{dx} \right) \, , \\
\nonumber
  & = - \frac{P_0 \, {\cal M} \left( \gamma + 1 \right)}{2 \, \rho \, T \left( {\cal M}^2 - 1 \right)} \\
  & \times \left[ \frac{d {\cal P}}{dx} - \frac{\left( \gamma - 1 \right) \left( \gamma \, {\cal M}^2 + 1 \right)}{\left( \gamma + 1 \right) {\cal M}_0} \, {\cal C} \, \rho \, S_{\textrm{\tiny rie}} \right] \, , \label{eq:dM_dx}
\end{align}
has a form which is similar to $d\rho / dx$ (\ref{eq:drho_dx}) since the $\left( \gamma \, {\cal M}^2 - 1 \right)$ term from $dT / dx$ (\ref{eq:dT_dx}) has disappeared.
The local Mach number is assumed to be monotonic so that equation (\ref{eq:dM_dx}) can be inverted:
\begin{subequations}
\label{eqs:dxdM_dPdM}
  \begin{gather}
  \label{eq:dx_dM}
    \frac{d x}{d {\cal M}} = - \frac{2 \, \rho \, T}{{\cal M} \left( 2 \, T \, \frac{d \rho}{dx} + \rho \, \frac{d T}{dx} \right)} \, .
  \end{gather}
{\color{black}
There is currently no mathematical proof that ${\cal M}$ is monotonic.
In our experience this assumption holds when producing nonequilibrium-diffusion solutions, but may be violated while producing S$_{\text{n}}$-transport solutions when the value of ${\cal M}$ gets too close to the ASP.
The effect this has on our solution method is discussed in Subsections \ref{subsec:integrating_the_RH_ODEs}, \ref{subsec:continuity_conditions}, and \ref{subsec:solution_procedure}.}
The spatial derivative of the radiation pressure (\ref{eq:dP_dx}) can be multiplied by equation (\ref{eq:dx_dM}) to produce:
  \begin{gather}
  \label{eq:dP_dM}
    \frac{d {\cal P}}{d {\cal M}} = \frac{d {\cal P}}{dx} \frac{dx}{d {\cal M}} = \frac{d {\cal P}}{d {\cal M}} \left( {\cal M}, {\cal P} \right) \, .
  \end{gather}
\end{subequations}
Equations (\ref{eqs:dxdM_dPdM}) are the two simplified ODEs that represent equations (\ref{eqs:nondimensional_RH_ODEs}), in the sense that all of the unknowns associated with equations (\ref{eqs:nondimensional_RH_ODEs}) can be reconstructed from the solutions to equations (\ref{eqs:dxdM_dPdM}).
Of course, once the triplet $({\cal M}, {\cal P}, x)$ is known then all of the other variables may be constructed and the RH solve is considered complete.
\subsection{The Rankine-Hugoniot jump conditions}
\label{subsec:Rankine_Hugoniot_jump_conditions}
As mentioned in Section \ref{sec:problem_statement}, the pre-shock, upstream, equilibrium boundary condition is satisfied as $x \rightarrow - \infty$, and the post-shock, downstream, equilibrium boundary condition is satisfied as $x \rightarrow + \infty$.
We initially choose $\rho_0 = 1$, $T_0 = 1$, ${\cal E}_0 = 1$, and ${\cal P}_0 = 1/3$ for the pre-shock, upstream, equilibrium boundary values, but the post-shock, downstream, equilibrium boundary values must be determined.
As a reminder, the nondimensional equilibrium expressions for ${\cal E}$, ${\cal F}$, ${\cal P}$ and $f$ were given at the end of Section \ref{sec:governing_equations} in expressions (\ref{eqs:equilibrium_EFPf}).

The Rankine-Hugoniot conditions are determined by integrating equations (\ref{eq:mass_ODE})-(\ref{eq:energy_ODE}), while using the upstream and downstream equilibria (\ref{eqs:equilibrium_EFPf}) as the integration boundaries.
The conservation of mass statement (\ref{eq:mass_ODE}) gives
\begin{gather}
\label{eq:integrated_mass_equilibrium}
  \rho_1 \, u_1 = \rho_0 \, u_0 = u_0 = {\cal M}_0 \, ,
\end{gather}
where $\rho_0 = 1$ and $T_0 = 1$ have been used.
Integrating the conservation statements of total momentum (\ref{eq:momentum_ODE}) and total energy (\ref{eq:energy_ODE}) produces two equations that are algebraically similar to each other:
\begin{subequations}
\label{eqs:Rankine_Hugoniot_momentum_and_energy}
\vspace{-20pt}
  \begin{multline}
    3 \, T_1 \, \rho_1^2 + \left[ \gamma \, P_0 \left( T_1^4 - 1 \right) -  3 \, \left( \gamma \, {\cal M}_0^2 + 1 \right) \right] \, \rho_1 \\
    + 3 \, \gamma \, {\cal M}_0^2 = 0 \, ,
  \end{multline}
\vspace{-30pt}
  \begin{multline}
    \left[ 6 \, T_1 - 3 \, \left( \gamma - 1 \right) {\cal M}_0^2 - 6 - 8 \, \left( \gamma - 1 \right) P_0 \right] \rho_1^2 \\
     + 8 \, \left( \gamma - 1 \right) P_0 \, T_1^4 \, \rho_1 + 3 \, \left( \gamma - 1 \right) \, {\cal M}_0^2 = 0 \, .
  \end{multline}
\end{subequations}
Equations (\ref{eqs:Rankine_Hugoniot_momentum_and_energy}) are a $2 \times 2$ system in $\rho_1$ and $T_1$, and do not have a closed-form solution, so a root-solving method is used.
We make no approximations regarding the size of $P_0$, but instead recognize that equations (\ref{eqs:Rankine_Hugoniot_momentum_and_energy}) are quadratic in $\rho_1$, with coefficients that are functions of $T_1$ only:
\begin{subequations}
  \begin{gather}
    a_2(T_1) \rho_1^2 + a_1(T_1) \rho + a_0(T_1) = 0 \, , \\
    b_2(T_1) \rho_1^2 + b_1(T_1) \rho + b_0(T_1) = 0 \, .
  \end{gather}
\end{subequations}
Solving these quadratic equations for $\rho_1$,
\begin{subequations}
\label{eqs:quadratic_solutions}
  \begin{gather}
    \rho_1 = \frac{- a_1(T_1) \pm \sqrt{a_1^2(T_1) - 4 a_2(T_1) a_0(T_1)}}{2 a_2(T_1)} \, , \\
    \rho_1 = \frac{- b_1(T_1) \pm \sqrt{b_1^2(T_1) - 4 b_2(T_1) b_0(T_1)}}{2 b_2(T_1)} \, ,
  \end{gather}
\end{subequations}
\begin{figure}[t!]
  \vspace{-20pt}
  \hspace{-20pt}
  \includegraphics[width = 1.1 \columnwidth]{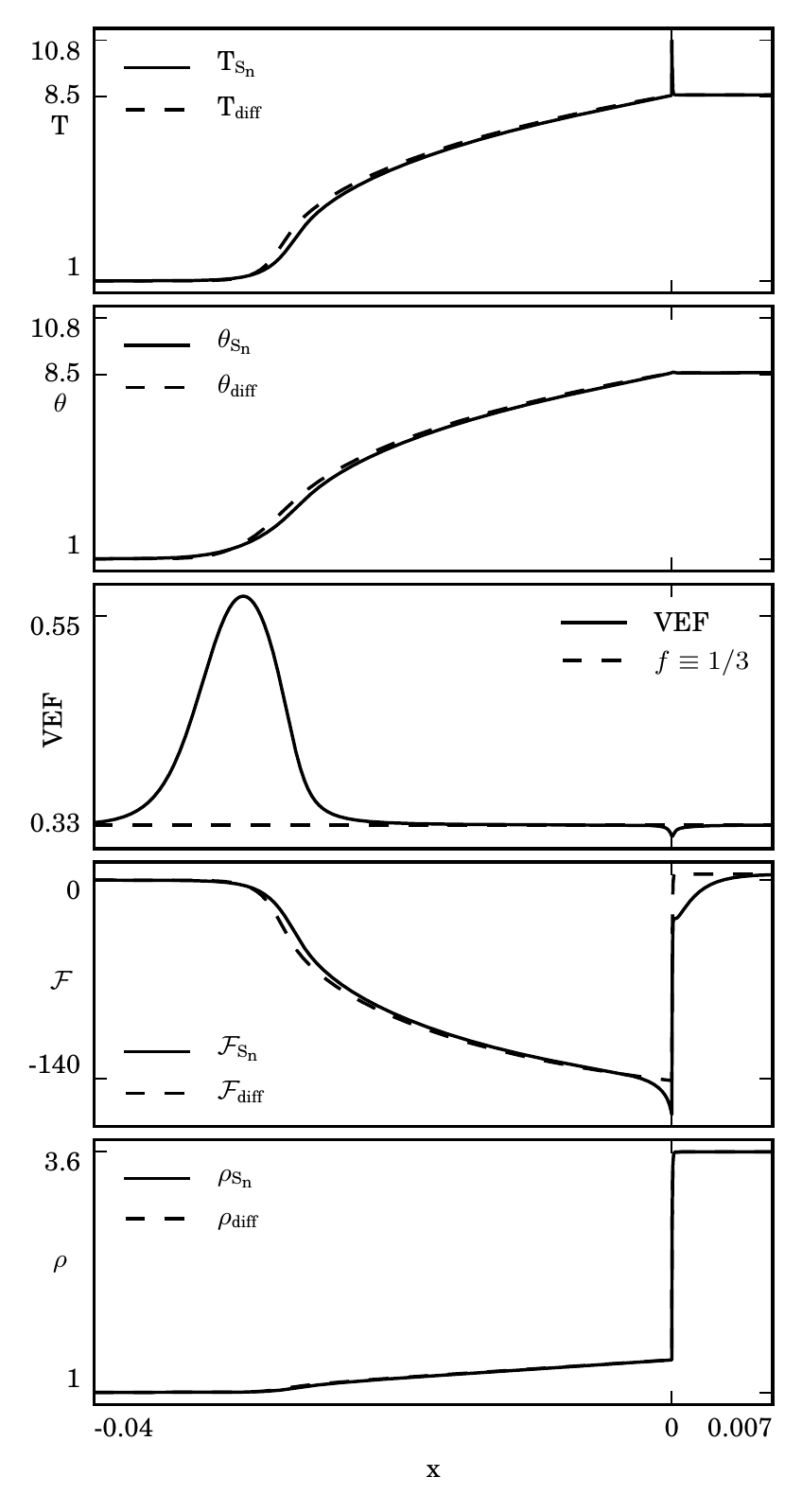}
\caption{
Radiative shock structures for ${\cal M}_0 = 5$ and $\sigma_{\textrm{t}} = 577.35 = \sigma_{\textrm{a}}$, but otherwise the same as in Figure \ref{fig:M3_const}.
The nonequilibrium-diffusion and S$_{\textrm{n}}$-transport solutions for $T$, $\theta$, and $\rho$ are in good agreement, and the VEF is one-third over a long distance in the precursor region and in the relaxation region.
The results for ${\cal F}$ generally agree in the precursor region, but near the embedded hydrodynamic shock ${\cal F}_{\textrm{\tiny S}_{\textrm{\tiny n}}}$ takes a dip not observed for ${\cal F}_{\textrm{\tiny diff}}$, and ${\cal F}_{\textrm{\tiny S}_{\textrm{\tiny n}}}$ also takes longer to relax to its downstream equilibrium state in the relaxation region than ${\cal F}_{\textrm{\tiny diff}}$.
This is due to the contribution of $I(\left| \mu \right| \sim 0)$ to ${\cal F}_{\textrm{\tiny S}_{\textrm{\tiny n}}}$ which is absent for ${\cal F}_{\textrm{\tiny diff}}$.
See the bottom two plots of Figure \ref{fig:M5_polarplots}, and the discussions in Subsections \ref{subsec:describe_radiation_flow} and \ref{subsec:analyze_radiation_flow}.
\label{fig:M5_const}}
\end{figure}
\begin{figure}[t!]
  \vspace{-20pt}
  \hspace{-20pt}
  \includegraphics[width = 1.1 \columnwidth]{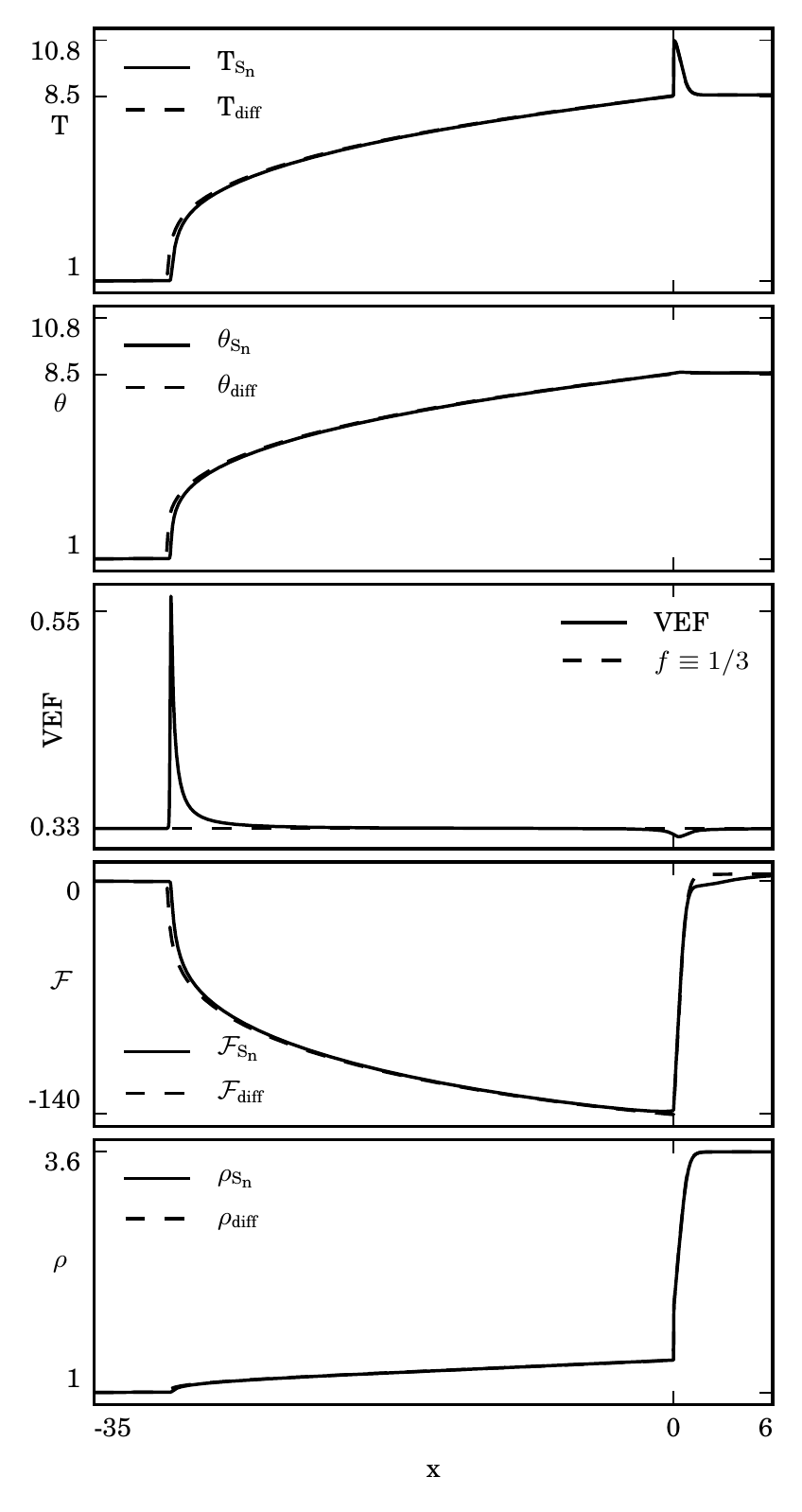}
\caption{
The same as Figure \ref{fig:M5_const}, but with Thomson scattering and Bremsstrahlung emission as given in Figure \ref{fig:M3_Trho}.
The nonequilibrium-diffusion and S$_{\textrm{n}}$-transport solutions for $T$, $\theta$, and $\rho$ are in good agreement.
The VEF is one-third over a long distance in the precursor region and in the relaxation region, and only significantly deviates from one-third over a narrow distance at the front of the upstream precursor region.
The results for ${\cal F}$ generally agree in the precursor region and near the embedded hydrodynamic shock, although ${\cal F}_{\textrm{\tiny S}_{\textrm{\tiny n}}}$ takes longer to relax to its downstream equilibrium state in the relaxation region than ${\cal F}_{\textrm{\tiny diff}}$, for reasons described in the caption to Figure \ref{fig:M5_const}.
\label{fig:M5_Trho}}
\end{figure}
we see that the discriminants of these solutions are functions of $T_1$ only.
To avoid choosing between the positive and negative roots of these solutions, we determine where each discriminant is zero by using a root-solving method, which gives two separate values of $T_1$.
These two values of $T_1$ are used as an initial guess for the values of $\rho_1$ and $T_1$ in equations (\ref{eqs:Rankine_Hugoniot_momentum_and_energy}), which are then solved using a root-solving method.
There are only two solutions to the quadratic equation for $\rho_1$, and we know one solution must exist at $(\rho_0, T_0) = (1,1)$.
Thus, we know the solutions cannot be complex, and if the root-solving method for equations (\ref{eqs:Rankine_Hugoniot_momentum_and_energy}) returns the solution at $(\rho_0, T_0) = (1,1)$ then we increase our initial guess for $\rho_1$ and $T_1$ until we obtain the other solution.
Once we know $\rho_1$ and $T_1$ we can construct the pair $({\cal P}_1, {\cal M}_1)$, and all of the other variables at the downstream equilibrium state as well.
It is worth pointing out that in general ${\cal P}_0 < {\cal P}_1$ and ${\cal M}_0 > 1 > {\cal M}_1$.
\subsection{Linearization near the RH equilibrium states}
\label{subsec:linearization_away_from_equilibrium}
At the equilibrium positions $({\cal M}_0, {\cal P}_0)$ and $({\cal M}_1, {\cal P}_1)$ equations (\ref{eqs:dxdM_dPdM}) evaluate to zero, so the integration cannot proceed from these equilibrium states.
Typically, L'hopital's rule would be used near the equilibrium states, but we found that using the linearization method described in this subsection was simpler, and more robust in the sense that it provided solutions over a larger portion of parameter space.

We assume that there is a region near each equilibrium position in which equations (\ref{eqs:dxdM_dPdM}) are reasonably represented by their linearizations.
Then, given the value of ${\cal M}_{0 \epsilon} = {\cal M}_0 - \epsilon$, where $\epsilon \ll 1$, near the state at $({\cal M}_0, {\cal P}_0)$, and the value of ${\cal M}_{1 \epsilon} = {\cal M}_1 + \epsilon$, near the state at $({\cal M}_1, {\cal P}_1)$, we can determine the appropriate values of ${\cal P}_{0 \epsilon}$ and ${\cal P}_{1 \epsilon}$ associated with ${\cal M}_{0 \epsilon}$ and ${\cal M}_{1 \epsilon}$, respectively.
We use a root-solving method to determine ${\cal P}_{0 \epsilon}$ and ${\cal P}_{1 \epsilon}$ by solving the linear Taylor expansion of equation (\ref{eq:dP_dM}):
\begin{gather}
  {\cal P}_{\textrm{\tiny i}} = {\cal P}_{\tiny i \epsilon} \pm \epsilon \left. \frac{d {\cal P}}{d {\cal M}} \right|_{\left( {\cal M}_{\tiny i \epsilon}, {\cal P}_{\tiny i \epsilon} \right)} \, ,
\end{gather}
where $i = 0$ or $1$, and the sign in front of $\epsilon$ is ``+'' for $i = 0$ and ``-'' for $i = 1$.
Since the problem is shift invariant the values of $x_{\textrm{\tiny i} \epsilon}$ are arbitrary.
\subsection{Integrating the RH ODEs}
\label{subsec:integrating_the_RH_ODEs}
The RH ODEs (\ref{eqs:dxdM_dPdM}) are integrated in Mach-space.
{\color{black} In the precursor region ${\cal M} > 1$ and the integration proceeds from the state $({\cal M}_{0 \epsilon}, {\cal P}_{0 \epsilon})$ to a state $({\cal M}_{\text{\tiny L}}, {\cal P}_{\text{\tiny L}})$ near ${\cal M} = 1$.
In the relaxation region ${\cal M} < 1$ and the integration proceeds from the state $({\cal M}_{1 \epsilon}, {\cal P}_{1 \epsilon})$ to a state $({\cal M}_{\text{\tiny R}}, {\cal P}_{\text{\tiny R}})$ near ${\cal M} = 1$.
The values ${\cal M}_{\text{\tiny L}} = 1 + \epsilon_{\text{\tiny ASP}}$ and ${\cal M}_{\text{\tiny R}} = 1 - \epsilon_{\text{\tiny ASP}}$ represent the integration endpoints, and the subscript-``ASP'' refers to the adiabatic sonic point discussed by Lowrie and Edwards \cite{LE2008}. 
The values of $\epsilon_{\textrm{\tiny ASP}}$ do not have to be the same in the precursor and relaxation regions.
Successful integration of the RH ODEs produces a curve
in the precursor region between the upstream equilibrium state, $({\cal M}_0, {\cal P}_0)$, and the integration endpoint, $({\cal M}_{\textrm{\tiny L}}, {\cal P}_{\textrm{\tiny L}})$,
and a similar curve in the relaxation region between the downstream equilibrium state, $({\cal M}_1, {\cal P}_1)$, and the other integration endpoint, $({\cal M}_{\textrm{\tiny R}}, {\cal P}_{\textrm{\tiny R}})$.
See the top plot in Figure \ref{fig:solution_procedure_PFM}.
In the next subsection we discuss how to transform successful integration curves over the precursor and relaxation regions into the solution for the RH solve.}

{\color{black} When integrating the precursor and relaxation regions there is an assumption that as ${\cal M}_{\text{\tiny L}}$ and ${\cal M}_{\text{\tiny R}}$ approach ${\cal M} = 1$ then the values of ${\cal P}_{\text{\tiny L}}$ and ${\cal P}_{\text{\tiny R}}$ will either approach each other so that $({\cal P}_{\text{\tiny R}} - {\cal P}_{\text{\tiny L}}) / {\cal P}_{\text{\tiny R}} \approx {\cal O}(\epsilon_{\text{\tiny ASP}})$ or overlap so that ${\cal P}_{\text{\tiny L}} > {\cal P}_{\text{\tiny R}}$.}
{\color{black} For some values of ${\cal M}_0$ the assumption that the local Mach number is monotonic is violated in the precursor region, and the integration fails before ${\cal M}$ reaches its original integration endpoint, ${\cal M}_{\text{\tiny L}}$.
When this happens ${\cal M}_{\text{\tiny L}}$ is redefined as the last successful point of integration.
As discussed in Subsection \ref{subsec:solution_procedure}, this can have the effect of causing the solution method to fail since it may happen that neither case, $({\cal P}_{\text{\tiny R}} - {\cal P}_{\text{\tiny L}}) / {\cal P}_{\text{\tiny R}} \approx {\cal O}(\epsilon_{\text{\tiny ASP}})$ nor ${\cal P}_{\text{\tiny L}} > {\cal P}_{\text{\tiny R}}$, occurs.
Once the solution method fails we do not continue trying to construct a converged solution.}
\subsection{Connecting the precursor and relaxation regions}
\label{subsec:continuity_conditions}
Having successfully integrated the RH ODEs (\ref{eqs:dxdM_dPdM}) it is necessary to determine if the solution is continuous in all variables, or if there is an embedded hydrodynamic shock.
{\color{black} This determination is made by comparing the values of ${\cal P}_{\text{\tiny L}}$ and ${\cal P}_{\text{\tiny R}}$.
If $({\cal P}_{\text{\tiny R}} - {\cal P}_{\text{\tiny L}}) / {\cal P}_{\text{\tiny R}} < \epsilon_{\text{\tiny tol}}$, where $\epsilon_{\text{\tiny tol}} \ll 1$ is of the same order as $\epsilon_{\text{\tiny ASP}}$, then the solution is continuous.
However, if ${\cal P}_{\textrm{\tiny L}} > {\cal P}_{\textrm{\tiny R}}$ then the integration curves overlap and the solution contains an embedded hydrodynamic shock.
We discuss below how the curves are modified to produce the solution.
See Figures \ref{fig:solution_procedure_Tx} and \ref{fig:solution_procedure_PFM}.}

In the case of a continuous solution we need to determine the value of ${\cal P}$ associated with ${\cal M} = 1$, and we need to shift the $x$-values in the precursor and relaxation regions so that $x = 0$ at ${\cal M} = 1$.
In order to determine the value of ${\cal P}$ at ${\cal M} = 1$ we linearly interpolate between the two states $({\cal M}_{\textrm{\tiny L}}, {\cal P}_{\textrm{\tiny L}})$ and $({\cal M}_{\textrm{\tiny R}}, {\cal P}_{\textrm{\tiny R}})$.
In order to shift the $x$-values correctly we determine the appropriate values of $x_{\textrm{\tiny L}} < 0$ and $x_{\textrm{\tiny R}} > 0$ by using the following relations:
\begin{subequations}
  \begin{gather}
    x_{\textrm{\tiny L}} = \epsilon_{\textrm{\tiny ASP}} \left. \frac{d x}{d {\cal M}} \right|_{\textrm{\tiny L}} \, , \\
    x_{\textrm{\tiny R}} = - \epsilon_{\textrm{\tiny ASP}} \left. \frac{d x}{d {\cal M}} \right|_{\textrm{\tiny R}} \, ,
  \end{gather}
\end{subequations}
which are accurate to ${\cal O}(\epsilon_{\textrm{\tiny ASP}}^2)$.
{\color{black} The $x$-values in the precursor region are now shifted so as to match the value of $x_{\text{\tiny L}}$ at ${\cal M}_{\text{\tiny L}}$, and the $x$-values in the relaxation region are similarly shifted to match the value of $x_{\text{\tiny R}}$ at ${\cal M}_{\text{\tiny R}}$.
Once this is done the triplet of points $({\cal P}, {\cal M}, x)$ is known and the continuous solution has been created.}

In the case of an embedded hydrodynamic shock we enforce continuity of the radiation flux and radiation pressure in order to determine the values of ${\cal F}_{\textrm{\tiny ps}}$ and ${\cal P}_{\textrm{\tiny ps}}$ at the embedded hydrodynamic shock, where ${\cal P}_{\textrm{\tiny R}} < {\cal P}_{\textrm{\tiny PS}} < {\cal P}_{\textrm{\tiny L}}$.
See Figure \ref{fig:solution_procedure_PFM}.
Then, we determine the value of ${\cal M}_{\textrm{\tiny p}}$ associated with ${\cal P}_{\textrm{\tiny ps}}$ in the precursor region, and the value of ${\cal M}_{\textrm{\tiny s}}$ associated with ${\cal P}_{\textrm{\tiny ps}}$ in the relaxation region.
The precursor {\color{black} solution} is now defined as existing between the equilibrium state at $({\cal M}_0, {\cal P}_0)$ and the downstream precursor state at $({\cal M}_{\textrm{\tiny p}}, {\cal P}_{\textrm{\tiny ps}})$, and the relaxation {\color{black} solution} is now defined as existing between the equilibrium state at $({\cal M}_1, {\cal P}_1)$ and the upstream relaxation state $({\cal M}_{\textrm{\tiny s}}, {\cal P}_{\textrm{\tiny ps}})$.
{\color{black} Those portions of the curves between $({\cal M}_{\text{\tiny p}}, {\cal P}_{\text{\tiny ps}})$ and $({\cal M}_{\text{\tiny L}}, {\cal P}_{\text{\tiny L}})$ in the precursor region, and $({\cal M}_{\text{\tiny s}}, {\cal P}_{\text{\tiny ps}})$ and $({\cal M}_{\text{\tiny R}}, {\cal P}_{\text{\tiny R}})$ in the relaxation region, are discarded.
The $x$-values in the precursor and relaxation regions are shifted by placing $x = 0$ at the embedded hydrodynamic shock.
Once this is done the triplet of points $({\cal P}, {\cal M}, x)$ is known and the solution containing an embedded hydrodynamic shock has been created.}

Continuity of the radiation variables is required because they are the angular moments of the radiation intensity, which itself is assumed to be a continuous solution of the 1D, planar, Boltzmann RT equation.
Similarly, we know that mass, momentum and energy are conserved when connecting the precursor and relaxation regions because they represent the velocity moments for the Maxwell-Boltzmann distribution function of material particles.
Inspection of equations (\ref{eqs:nondimensional_RH_ODEs}) shows that continuity of the radiation pressure implies continuity of the material-momentum flux, and continuity of the radiation flux implies continuity of the material-energy flux.
After integrating the total-momentum conservation equation (\ref{eq:momentum_ODE}) across the embedded hydrodynamic shock, imposing continuity of the radiation flux (\ref{eq:nondimensional_radiation_flux}) and the radiation pressure, and using equations (\ref{eq:integrated_mass}) and (\ref{eqs:material_density_and_temperature}), a function that must be satisfied as it crosses from state-``p'' to state-``s'' is produced \cite{Lamb1945}:
\begin{gather}
  \left. \frac{{\cal M}^2 \left[ \left( \gamma - 1 \right) {\cal M}^2 + 2 \right]}{\left( \gamma {\cal M}^2 + 1 \right)^2} \right|_{\textrm{\tiny p}}^{\textrm{\tiny s}} = 0 \, .
\end{gather}
This expression provides a convenient check that continuity of the radiation flux and the radiation pressure has been properly imposed across the embedded hydrodynamic shock.
\begin{figure}[t!]
  \vspace{-30pt}
  \hspace{-15pt}
  \includegraphics[width = 1.1 \columnwidth]{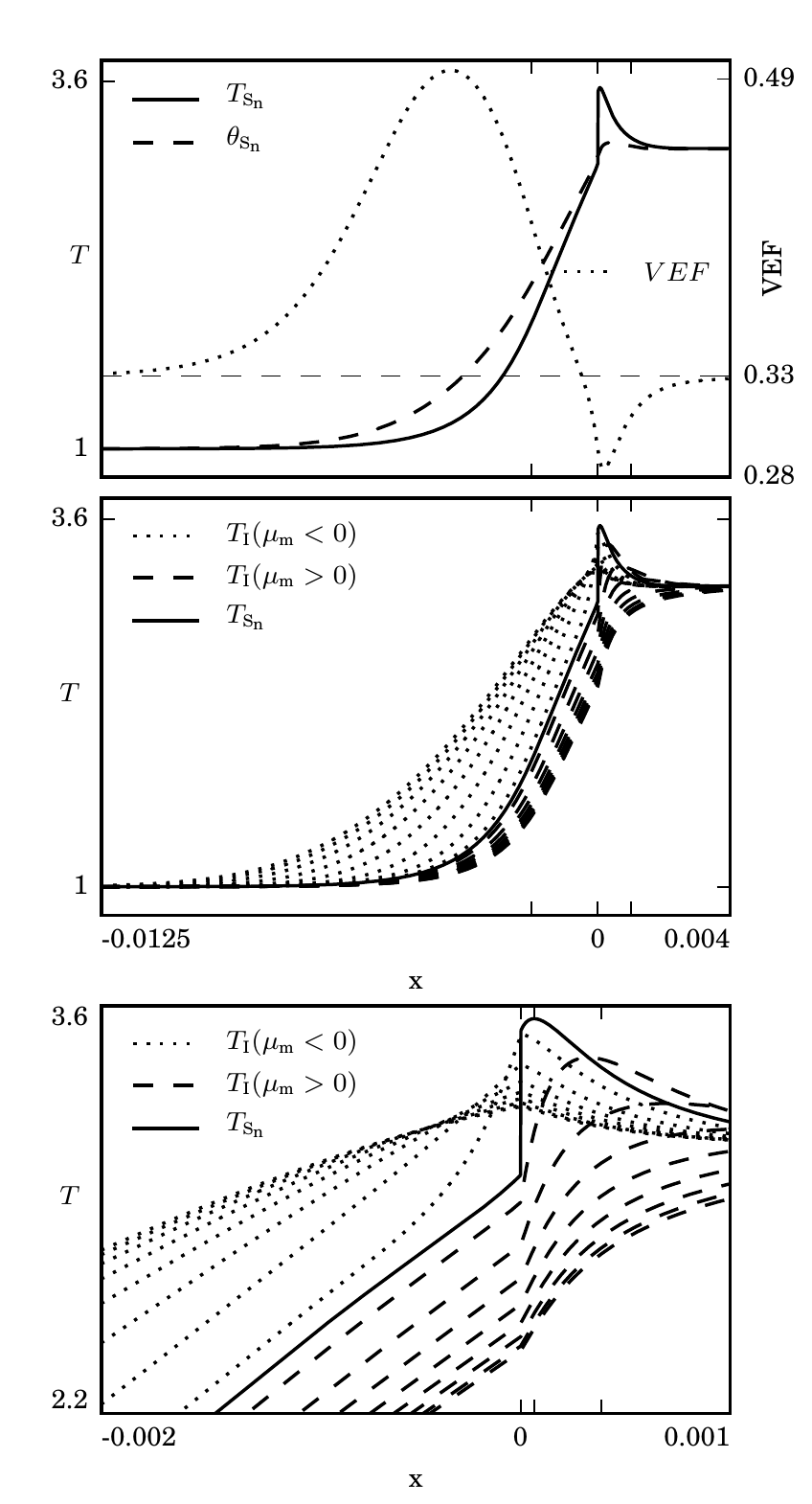}
\caption{
The same values as given in Figure \ref{fig:M1p05_Sn} but for ${\cal M}_0 = 2.7$.
The S$_{\textrm{n}}$ solutions for $T$, $\theta$ and the VEF are shown together in the top plot.
In order to compare the radiation intensity to the material temperature, we define a new variable, $T_{\textrm{\tiny I}} (\mu) = [4 \pi I(\mu)]^{1/4}$, which we refer to as the intensity temperature, where $I$ is given in equation (\ref{eq:analytic_radiation_intensity}), as explained at the end of Subsection \ref{subsec:analytic_solution_for_I}.
The S$_{16}$ intensity temperature solutions, $T_{I}(\mu_{\textrm{\tiny m}})$, are shown in the middle plot, along with $T_{\textrm{\tiny S}_{\textrm{\tiny n}}}$ to serve as a fiducial curve.
The bottom plot is a zoomed-in view of the middle plot around the embedded hydrodynamic shock.
The S$_{16}$ curves nearest the curve for $T_{\textrm{\tiny S}_{\textrm{\tiny n}}}$, in the precursor region, are most affected by the embedded hydrodynamic shock.
The intensity temperatures are continuous across the embedded hydrodynamic shock, but their derivatives are not.
The tickmarks in the bottom plot correspond to the locations where $T_{\textrm{\tiny max}}$ and $\theta_{\textrm{\tiny max}}$ occur.
See Subsections \ref{subsec:describe_radiation_flow} and \ref{subsec:analyze_radiation_flow}.
\label{fig:M2p7_with_intensities_zoomed}}
\end{figure}
\begin{figure}[t!]
  \vspace{-20pt}
  \hspace{-20pt}
  \includegraphics[width = 1.1 \columnwidth]{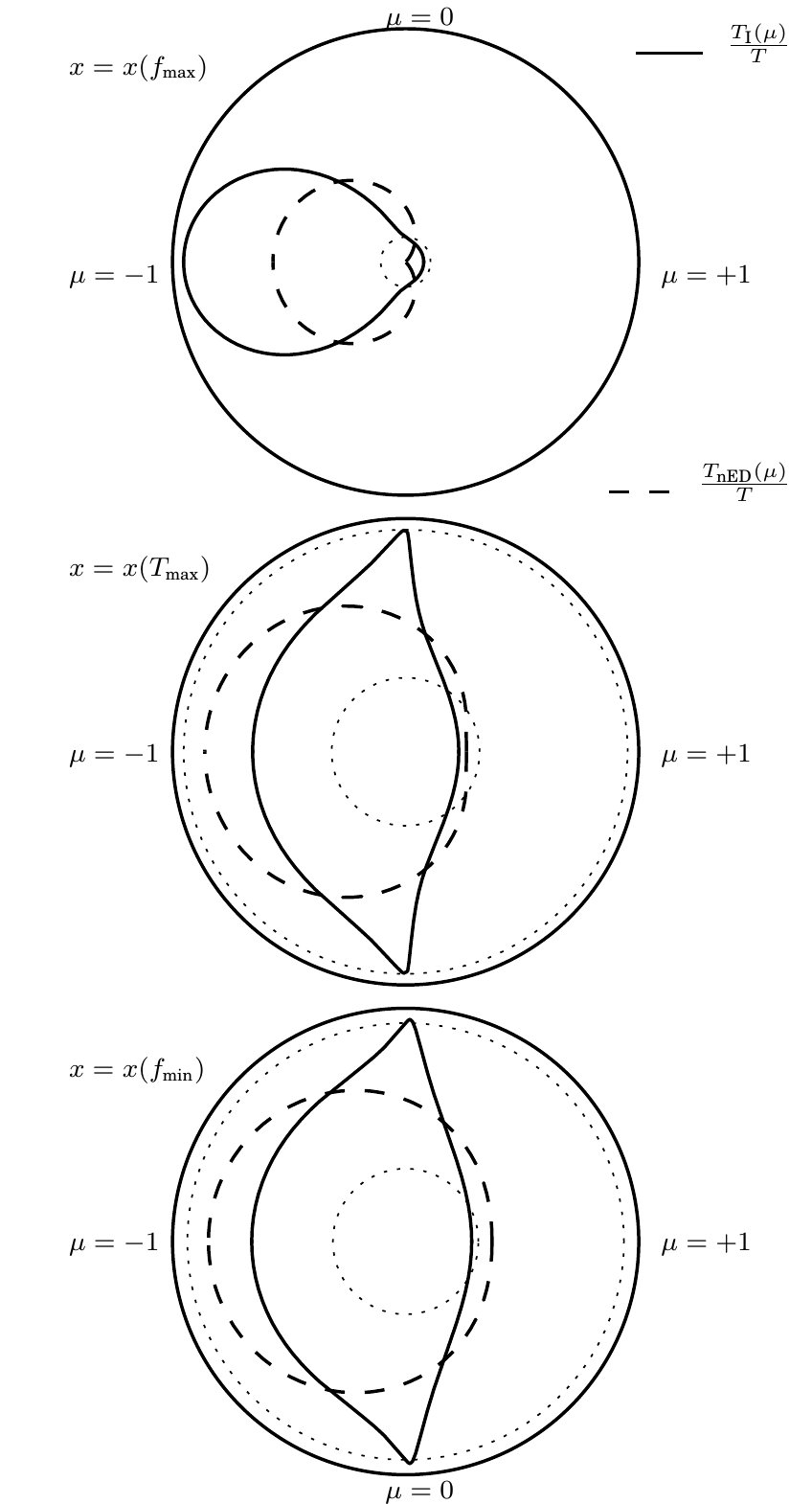}
\caption{
The same values given in Figure \ref{fig:M2p7_with_intensities_zoomed}.
Polar plots, in angle, comparing the normalized intensity temperature, $T_{\textrm{\tiny I}} / T$ (solid lines), with the normalized diffusion temperature, $T_{\textrm{\tiny nED}} / T$ (dashed lines), as described in Subsections \ref{subsec:analytic_solution_for_I} and \ref{subsec:simplified_radiation_diffusion_models}, at the $x$-locations where $f_{\textrm{\tiny max}}$, $T_{\textrm{\tiny max}}$, and $f_{\textrm{\tiny min}}$ occur, in descending order, respectively.
In the center of the top plot, the dotted circle has unit radius, while in the bottom two plots the outer dotted circle has unit radius and the inner dotted circle has radius one-third.
For most of the precursor region the diffusion temperature, $T_{\textrm{\tiny nED}}$, underestimates the value of the intensity temperature, $T_{\textrm{\tiny I}}$, along $\mu \sim \pm 1$, setting the value nearly to zero along $\mu = 1$, and overestimates its value along $\left| \mu \right| \sim 0$, as seen in the top plot.
The opposite occurs in the relaxation region, as seen in the bottom two plots.
See Subsections \ref{subsec:analytic_solution_for_I}, \ref{subsec:describe_radiation_flow}, and \ref{subsec:analyze_radiation_flow}.
\label{fig:M2p7_polarplots}}
\end{figure}
\subsection{The ODEs for the RT solve}
\label{subsec:SN_ODE}
The RT equation (\ref{eq:transport_equation}) can be directionally discretized along n directions, such that $\mu \mapsto \mu_{\textrm{\tiny m}}$, where m is an integer between 1 and n.
This is called the discrete-ordinates in angle (``S$_{\textrm{n}}$'') method.
These discrete directions, $\mu_{\textrm{\tiny m}}$, can be chosen as Gauss-Legendre quadrature roots with associated weights, $w_{\textrm{\tiny m}}$, which allows exact quadrature integration of a polynomial of order $2 n - 1$. 
By discretizing the direction variable, the S$_{\textrm{n}}$-transport method also directionally discretizes the radiation intensity, $I(\mu) \mapsto I_{\textrm{\tiny m}} \equiv I(\mu_{\textrm{\tiny m}})$, and generates n directionally-discretized RT ODEs:
\begin{multline}
\label{eq:Sn_transport_broken}
  \mu_{\textrm{\tiny m}} \, \frac{d I_{\textrm{\tiny m}}}{dx} = - \sigma_{\textrm{t}} \, I_{\textrm{\tiny m}} + \frac{\sigma_{\textrm{s}}}{4 \pi} \, {\cal E} + \frac{\sigma_{\textrm{a}}}{4 \pi} \, T^4 \\
  - 2 \, \frac{\sigma_{\textrm{s}}}{4 \pi} \, \beta \, {\cal F} + \beta \, \mu_{\textrm{\tiny m}} \left( \sigma_{\textrm{t}} \, I_{\textrm{\tiny m}} + \frac{3 \sigma_{\textrm{s}}}{4 \pi} \, {\cal E} + \frac{3 \sigma_{\textrm{a}}}{4 \pi} \, T^4 \right) \\
  + \frac{1}{\pi} \left[ \beta^2 \left( 2 \, \sigma_{\textrm{s}} - 3 \, \sigma_{\textrm{t}} \, \mu_{\textrm{\tiny m}}^2 \right) {\cal P} \right]_{\textrm{\tiny eq}} \, .
\end{multline}
Now, $I_{\textrm{\tiny m}}$ is the directionally-discrete radiation intensity for which equation (\ref{eq:Sn_transport_broken}) is solved.
All other terms on the right-hand side are assumed to be known from the RH solve, which is described in Subsections \ref{subsec:nondimensional_EOS} - \ref{subsec:continuity_conditions}.
As a reminder, the equilibrium term on the last line ensures that when the RT ODEs are evaluated at an equilibrium state they are identically zero.
When the RT ODEs are evaluated anywhere else along the shock, if $x \leq 0$ then the equilibrium term is evaluated using the upstream equilibrium values, and if $x > 0$ then it is evaluated using the downstream equilibrium values.
Additionally, for the sake of computational speed and accuracy, we derive the Jacobian of the RT ODEs (\ref{eq:Sn_transport_broken}):
  \begin{gather}
    \frac{d}{d I_{\textrm{\tiny m}}} \frac{d I_{\textrm{\tiny m}}}{d x} = - \frac{\sigma_{\textrm{t}}}{\mu_{\textrm{\tiny m}}} + \beta \, \sigma_{\textrm{t}} \, .
  \end{gather}
where only the terms containing $I_{\textrm{\tiny m}}$ contribute to the Jacobian because all of the other terms are assumed to be known from the RH solve.
\subsection{The initial values for the RT solution, and linearizing away from the initial values}
\label{subsec:RT_initial_values_and_linearization}
The initial values for the radiation intensities in equilibrium are determined from equations (\ref{eqs:equilibrium_EFPf}).
If $\mu_{\textrm{\tiny m}} > 0$ then the equilibrium values in those expressions are taken from the upstream equilibrium state, $({\cal M}_0, {\cal P}_0)$, and if $\mu_{\textrm{\tiny m}} < 0$ then those values are taken from the downstream equilibrium state, $({\cal M}_1, {\cal P}_1)$.
To begin integrating the RT ODEs (\ref{eq:Sn_transport_broken}) each RT solution must be moved away from the equilibrium state.
We do this by using a linearization procedure analogous to what is done in Subsection \ref{subsec:linearization_away_from_equilibrium}.
Given a position near either equilibrium state, $x_{\epsilon} = x_{\textrm{\tiny eq}} + \epsilon$, where ${\textrm{sign}}(\epsilon) \equiv {\textrm{sign}}(\mu_{\textrm{\tiny m}})$, we use a root-solving method to determine $I_{\textrm{\tiny m}, \epsilon}$ by solving the linear Taylor expansion of the RT ODEs (\ref{eq:Sn_transport_broken}):
  \begin{gather}
  \label{eq:linearized_transport}
    I_{\textrm{\tiny m, eq}} = I_{\textrm{\tiny m}, \epsilon} + \left. \epsilon \frac{dI_{\textrm{\tiny m}}}{d x} \right|_{x_{\epsilon}, I_{\textrm{\tiny m}, \epsilon}} \, .
  \end{gather}
The state $(x_{\epsilon}, I_{\textrm{\tiny m}, \epsilon})$ is an appropriate starting point for the integration of the RT ODEs (\ref{eq:Sn_transport_broken}).
Since each ODE represents an initial-value problem, integrating them across the shock's spatial domain causes them to naturally arrive at their other equilibrium boundary.
\subsection{Constructing the RT solutions}
\label{subsec:construct_RT_solutions}
Integration of the n separate RT ODEs (\ref{eq:Sn_transport_broken}) produces n directionally-discrete radiation-intensity solutions, $I_{\textrm{\tiny m}}$, corresponding to the n directionally-discrete coordinates, $\mu_{\textrm{\tiny m}}$.
Quadrature integration of the first three angular moments of these radiation intensities, which were originally defined in equations (\ref{eq:nondimensional_radiation_variable_definitions}) for the continuous variable $\mu$, produces new values of the radiation energy density, radiation flux, and radiation pressure,
\begin{subequations}
\label{eq:transport_RH}
  \begin{align}
    {\cal E} & \approx 2 \, \pi \sum_{\textrm{\tiny m} = 1}^{\textrm{\tiny n}} w_{\textrm{\tiny m}} \, I_{\textrm{\tiny m}} \label{eq:quadrature_E} \, , \\
    {\cal F} & \approx 2 \, \pi \sum_{\textrm{\tiny m} = 1}^{\textrm{\tiny n}} \mu_{\textrm{\tiny m}} \, w_{\textrm{\tiny m}} \, I_{\textrm{\tiny m}} \label{eq:quadrature_F} \, , \\
    {\cal P} & \approx 2 \, \pi \sum_{\textrm{\tiny m} = 1}^{\textrm{\tiny n}} \mu_{\textrm{\tiny m}}^2 \, w_{\textrm{\tiny m}} \, I_{\textrm{\tiny m}} \label{eq:quadrature_P} \, ,
  \end{align}
respectively, from which the new VEF is then constructed:
  \begin{gather}
    f = \frac{{\cal P}}{{\cal E}} = \frac{\sum_{\textrm{\tiny m} = 1}^{\textrm{\tiny n}} \mu_{\textrm{\tiny m}}^2 \, w_{\textrm{\tiny m}} \, I_{\textrm{\tiny m}}}{\sum_{\textrm{\tiny m} = 1}^{\textrm{\tiny n}} w_{\textrm{\tiny m}} \, I_{\textrm{\tiny m}}} \label{eq:VEF_Sn} \, .
  \end{gather}
\end{subequations}
This new VEF is used in the next RH solve to reconstruct the radiative-shock structure.
\begin{figure}[t!]
  \vspace{-30pt}
  \hspace{-20pt}
  \includegraphics[width = 1.1 \columnwidth]{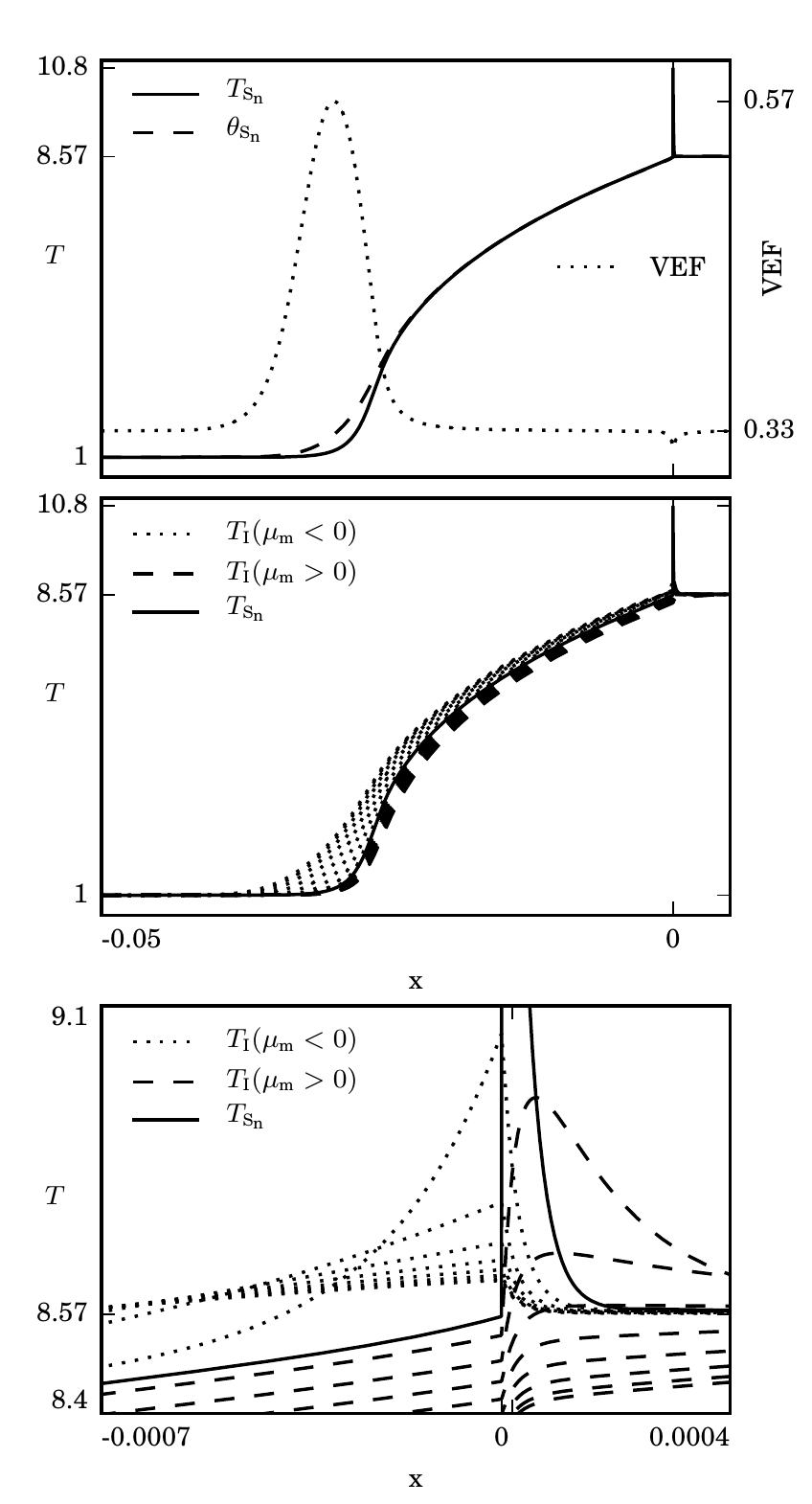}
\caption{
The same conditions as given in Figure \ref{fig:M1p05_Sn}, but for ${\cal M}_0 = 5$.
The S$_{\textrm{n}}$ solutions for $T$, $\theta$ and the VEF are shown together in the top plot.
The S$_{16}$ intensity temperature solutions, $T_{\textrm{\tiny I}}(\mu_{\textrm{\tiny m}})$, are shown in the middle plot, along with $T_{\textrm{\tiny S}_{\textrm{\tiny n}}}$ as a fiducial curve.
The bottom plot shows the same solutions zoomed-in around the embedded hydrodynamic shock.
The S$_{16}$ curves nearest the $T_{\textrm{\tiny S}_{\textrm{\tiny n}}}$ curve, in the precursor region, are most affected by the embedded hydrodynamic shock, where the intensity temperatures are continuous but their derivatives are not.
Despite how narrow the Zel'dovich spike is, it does significantly affect the radiation intensities over a distance larger than its own thickness, especially those intensity temperatures traveling along $\left| \mu \right| \sim 0$.
See also Figure \ref{fig:M5_polarplots}, and Subsections \ref{subsec:describe_radiation_flow} and \ref{subsec:analyze_radiation_flow}.
\label{fig:M5_with_intensities_zoomed}}
\end{figure}
\begin{figure}[t!]
  \vspace{-20pt}
  \hspace{-20pt}
  \includegraphics[width = 1.1 \columnwidth]{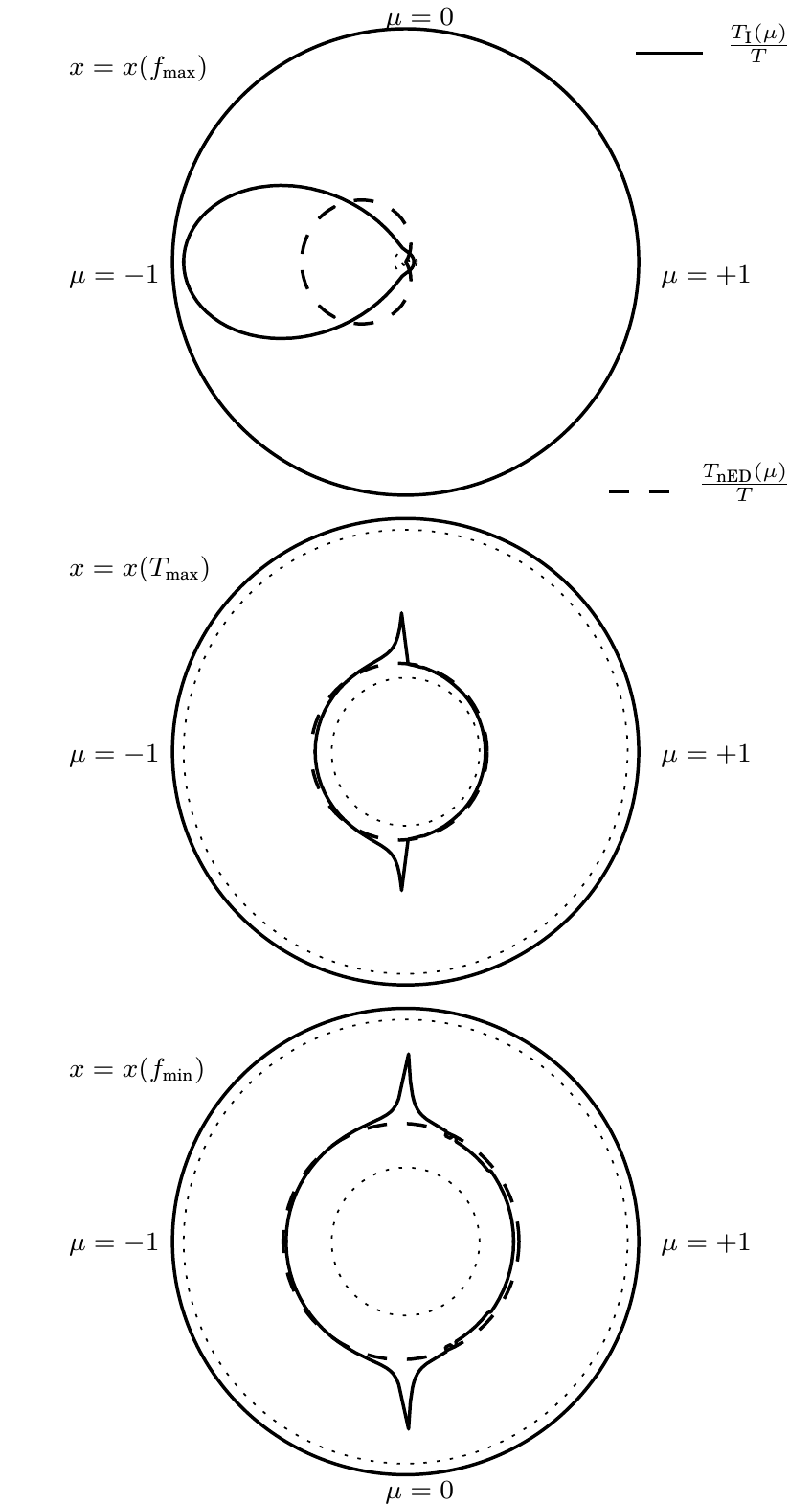}
\caption{
The same values as given in Figure \ref{fig:M5_with_intensities_zoomed} (${\cal M}_0 = 5$), at the x-locations where $f_{\textrm{\tiny max}}$, $T_{\textrm{\tiny max}}$, and $f_{\textrm{\tiny min}}$ occur, in descending order, respectively, but with the same dotted lines as in Figure \ref{fig:M2p7_polarplots}.
The top plot provides an example of what is meant by saying that the radiation is forward-peaked.
In the middle plot, the radiation flowing along $\left| \mu \right| \sim 0$ is strictly from the $\mu < 0$ directions, which is being emitted from the Zel'dovich spike and traveling into the upstream precursor region.
Further into the relaxation region, in the bottom plot, the radiation flow along $\left| \mu \right| \sim 0$ is fairly equal.
Thus, while the radiation flow is not isotropic at this location it is symmetric.
See Subsections \ref{subsec:describe_radiation_flow} and \ref{subsec:analyze_radiation_flow}.
\label{fig:M5_polarplots}}
\end{figure}
\subsection{Semi-analytic solution procedure}
\label{subsec:solution_procedure}
The solution procedure is a two-step iteration method: 1) the RH ODEs (\ref{eqs:dxdM_dPdM}) to be integrated represent the ``RH solve'', and 2) the RT ODEs (\ref{eq:Sn_transport_broken}) to be integrated represent the ``RT solve.''
After each RH solve and before an RT solve, we find it helpful to increase the spatial resolution near the equilibrium regions.
This is because the relaxation length for the radiation intensities is longer than the relaxation length for all other variables.
At the beginning of the next RH solve, we use the radiation pressure (\ref{eq:quadrature_P}) from the previous RT solve and the value of ${\cal M}_{\epsilon}$ near the equilibrium state to determine the new value of ${\cal P}_{\epsilon}$.
Occasionally we find that this initialization leads to a nonphysical solution in the relaxation region, such that the radiation pressure inappropriately grows exponentially as ${\cal M} = 1$ is approached.
When this happens we slowly decrease the value of ${\cal P}_{\epsilon}$ until the nonphysical solution disappears.

The two-step procedure defines one iteration.
At the end of each iteration, $L_2$ and relative $L_{\infty}$ norms for the differences between the RH and RT solutions for the radiation variables are computed, as are $L_2$ and relative $L_{\infty}$ norms for the differences between the VEFs between iterations.
For the plots shown, the VEF relative difference is less than $10^{-4}$: $\left| f_{\textrm{new}} - f_{\textrm{old}} \right| / f_{\textrm{new}} < 10^{-4}$.

Figure \ref{fig:solution_procedure_Tx} illustrates the general method for constructing a radiative-shock solution for the RH solve.
Figure \ref{fig:solution_procedure_PFM} illustrates the method of enforcing continuity of the radiation pressure and radiation flux for the same radiative-shock solution.
{\color{black}
We find that for a small range of initial Mach numbers, ${\cal M}_0 = 3.37-4.12$, our solution method fails.
This region is shown in Figure \ref{fig:TvM} as existing between the vertical dashed lines.
The plotted values in this region are taken from the failed solution's last completed iteration.
The reason for this failure was touched on in Subsection \ref{subsec:integrating_the_RH_ODEs}.
Our solution method relies on the assumption that ${\cal M}$ is monotonic across the shock's structure.
Our experience is that this assumption is valid when producing nonequilibrium-diffusion solutions which use a constant Eddington factor.
However, when producing S$_{\text{\tiny n}}$-transport solutions our experience is that ${\cal M}$ may become nonmonotonic while integrating the precursor region at a value that is quite far from ${\cal M}_{\text{\tiny L}}$, causing the integrator to stop.
When the integrator stops we define the last successfully integrated value of ${\cal M}$ as ${\cal M}_{\text{\tiny L}}$, and the associated value of ${\cal P}$ as ${\cal P}_{\text{\tiny L}}$.
What matters to the success of the integration method is whether this value of ${\cal P}_{\text{\tiny L}}$ meets the criteria for the solution to either be continuous or to contain an embedded hydrodynamic shock, as defined in Subsection \ref{subsec:continuity_conditions}.
If neither criterion is met then the solution method fails.}

While we are able to obtain solutions for values of ${\cal M}_0 > 5$, they generally have the same features as seen for the ${\cal M}_0 = 5$ results presented in the next section.
\subsection{Analytic solution for the radiation intensity}
\label{subsec:analytic_solution_for_I}
The RT solve depends on the directionally-discretized RT ODEs (\ref{eq:Sn_transport_broken}).
However, the RT equation (\ref{eq:transport_equation}) is continuous in angle, and once the shock structure is known this equation can be solved analytically \cite{Chandrasekhar1960, MM1999}.
For notational convenience we rewrite equation (\ref{eq:transport_equation}) with all terms involving the radiation intensity moved to the left-hand side, and all other terms collected on the right-hand side and denoted by $Q$:
\begin{subequations}
  \begin{gather}
    \mu \, \frac{d I}{dx} + \sigma_{\textrm{t}} \left( 1 - \beta \, \mu \right) I = Q \, . \label{eq:dIdx}
  \end{gather}
Equation (\ref{eq:dIdx}) has the solution \cite{Chandrasekhar1960, MM1999}
  \begin{multline}
  \label{eq:analytic_radiation_intensity}
    I( x, \, \mu ) = I_{\textrm{\tiny eq}}( x_{\textrm{\tiny eq}}, \, \mu ) e^{- \tau \left( x_{\textrm{\tiny eq}}, \, x \right)} \\
                            + \int_{x_{\textrm{\tiny eq}}}^x \frac{Q( z, \, \mu )}{\mu} \, e^{- \tau \left( z, \, x \right)} \, dz \, ,
  \end{multline}
where
  \begin{gather}
    \tau \left( x^{\prime}, \, x^{\prime \prime} \right) = \int_{x^{\prime}}^{x^{\prime \prime}} \frac{\sigma_{\textrm{t}}( y )}{\mu} \, \left[ 1 - \beta( y ) \mu \right] \, dy \, . \label{eq:tau_z_x}
  \end{gather}
\end{subequations}
The terms with subscript-``eq''are to be evaluated at the equilibrium state.
If $\mu < 0$ then this is the downstream, post-shock, equilibrium state where $T_{\textrm{\tiny eq}} = T_1$, $\beta_{\textrm{\tiny eq}} = u_1 / {\cal C}$, and $x_{\textrm{\tiny eq}} \gg 0$; if $\mu > 0$ then this is the upstream, pre-shock, equilibrium state where $T_{\textrm{\tiny eq}} = 1$, $\beta_{\textrm{\tiny eq}} = u_0 / {\cal C} = {\cal M}_0 / {\cal C}$, and $x_{\textrm{\tiny eq}} \ll 0$.
The first term in the solution (\ref{eq:analytic_radiation_intensity}) represents the contribution from the radiation intensity at equilibrium which exponentially decays as it moves from the equilibrium state, $x_{\textrm{\tiny eq}}$, to the state at point $x$.
The integral term in the solution (\ref{eq:analytic_radiation_intensity}) represents the directionally-modified contribution of radiation emitted and absorbed by the material at all points $z$ between $x_{\textrm{\tiny eq}}$ and $x$, and which exponentially decays as it moves through the material from point $z$ to point $x$.
The expressions for $\tau$ (\ref{eq:tau_z_x}) describes how the radiation exponentially decays as it flows through the material, from either of points $x_{\textrm{\tiny eq}}$ or $z$ to the point $x$.

In this paper, we use the analytic solution of the radiation intensity (\ref{eq:analytic_radiation_intensity}) to understand the structure of the S$_{\textrm{n}}$-transport solutions, as discussed at the end of Subsection \ref{subsec:describe_radiation_flow}, and to produce directionally-dependent polar plots of the radiation-intensity solution at specified locations along the shock structure; see Figures \ref{fig:M2p7_polarplots} and \ref{fig:M5_polarplots}, which are described in Subsections \ref{subsec:describe_radiation_flow} and \ref{subsec:analyze_radiation_flow}.
In these figures, in order to compare the analytic radiation intensity (\ref{eq:analytic_radiation_intensity}) to the material temperature, we define a new variable, $T_{\textrm{\tiny I}} = [4 \pi I(\mu)]^{1/4}$, which we call the intensity temperature.
\subsection{The nonequilibrium-diffusion radiation model}
\label{subsec:simplified_radiation_diffusion_models}
Due to the complexity of solving the RT equation (\ref{eq:transport_equation}) model simplifications have been developed which attempt to accurately mimic RT solutions but at a lower computational cost.
The nonequilibrium-diffusion radiation model is commonly used for physical problems when the material system is optically thick.
Nonequilibrium-diffusion assumes that the radiation intensity is linearly anisotropic,
\begin{subequations}
\begin{gather}
\label{eq:nonequilibrium_diffusion_I}
  I( \mu ) = \frac{1}{4 \pi} \left( {\cal E} + 3 \, \mu \, {\cal F} \right) \, ,
\end{gather}
where,
\begin{gather}
  {\cal E} = \theta^4 \, ,
\end{gather}
\vspace{-25pt}
\begin{multline}
  {\cal F} = - \frac{1}{3 \, \sigma_{\textrm{t}}} \frac{d {\cal E}}{dx} \\
  + \frac{1}{\sigma_{\textrm{t}}} \, \beta \, \left( \frac{1}{3} \, \sigma_{\textrm{t}} \, {\cal E} + \sigma_{\textrm{s}} \, {\cal E} + \sigma_{\textrm{a}} \, T^4 \right) \, ,
\end{multline}
\end{subequations}
and the radiation temperature, $\theta$, is not in thermal equilibrium with the material temperature, $T$.
The first three angular moments of equation (\ref{eq:nonequilibrium_diffusion_I}) are consistent with the definitions for ${\cal E}$, ${\cal F}$, and the Eddington approximation, respectively,
\begin{subequations}
  \begin{gather}
    {\cal E} \equiv 2 \, \pi \int_{-1}^1 I( \mu ) \, d\mu = {\cal E} \, , \\
    {\cal F} \equiv 2 \, \pi \int_{-1}^1 \mu \, I( \mu ) \, d\mu = {\cal F} \, , \\
    {\cal P} \equiv 2 \, \pi \int_{-1}^1 \mu^2 \, I( \mu ) \, d\mu = \frac{{\cal E}}{3} \, .
  \end{gather}
\end{subequations}
In this way, the nonequilibrium-diffusion model provides an approximate description of the radiation flow which assumes that the Eddington factor is a constant value of one-third.
Comparisons of the analytic directionally-dependent radiation-transport solutions (\ref{eq:analytic_radiation_intensity}) and the nonequilibrium-diffusion radiation model, described here, are shown in the polar plots in Figures \ref{fig:M2p7_polarplots} and \ref{fig:M5_polarplots}, and discussed in Subsection \ref{subsec:describe_radiation_flow}.
In the polar plots, we define a new variable, $T_{\textrm{\tiny nED}} = [{\cal E} + 3 \mu {\cal F}]^{1/4}$, which we call the directionally-dependent nonequilibrium-diffusion intensity temperature.
\begin{figure}[t!]
  \vspace{-21pt}
  \hspace{-15pt}
  \includegraphics[width = 1.1 \columnwidth]{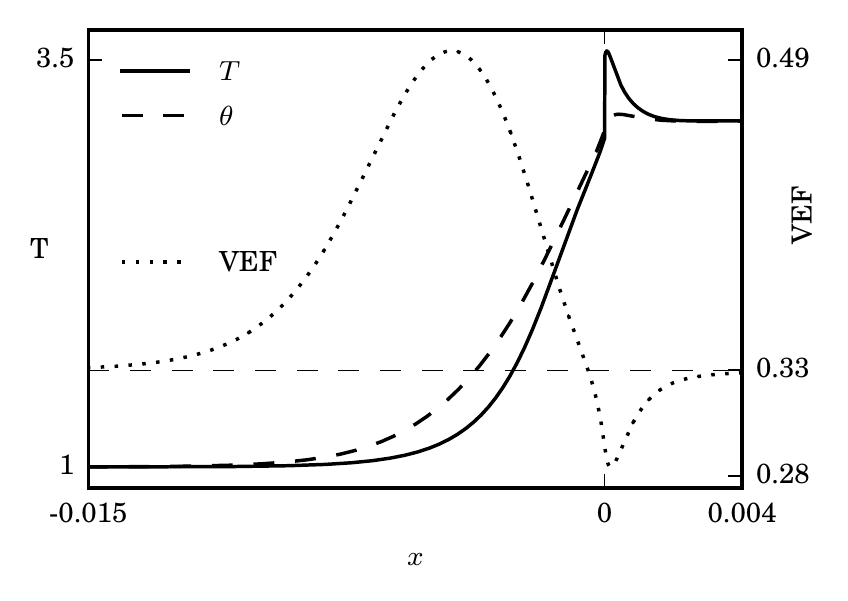}
\caption{
The radiative shock solution for ${\cal M}_0 = 2.7$, but with all other values being the same as in Figure \ref{fig:M1p05_Sn}.
There is an embedded hydrodynamic shock but the maximum material temperature is separated from it so that $T_{\textrm{\tiny max}} > T_{\textrm{\tiny s}}$, $\theta$ is nonmonotonic so that $\theta_{\textrm{\tiny max}} > T_1$, although $\theta_{\textrm{\tiny ps}} < T_1$ and $T_{\textrm{\tiny p}} < T_1$; see Figures \ref{fig:TvM} and \ref{fig:M2p7_Sn_zoomed}.
The VEF deviates considerably from 1/3, except at the equilibria end-states.
\label{fig:M2p7_Sn}}
\end{figure}
\begin{figure}[t!]
  \vspace{-20pt}
  \hspace{-15pt}
  \includegraphics[width = 1.1 \columnwidth]{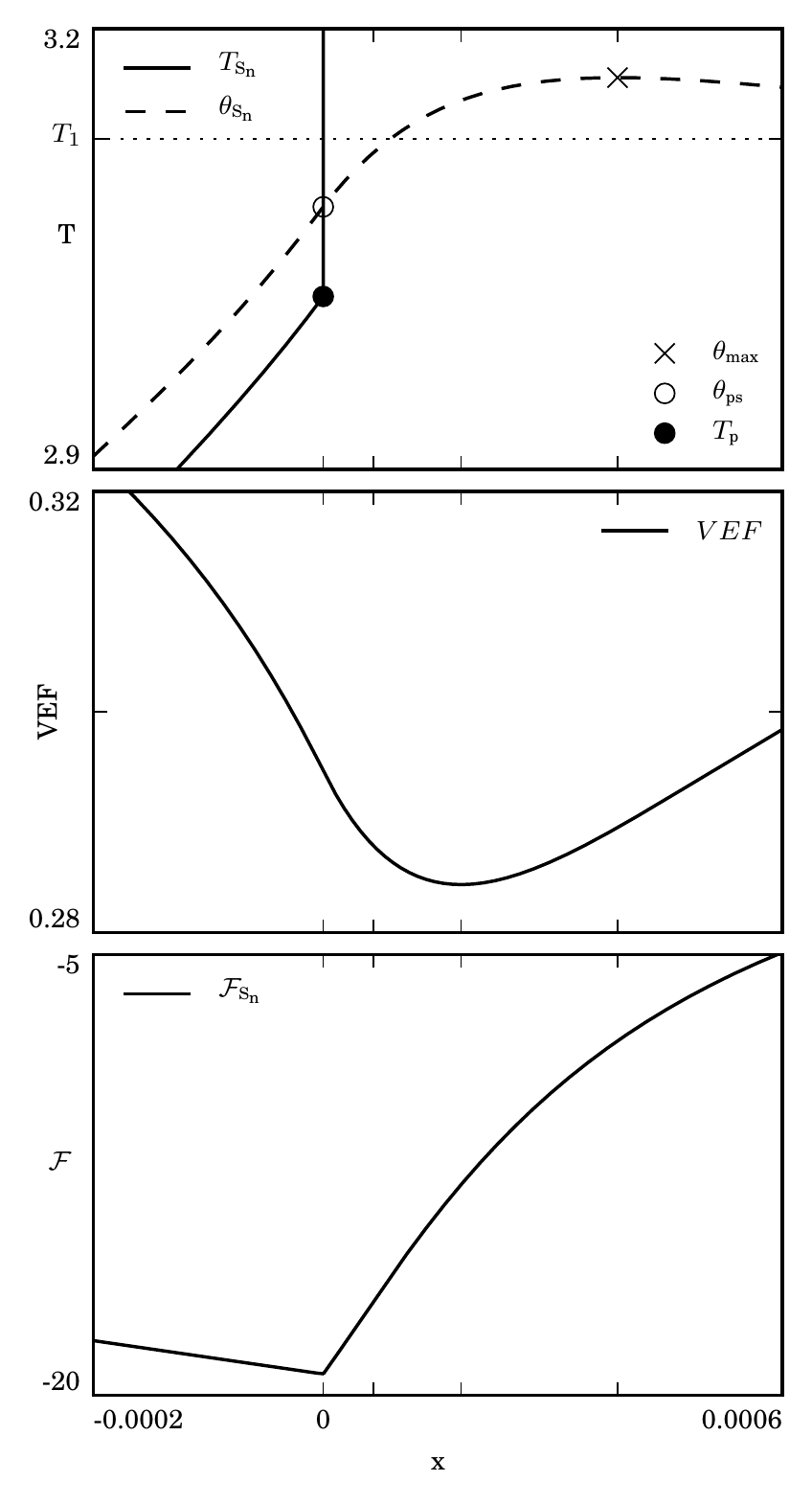}
\caption{
The same values as given in Figure \ref{fig:M2p7_Sn}, but zoomed into the spatial region around $\theta_{\textrm{\tiny max}}$, showing $T$ and $\theta$ in the top plot, the VEF in the middle plot, and the ${\cal F}$ in the bottom plot.
In all three plots, the tickmarks to the right of $x=0$ correspond to the locations of $T_{\textrm{\tiny max}}$, $f_{\textrm{\tiny min}}$, and $\theta_{\textrm{\tiny max}}$, respectively, as $x$ increases.
The horizontal dotted line in the top plot denotes where $T = T_1$, and it can be seen that $\theta_{\textrm{\tiny max}} > T_1$.
The location of $\theta_{\textrm{\tiny max}}$ is not associated with $f_{\textrm{\tiny min}}$, and at the location where $\theta_{\textrm{\tiny max}}$ occurs ${\cal F} \neq {\cal F}_{\textrm{\tiny eq}}$, where ${\cal F}_{\textrm{\tiny eq}} > 0$, as discussed in Subsections \ref{subsec:describe_radiation_flow} and \ref{subsec:analyze_radiation_flow}.
\label{fig:M2p7_Sn_zoomed}}
\end{figure}
\begin{figure}[t!]
  \vspace{-21pt}
  \hspace{-15pt}
  \includegraphics[width = 1.1 \columnwidth]{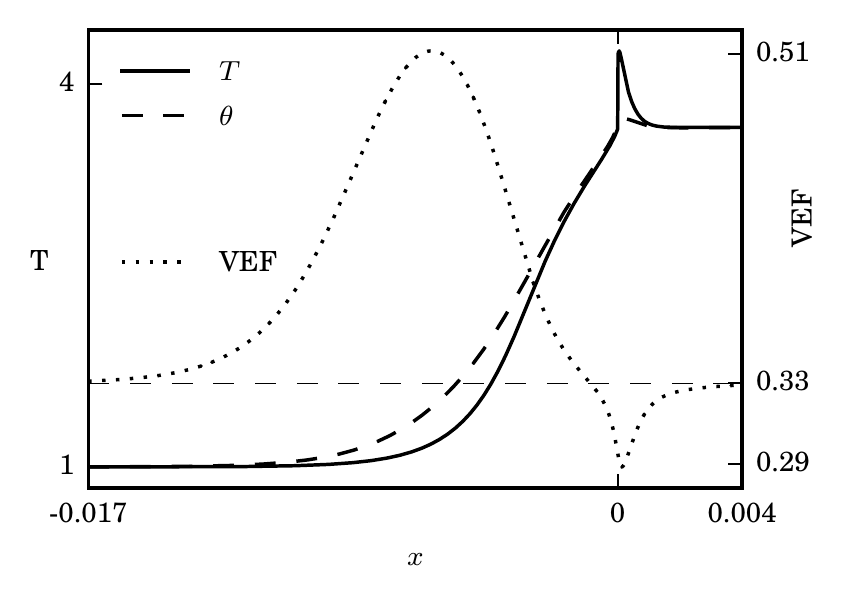}
\caption{
The radiative shock solution for ${\cal M}_0 = 3$, but with all other values being the same as in Figure \ref{fig:M1p05_Sn}.
There is an embedded hydrodynamic shock but the maximum material temperature is separated from it so that $T_{\textrm{\tiny max}} > T_{\textrm{\tiny s}}$, $\theta$ is nonmonotonic so that $\theta_{\textrm{\tiny max}} > T_1$, and now $\theta_{\textrm{\tiny ps}} > T_1$, but $T_{\textrm{\tiny p}} < T_1$; see Figures \ref{fig:TvM} and \ref{fig:M3_Sn_zoomed}.
The VEF deviates considerably from one-third, except at the equilibria end-states.
\label{fig:M3_Sn}}
\end{figure}
\begin{figure}[t!]
  \vspace{-20pt}
  \hspace{-15pt}
  \includegraphics[width = 1.1 \columnwidth]{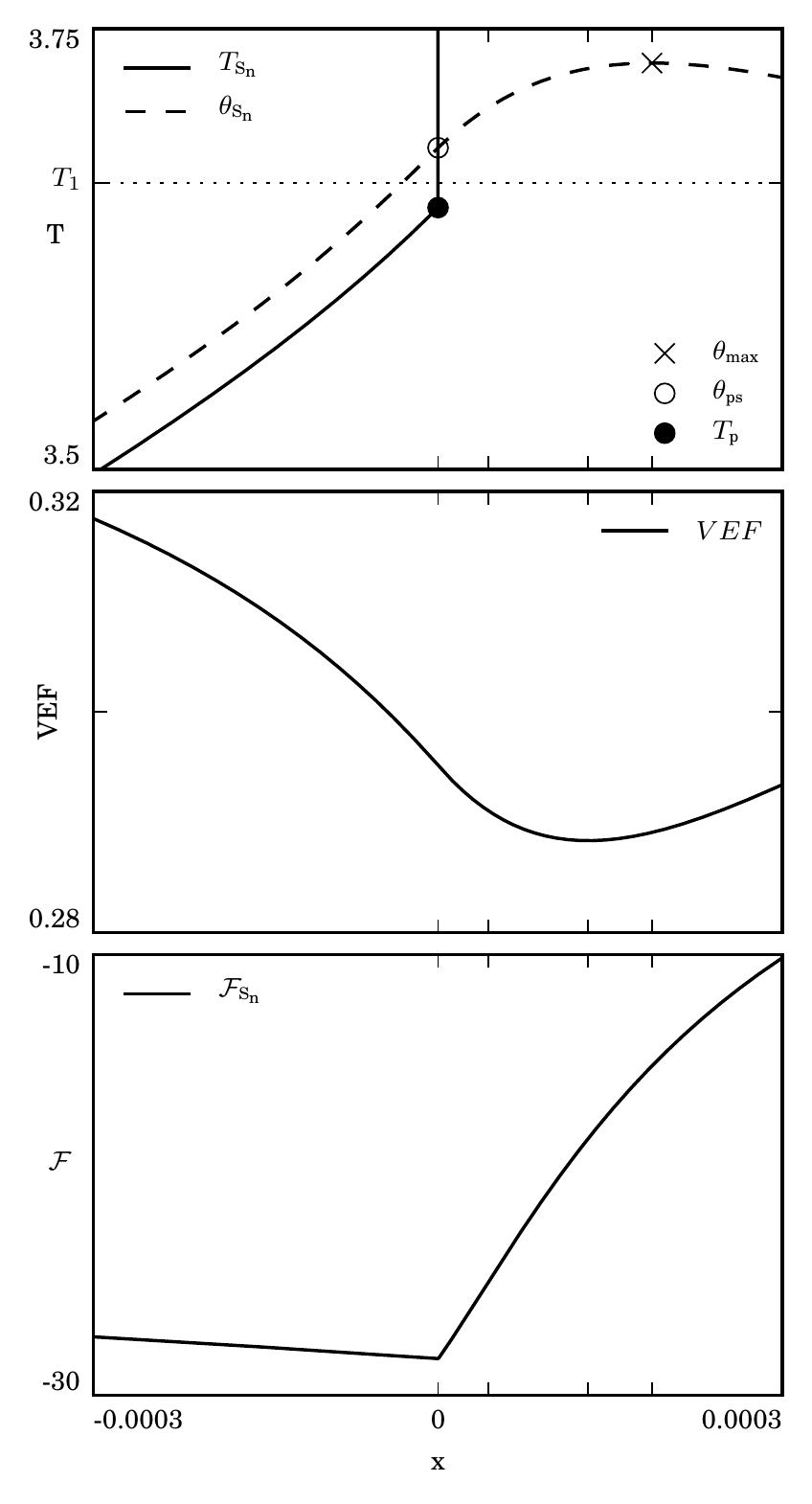}
\caption{
The same values as given in Figure \ref{fig:M3_Sn}, with the same variables as plotted in Figure \ref{fig:M2p7_Sn_zoomed}, including the tickmarks.
Now, $\theta_{\textrm{\tiny max}} > T_1$ and $\theta_{\textrm{\tiny ps}} > T_1$, although $T_{\textrm{\tiny p}} < T_1$.
The locations of $\theta_{\textrm{\tiny max}}$ and $f_{\textrm{\tiny min}}$ are separated, and at the location where $\theta_{\textrm{\tiny max}}$ occurs ${\cal F} \neq {\cal F}_{\textrm{\tiny eq}}$, as discussed in Subsections \ref{subsec:describe_radiation_flow} and \ref{subsec:analyze_radiation_flow}.
\label{fig:M3_Sn_zoomed}}
\end{figure}
\begin{figure}[t!]
  \vspace{-21pt}
  \hspace{-15pt}
  \includegraphics[width = 1.1 \columnwidth]{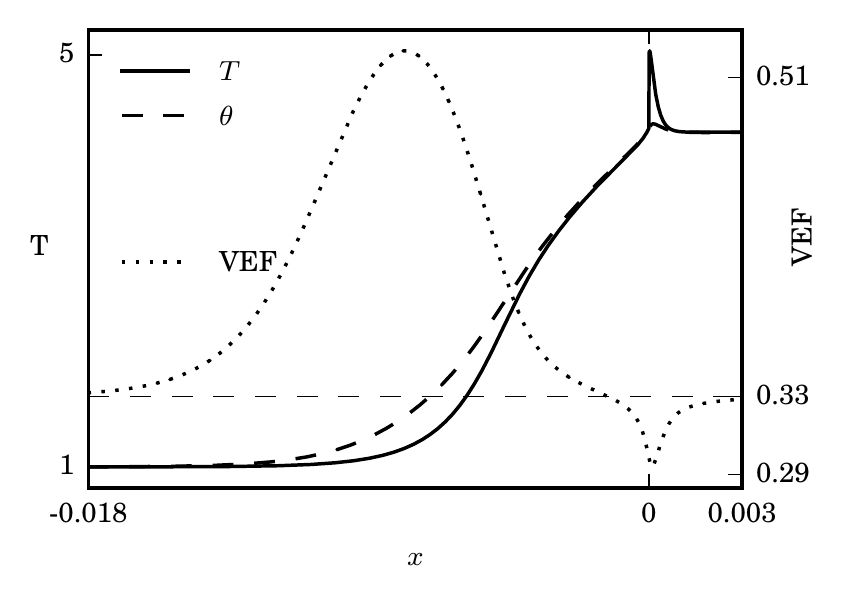}
\caption{
The radiative shock solution for ${\cal M}_0 = 3.3$, but with all other values being the same as in Figure \ref{fig:M1p05_Sn}.
There is an embedded hydrodynamic shock but the maximum material temperature is separated from it so that $T_{\textrm{\tiny max}} > T_{\textrm{\tiny s}}$, $\theta$ is nonmonotonic so that $\theta_{\textrm{\tiny max}} > T_1$, and now $\theta_{\textrm{\tiny ps}} > T_1$ as well as $T_{\textrm{\tiny p}} > T_1$; see Figures \ref{fig:TvM} and \ref{fig:M3_Sn_zoomed}.
The VEF deviates considerably from one-third, except at the equilibria end-states.
\label{fig:M3p3_Sn}}
\end{figure}
\begin{figure}[t!]
  \vspace{-20pt}
  \hspace{-15pt}
  \includegraphics[width = 1.1 \columnwidth]{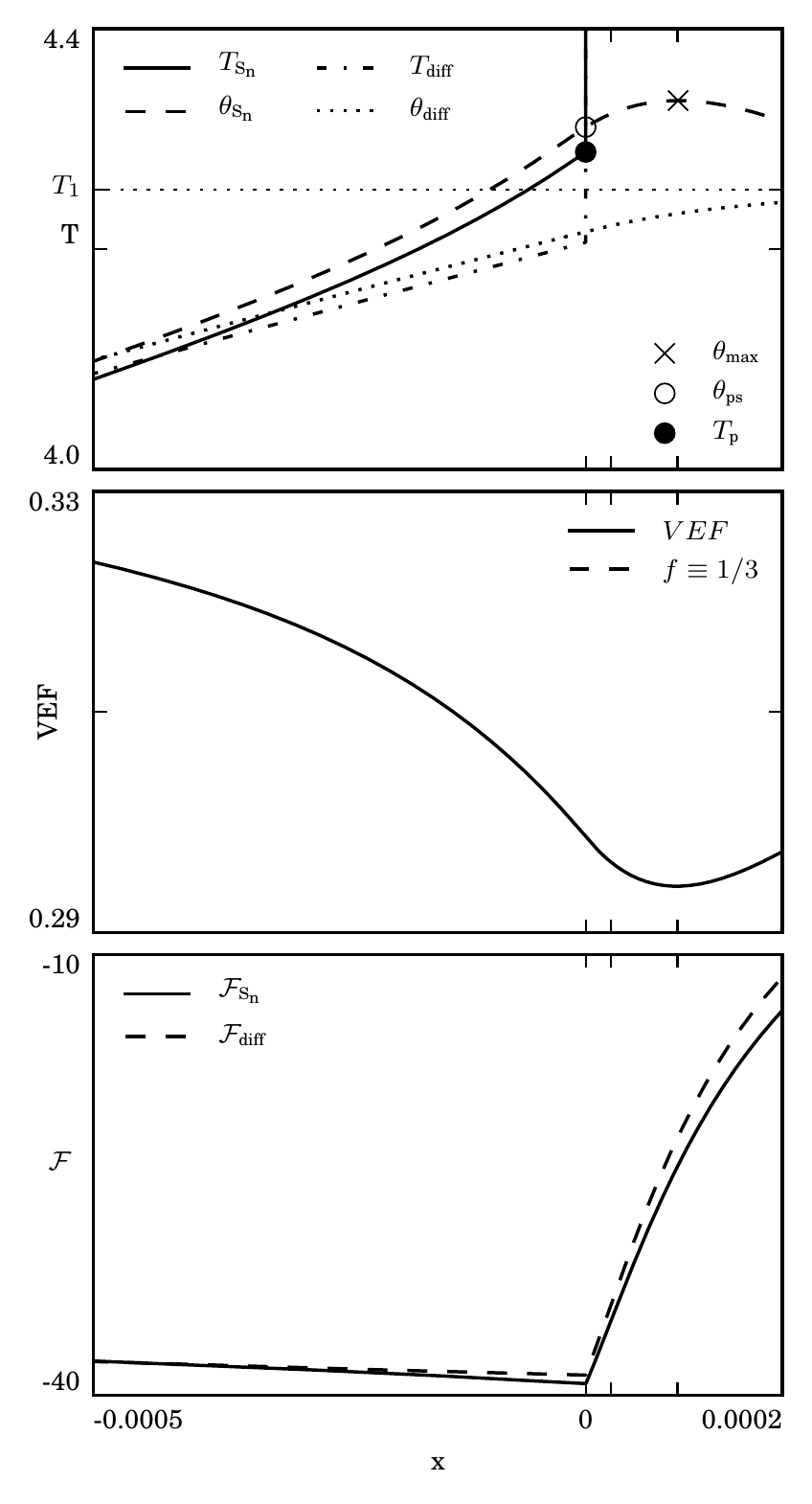}
\caption{
The same values as given in Figure \ref{fig:M3p3_Sn}, with the same variables as plotted in Figure \ref{fig:M2p7_Sn_zoomed}, including the tickmarks.
Now, $\theta_{\textrm{\tiny max}} > T_1$, $\theta_{\textrm{\tiny ps}} > T_1$, and $T_{\textrm{\tiny p}} > T_1$.
The locations of $\theta_{\textrm{\tiny max}}$ and $f_{\textrm{\tiny min}}$ are not well separated, and at the location where $\theta_{\textrm{\tiny max}}$ occurs ${\cal F} \neq {\cal F}_{\textrm{\tiny eq}}$, as discussed in Subsections \ref{subsec:describe_radiation_flow} and \ref{subsec:analyze_radiation_flow}.
The temperature solutions for nonequilibrium-diffusion, $T_{\textrm{\tiny diff}}$ and $\theta_{\textrm{\tiny diff}}$, are included in the top plot to show the effects of the adaptation zone conjectured by Drake \cite{Drake2007a, Drake2007b}.
See Subsections \ref{subsec:describe_radiation_flow} and \ref{subsec:analyze_radiation_flow}.
\label{fig:M3p3_Sn_zoomed}}
\end{figure}
\section{Results}
\label{sec:results}
All of the radiative-shock solutions presented in this paper use S$_{16}$-transport, $\gamma = 5 / 3$ for a monatomic ideal-gas, and an upstream, unshocked reference temperature of $\tilde{T}_0 = 100 \, \textrm{eV}$ and reference density of $\tilde{\rho}_0 = 1 \, \textrm{g} / \textrm{cm}^3$, which results in $P_0 \approx 8.53\times10^{-5}$ and ${\cal C} \approx 2364$.
We chose S$_{16}$ because it does show enhanced accuracy as compared to S$_8$, whereas S$_{32}$ is not noticeably more accurate than S$_{16}$ in our figures.
Further, all of the solutions presented herein use a constant nondimensional cross section of $\sigma_{\textrm{t}} = 577.35 = \sigma_{\textrm{a}}$ with no scattering, except for Figures \ref{fig:M3_Trho} and \ref{fig:M5_Trho} which represent solutions containing both Bremsstrahlung emission, $\sigma_{\textrm{a}} \approx 44.88 \, \rho^2 / T^{7/2}$, and Thomson scattering, $\sigma_{\textrm{s}} = 0.4006 \, \rho$.
The values for $P_0$, ${\cal C}$, and $\sigma_{\textrm{t}} = \sigma_{\textrm{a}} + \sigma_{\textrm{s}}$, as well as the initial Mach numbers of 1.05, 1.2, 2, 3, and 5, are chosen so as to aide comparison with figures in the paper by Lowrie and Edwards \cite{LE2008}.

In this section, details regarding the grey S$_{\textrm{n}}$-transport radiative-shock solutions presented in Figures \ref{fig:TvM}-\ref{fig:M5_Trho} are discussed.
A general description of the shock solutions are given in Subsection \ref{subsec:general_description}.
In Subsection \ref{subsec:compare_shock_structures}, we compare the shock structures for the solutions from nonequilibrium-diffusion and S$_{\textrm{n}}$-transport, both for constant cross sections and for cross sections that depend on material density and temperature, which are shown in Figures \ref{fig:M3_const}-\ref{fig:M5_Trho}.
In Subsection \ref{subsec:describe_radiation_flow}, we examine the S$_{\textrm{n}}$-transport radiation intensities across the shock structure, shown in Figures \ref{fig:M2p7_with_intensities_zoomed} and \ref{fig:M5_with_intensities_zoomed}.
We also compare the angular dependence of the radiation intensity computed from nonequilibrium-diffusion and angle-dependent radiation transport in polar plots at specific spatial locations across the shock structure, as seen in Figures \ref{fig:M2p7_polarplots} and \ref{fig:M5_polarplots}.
Conjectures by Drake \cite{Drake2007a, Drake2007b} that the temperatures in the precursor region very near the Zel'dovich spike could have values that exceed the value of the downstream equilibrium temperature are confirmed and shown in Figures \ref{fig:M2p7_Sn_zoomed}, \ref{fig:M3_Sn_zoomed}, and \ref{fig:M3p3_Sn_zoomed}, as well as Drake's prediction of an adaptation zone adjacent to the Zel'dovich spike, which is shown in Figure \ref{fig:M3p3_Sn_zoomed}.
Predictions by McClarren and Drake \cite{McD2010} of anti-diffusive radiation under the Zel'dovich spike are confirmed in Figures \ref{fig:TvM}, \ref{fig:M3_const}, \ref{fig:M3_Trho}, and \ref{fig:M2p7_Sn}-\ref{fig:M3p3_Sn_zoomed}, where the radiation temperature has a local maximum which is greater than the downstream equilibrium temperature, $\theta_{\textrm{\tiny max}} > T_1$, and where the radiation flux does not go through its equilibrium value, ${\cal F}_{\textrm{\tiny eq}}$, at the location of $\theta_{\textrm{\tiny max}}$.
We also show that anti-diffusive radiation occurs for a range of initial Mach numbers.
In Subsection \ref{subsec:analyze_radiation_flow}, we close this section by using the analytic solution for the directionally-dependent radiation intensity (\ref{eq:analytic_radiation_intensity}) to explain the adaptation zone and the structure of the S$_{\textrm{n}}$-transport radiation intensities across the radiative shock.
\subsection{Brief description of the solutions}
\label{subsec:general_description}
The terminology used in this subsection is illustrated in Figures \ref{fig:generic_M2_shock} and \ref{fig:generic_M5_shock}, and largely follows terminology introduced by Drake \cite{Drake2007a, Drake2007b}, and Lowrie and Edwards \cite{LE2008}.
The VEF, $f$, provides a generic metric for the angular distribution of the radiation.
If the radiation intensity is dominant along $\mu \sim -1$ then $f \approx 1$, whereas if the radiation intensity is dominant along $\left| \mu \right| \sim 0$ then $f \approx 0$.

Figures \ref{fig:generic_M2_shock} and \ref{fig:generic_M5_shock} provide an annotated illustration of the VEF over the spatial domain of the radiative shock.
In the transmissive region of the upstream precursor, where $f > 1/3$, the angular distribution of radiation is directed along $\mu \sim -1$, such that radiation in this region is dominantly traveling along the axis of the shock; this radiation is described as being ``forward-peaked''.
In the diffusive region of the upstream precursor, where $f \approx 1/3$ over a considerable spatial extent adjacent to the embedded hydrodynamic shock, the angular distribution of radiation is isotropic.
It is important to point out that there is no reason in general to expect the radiation to be isotropic at the location where the VEF passes through one-third.
In the oblique region of the downstream relaxation region, where $f < 1/3$, the angular distribution of radiation contains dominant angular components along $\left| \mu \right| \sim 0$, so that $I( x, \left| \mu \right| \sim 0 ) > I( x, \left| \mu \right| \sim 1 )$.

For sufficiently low values of ${\cal M}_0$, all physical properties of the radiative shock are continuous and monotonic, as seen in Figure \ref{fig:M1p05_Sn}, except for the VEF, which has the general form described in the preceding paragraph.
As ${\cal M}_0$ is increased an embedded hydrodynamic shock forms separating the upstream precursor and downstream relaxation regions by a discontinuity in the material properties, as seen in Figure \ref{fig:M1p2_Sn}.
The value of the material properties at the downstream precursor state of the discontinuity are labeled with a subscripted-``p'', and those values at the upstream relaxation state are labeled with a subscripted-``s'', as illustrated in Figure \ref{fig:generic_M2_shock}.
As mentioned in Subsection \ref{subsec:continuity_conditions}, the radiation variables are continuous across the embedded hydrodynamic shock, and the radiation values there are labeled with a subscripted-``ps''.
For sufficiently weak shocks containing an embedded hydrodynamic shock the material temperature at the downstream precursor state, $T_{\textrm{\tiny p}}$, is consistently less than the downstream equilibrium temperature, $T_1$, as shown in Figures \ref{fig:TvM}, \ref{fig:M1p2_Sn} and \ref{fig:M2_Sn}.
Figure \ref{fig:M2_Sn} also shows that if the maximum material temperature, $T_{\textrm{\tiny max}}$, is greater than $T_1$ then a Zel'dovich temperature spike exists in the relaxation region.
It is also possible that $T_{\textrm{\tiny max}}$ is greater than the value of the material temperature at the upstream relaxation state of the embedded hydrodynamic shock, $T_{\textrm{\tiny s}}$, such that the maximum material temperature occurs downstream of the embedded hydrodynamic shock.
These characteristics of the material and radiation temperatures were established by Lowrie and Edwards \cite{LE2008}.
\subsection{Comparison of the nonequilibrium diffusion and S$_{\textrm{n}}$-transport radiative shock structures}
\label{subsec:compare_shock_structures}
The nonequilibrium-diffusion solutions by Lowrie and Edwards \cite{LE2008} used the constant Eddington approximation, ${\cal P} = {\cal E} / 3$, whereas the solutions presented herein use a VEF, ${\cal P} = f {\cal E}$, allowing $f$ to vary spatially.
This has the effect of changing the expression for the radiation flux for nonequilibrium-diffusion,
\begin{multline}
\label{eq:nondimensional_radiation_flux_diffusion}
  {\cal F} = - \frac{1}{3 \, \sigma_{\textrm{t}}} \frac{d {\cal E}}{d x} \\
          + \frac{1}{\sigma_{\textrm{t}}} \beta \left( \frac{1}{3} \, \sigma_{\textrm{t}} {\cal E} + \sigma_{\textrm{s}} \, {\cal E} + \sigma_{\textrm{a}} \, T^4 \right) \, ,
\end{multline}
to that for S$_\textrm{n}$-transport (\ref{eq:nondimensional_radiation_flux}),
\begin{multline}
  {\cal F} = - \frac{1}{\sigma_{\textrm{t}}} \frac{d {\cal P}}{dx} \\
           + \frac{1}{\sigma_{\textrm{t}}} \beta \left( \sigma_{\textrm{t}} \, {\cal P} + \sigma_{\textrm{s}} \, {\cal E} + \sigma_{\textrm{a}} \, T^4 \right) \, .
\end{multline}
Thus, where $f \approx 1/3$ and almost spatially constant, the radiative-shock solutions for nonequilibrium diffusion and S$_{\textrm{n}}$-transport closely agree.
Such minimal change in the VEF is seen in Figure \ref{fig:M1p05_Sn}, which represents a weak and continuous shock for ${\cal M}_0 = 1.05$.
In Figures \ref{fig:M3_const} and \ref{fig:M5_const}, we compare the nonequilibrium-diffusion and S$_{\textrm{n}}$-transport radiative-shock solutions for a constant, purely absorbing cross section, $\sigma_{\textrm{t}} = 577.35 = \sigma_{\textrm{a}}$, for initial Mach numbers of 3 and 5, respectively.
In Figures \ref{fig:M3_Trho} and \ref{fig:M5_Trho}, we compare the nonequilibrium-diffusion and S$_{\textrm{n}}$-transport radiative-shock solutions which include both Thomson scattering, $\sigma_{\textrm{s}} = 0.4006 \, \rho$, and Bremsstrahlung emission, $\sigma_{\textrm{a}} = 44.78 \, \rho^2 / \, T^{7/2}$, also for initial Mach numbers of  3 and 5, respectively.

As seen in Figures \ref{fig:M3_const} and \ref{fig:M5_const}, for the constant cross section cases, the material and radiation temperature solutions for S$_{\textrm{n}}$-transport, $T_{\textrm{\tiny S}_{\textrm{\tiny n}}}$ and $\theta_{\textrm{\tiny S}_{\textrm{\tiny n}}}$, respectively, consistently rise earlier than the nonequilibrium-diffusion material and radiation temperature solutions, $T_{\textrm{\tiny diff}}$ and $\theta_{\textrm{\tiny diff}}$, respectively, as the VEF rises above one-third.
Near the apex of the VEF, $T_{\textrm{\tiny diff}}$ becomes greater than $T_{\textrm{\tiny S}_{\textrm{\tiny n}}}$, and $\theta_{\textrm{\tiny diff}}$ likewise becomes greater than $\theta_{\textrm{\tiny S}_{\textrm{\tiny n}}}$, resulting in a slight separation between the transport and diffusion solutions as they move downstream and the VEF approaches one-third near the embedded hydrodynamic shock.
The shape of the VEF in the transmissive region becomes increasingly symmetric as ${\cal M}_0$ increases.
The VEF passes below one-third very near the embedded hydrodynamic shock, by which point $\theta_{\textrm{\tiny S}_{\textrm{\tiny n}}} > \theta_{\textrm{\tiny diff}}$.
The value of $T_{\textrm{\tiny S}_{\textrm{\tiny n}}}$ at the downstream precursor state-``p'', may be less than or greater than the value of $T_{\textrm{\tiny diff}}$ there.
As the VEF passes through its minimum, and then asymptotically increases back to one-third, $T_{\textrm{\tiny S}_{\textrm{\tiny n}}}$ relaxes to the downstream equilibrium temperature, $T_1$, more slowly than $T_{\textrm{\tiny diff}}$.
The same cannot always be said for the radiation temperatures because when $\theta_{\textrm{\tiny S}_{\textrm{\tiny n}}}$ has a maximum then it has already passed above $T_1$, and if it barely passes above $T_1$ then it may relax faster to $T_1$ than is possible for $\theta_{\textrm{\tiny diff}}$.
The description of the radiation fluxes in these figures follows a similar description as the material temperatures, although a longer distance in the relaxation region is needed to return to ${\cal F}_{\textrm{\tiny eq}}$.
This is because the radiation intensities are dominant along $\left| \mu \right| \sim 0$ under the Zel'dovich spike, as explained in Subsection \ref{subsec:describe_radiation_flow}, and since the radiation intensity is not generally an even or symmetric function of $\mu$ the integral defining the S$_{\textrm{n}}$-transport radiation flux (\ref{eq:nondimensional_radiation_flux}) appropriately adds these contributions, whereas they are neglected by the nonequilibrium-diffusion solution.
A qualitative distinction between the S$_{\textrm{n}}$-transport and nonequilibrium-diffusion radiation fluxes at the embedded hydrodynamic shock and in the relaxation region, is noticeable in Figure \ref{fig:M5_const}.

Consider now Figures \ref{fig:M3_Trho} and \ref{fig:M5_Trho}, for which the cross sections are represented by Bremsstrahlung emission, $\sigma_{\textrm{a}} = 44.78 \rho^2 \, / T^{7/2}$, and Thomson scattering, $\sigma_{\textrm{s}} = 0.4006 \rho$.
Comparing their results with those for the constant cross section, the first thing to notice is that as ${\cal M}_0$ increases the radiative shock looks more like a Marshak wave moving into cold material.
The structures of the nonequilibrium-diffusion and S$_{\textrm{n}}$-transport solutions for the ${\cal M}_0 = 3$ shock, shown in Figure \ref{fig:M3_Trho}, show considerable spatial separation which is due to the effect of the VEF.
The sharp growth and slow decay of the VEF in the upstream precursor region results in the separation of the S$_{\textrm{n}}$-transport and nonequilibrium-diffusion solutions for the temperatures and the radiation flux.
As the VEF reapproaches one-third from above the temperature solutions begin to agree, and where the VEF is below one-third in the relaxation region the S$_{\textrm{n}}$-transport temperature and radiation flux solutions relax to their equilibrium values over a longer distance than do the nonequilibrium-diffusion solutions.
For the ${\cal M}_0 = 5$ shock, in Figure \ref{fig:M5_Trho}, the spatial separation is not as noticeable and the nonequilibrium-diffusion and S$_{\textrm{n}}$-transport solutions are in good agreement, although the S$_{\textrm{n}}$-transport radiation flux in the relaxation region is noticeably different from the nonequilibrium-diffusion radiation flux.
The lack of spatial separation is because the VEF is different from one-third over a narrow spatial domain, even though it is considerably different from one-third over this domain.
As shown, as ${\cal M}_0$ increases the VEF becomes significantly narrowed and peaked.

Finally, it is interesting to note that for both shock strengths there are similarities in a few general features of the shock structure regardless of the functional dependence of the cross section.
Specifically, the values of $f_{\textrm{\tiny max}}$, $f_{\textrm{\tiny min}}$, $T_{\textrm{\tiny max}}$, $\theta_{\textrm{\tiny max}}$, ${\cal F}_{\textrm{\tiny min}}$, and the values at both state-``p'' and state-``s'' are the same to within 1 percent for a given value of ${\cal M}_0$ and do not seem to be significantly affected by the cross section.
\subsection{Description of the radiation flow}
\label{subsec:describe_radiation_flow}
This subsection uses the analytic descriptions provided in Subsections \ref{subsec:analytic_solution_for_I} and \ref{subsec:simplified_radiation_diffusion_models} to describe differences between the nonequilibrium-diffusion and S$_{\textrm{n}}$-transport radiation flows.
The results in Figures \ref{fig:M2p7_with_intensities_zoomed}-\ref{fig:M5_polarplots} are described first, which lay the groundwork for understanding why the directional-dependence of the radiation is able to qualitatively change the radiative-shock solutions.
The results in Figures \ref{fig:M2p7_Sn}-\ref{fig:M3p3_Sn_zoomed} are then described, which lead to a discussion of anti-diffusive radiation predicted by McClarren and Drake \cite{McD2010}, as well as the adaptation zone conjectured by Drake \cite{Drake2007a, Drake2007b}.

For ease of comparing the radiation-intensity solutions, from either S$_{\textrm{n}}$-transport or nonequilibrium-diffusion, with the material and radiation temperatures, we define a new variable as the radiation-intensity temperature, $T_{I}(\mu) \equiv [4 \pi I(\mu)]^{1/4}$.
In the middle plots of Figures \ref{fig:M2p7_with_intensities_zoomed} and \ref{fig:M5_with_intensities_zoomed}, curves for the S$_{16}$-transport intensity temperatures, $T_{I_{\textrm{\tiny m}}} = T_{I}(\mu_{\textrm{\tiny m}})$, are plotted along with a fiducial curve for the material temperature, $T$.
As a reminder, the radiation-intensity curves associated with $\mu < 0$ travel leftward and curves associated with $\mu > 0$ travel rightward.
In the precursor region the curves for $T_{I}(\mu_{\textrm{\tiny m}} < 0)$ lie above the curve for $T$, and the curves for $T_{I}(\mu_{\textrm{\tiny m}} > 0)$ lie below the curve for $T$.
Additionally, the curves for $T_{I}(\left| \mu_{\textrm{\tiny m}} \right| \sim 0)$ are typically closer to the curve for $T$, and the curves for $T_{I}(\left| \mu_{\textrm{\tiny m}} \right| \sim 1)$ typically lie farther from the curve for $T$.
At the locations where the value of the VEF is largest, the curve for $T_{I}(\mu_{\textrm{\tiny m}} \sim -1)$ is the top-most curve, while the curve for $T_{I}(\mu_{\textrm{\tiny m}} \sim 1)$ is the bottom-most curve, and the curves between these two extremes vary smoothly with $\mu$.
The top plots in Figures \ref{fig:M2p7_polarplots} and \ref{fig:M5_polarplots} show polar plots of $T_{\textrm{\tiny I}} / T$ (solid line), where the radiation intensity in the intensity temperature is determined from the analytic expression (\ref{eq:analytic_radiation_intensity}) which is continuous in angle, evaluated at the location where the value of the VEF is largest; the nonequilibrium-diffusion radiation intensity (\ref{eq:nonequilibrium_diffusion_I}) is used to show the nonequilibrium-diffusion intensity temperature (dashed line), which goes to zero at $\mu = 1$, and the dotted curve near the center is a fiducial circle of radius one.
The radiation at this spatial location, as computed by S$_{\textrm{n}}$-transport, is forward-peaked and dominantly travels along $\mu = -1$, whereas considerably less radiation is traveling along $\mu > 0 $, which is in agreement with the middle plots in Figures \ref{fig:M2p7_with_intensities_zoomed} and \ref{fig:M5_with_intensities_zoomed}.

The bottom plots in Figures \ref{fig:M2p7_with_intensities_zoomed} and \ref{fig:M5_with_intensities_zoomed} are zoomed in around the Zel'dovich spike.
The tickmarks in each bottom plot, to the right of $x = 0$, denote the spatial locations where $T_{\text{\tiny max}}$ and $\theta_{\text{\tiny max}}$ occur.
For both of these plots, the location where $f_{\text{\tiny min}}$ occurs is almost coincident with the location where $T_{\text{\tiny max}}$ occurs, and is indistinguishable in the axis labeling.
The polar plots in Figures \ref{fig:M2p7_polarplots} and \ref{fig:M5_polarplots} are constructed at the locations where $f_{\text{\tiny max}}$, $T_{\text{\tiny max}}$, and $f_{\text{\tiny min}}$ occur, in descending order.
In the top polar plots in Figures \ref{fig:M2p7_polarplots} and \ref{fig:M5_polarplots}, the interior dotted line is a fiducial circle of radius one.
In the bottom two polar plots, the outermost dotted curve represents a fiducial circle of radius one, and the innermost curve represents a fiducial circle of radius one-third.

The radiation intensities in the bottom plots of Figures \ref{fig:M2p7_with_intensities_zoomed} and \ref{fig:M5_with_intensities_zoomed} are continuous across the embedded hydrodynamic shock although their spatial derivative is discontinuous there.
In the bottom plot of Figure \ref{fig:M2p7_with_intensities_zoomed}, as the S$_{\textrm{n}}$-transport intensity temperatures approach the embedded hydrodynamic shock, those corresponding to $T_{I}(\left| \mu_{\textrm{\tiny m}} \right| \sim 0)$ are most effected by the Zel'dovich spike, and this is displayed in the bottom two polar plots of Figure \ref{fig:M2p7_polarplots} since the S$_{\textrm{n}}$-transport curves for $T_{I} / T$ (solid lines) are dominant along the $\left| \mu \right| \sim 0$ portions of the polar plot.
This is what is meant by oblique radiation.
It is interesting that in the bottom plot of Figures \ref{fig:M2p7_with_intensities_zoomed} and \ref{fig:M2p7_polarplots}, separately, the curves for $T_{I}(\left| \mu \right| \sim 0)$ are greater than $T$ over an extended spatial region adjacent to the embedded hydrodynamic shock.
This occurs in the downstream precursor region, and in the upstream relaxation region while the intensity temperatures relax to their downstream post-shock equilibrium values.
Referring back to the expression for the equilibrium radiation intensity (\ref{eq:equilibrium_EFPf_I}), as the post-shock downstream equilibrium state is approached, it is clear that this will become the case for the angularly-discrete intensity temperatures at all values of $\mu_{\textrm{\tiny m}} > 0$.

The radiation transport solutions (solid lines) and the nonequilibrium-diffusion solutions (dashed lines) presented in the polar plots in Figures \ref{fig:M2p7_polarplots} and \ref{fig:M5_polarplots} show that there is very little agreement for the radiation flow between these two radiation models.
In the top polar plots of Figures \ref{fig:M2p7_polarplots} and \ref{fig:M5_polarplots} the nonequilibrium-diffusion model underestimates the value of the intensity temperature along the $\mu = -1$ direction by almost a factor of two, and along the $\mu = 1$ direction it gives values near zero instead of the transport calculated values near one.
In the bottom polar plots of Figures \ref{fig:M2p7_polarplots} and \ref{fig:M5_polarplots} the nonequilibrium-diffusion radiation model obviously fails to capture the oblique radiation flow, but overestimates the radiation flow along the directions $\left| \mu \right| \sim 1$.
In the case of an experiment, like a thin shocktube, this could mean that more radiation energy would be deposited into the shock tube wall than would be predicted if the radiation is modeled with nonequilibrium-diffusion.
Or, if the material temperature were being inferred by looking across the shocktube, then the apparent temperature would be higher than the actual temperature because of the extra radiation fluence.

Finally, we describe Figures \ref{fig:M2p7_Sn}-\ref{fig:M3p3_Sn_zoomed}.
Figures \ref{fig:M2p7_Sn}, \ref{fig:M3_Sn}, and \ref{fig:M3p3_Sn} show the S$_{\textrm{n}}$-transport solutions for $T$, $\theta$, and the VEF, for ${\cal M}_0 =$ 2.7, 3, and 3.3, respectively.
Figures \ref{fig:M2p7_Sn_zoomed}, \ref{fig:M3_Sn_zoomed}, and \ref{fig:M3p3_Sn_zoomed} are zoomed-in around the embedded hydrodynamic shock and the location of $\theta_{\textrm{\tiny max}}$, from Figures \ref{fig:M2p7_Sn}, \ref{fig:M3_Sn}, and \ref{fig:M3p3_Sn}, respectively, and present $T$ and $\theta$ in the top plot, the VEF in the middle plot, and ${\cal F}$ in the bottom plot.
The top plot also annotates $\theta_{\textrm{\tiny max}}$, $\theta_{\textrm{\tiny ps}}$, and $T_{\textrm{\tiny p}}$.
The top plot in Figure \ref{fig:M3p3_Sn_zoomed} also displays the nonequilibrium-diffusion temperature solutions, in order to make the adaptation zone more apparent.
In the top plot of Figure \ref{fig:M2p7_Sn_zoomed}, only $\theta_{\textrm{\tiny max}}$ is greater than $T_1$, whereas in the top plot of Figure \ref{fig:M3_Sn_zoomed}, $\theta_{\textrm{\tiny ps}}$ is also greater than $T_1$, and finally, in the top plot of Figure \ref{fig:M3p3_Sn_zoomed}, $T_p$ is greater than $T_1$ as well.
The bottom plots show that at the location of $\theta_{\textrm{\tiny max}}$, the radiation flux is nowhere near its value of ${\cal F}_{\textrm{\tiny eq}}$, which is relevant to the prediction by McClarren and Drake \cite{McD2010} of anti-diffusive radiation.

\subsection{Analyzing the radiation flow}
\label{subsec:analyze_radiation_flow}
This subsection uses the analytic results provided in Subsection \ref{subsec:analytic_solution_for_I} to understand how the angular dependence of the radiation flow supports the adaptation zone, as well as the structure of the radiation intensities across the radiative shock.

Zel'dovich \cite{Zeldovich1957} claimed that the radiation temperature could nowhere exceed the downstream equilibrium temperature, so that $\theta_{\textrm{\tiny max}} \leq T_1$ everywhere, and also that $T_{\textrm{\tiny p}} \leq T_1$, regardless of the shock strength.
Raizer \cite{Raizer1957} agreed with both of these claims, stating that allowing $T_{\textrm{\tiny p}} \geq T_1$ would allow $\theta_{\textrm{\tiny max}} \geq T_1$, ``which clearly does not make sense.''
These arguments reappear with less qualification in the canonical text by Zel'dovich and Raizer \cite{ZR2002}.
Mihalas and Mihalas \cite{MM1999} cited the two papers above, stating that Zel'dovich's paper is ``a rigorous mathematical analysis of the radiation transport equation''.  
However, Zel'dovich chose to replace ``the [exponential integral] $2 E_2 \left( \xi \right)$ by the exponent $\exp \left( -\xi / \alpha \right)$, where $\alpha$ is a dimensionless number that differs little from unity'', and Raizer chose to neglect these integrals ``since the radiation generated in the heating zone contributes very little to the total flux and density''.
Another concern of those authors was that at the location where $\theta_{\textrm{\tiny max}}$ occurs, they expected ${\cal F} = {\cal F}_{\textrm{\tiny eq}}$, which is an implicit assumption of the nonequilibrium-diffusion radiation model.
In contradistinction, Drake and McClarren \cite{McD2010} retained the exponential integrals, and predicted that it is possible to have $\theta_{\textrm{\tiny max}} > T_1$ under a Zel'dovich spike, and that at that location ${\cal F} \neq {\cal F}_{\textrm{\tiny eq}}$, and called this ``anti-diffusive'' radiation.
In Figure \ref{fig:TvM}, values of $T_{\textrm{\tiny max}}$, $T_{\textrm{\tiny s}}$, $\theta_{\textrm{\tiny max}}$, $\theta_{\textrm{\tiny ps}}$, and $T_{\textrm{\tiny p}}$, normalized by $T_1$, are plotted against ${\cal M}_0$.
There, it is seen that $T_{\textrm{\tiny max}}$ separates from $T_{\textrm{\tiny s}}$ at ${\cal M}_0 \approx 2.2$, $\theta_{\textrm{\tiny max}}$ becomes greater than $T_1$ at ${\cal M}_0 \approx 2.3$, $\theta_{\textrm{\tiny ps}}$ becomes greater than $T_1$ at ${\cal M}_0 \approx 2.9$, and $T_{\textrm{\tiny p}}$ becomes greater than $T_1$ at ${\cal M}_0 \approx 3.1$.
This is because the S$_{\textrm{n}}$-transport radiation model correctly accounts for the angular distribution of the radiation flow.
The point of this subsection is to explain how the angular dependence of the radiation intensity causes this to happen.

Radiation diffusion defines the radiation flux (\ref{eq:nondimensional_radiation_flux_diffusion}) as being proportional to the negative gradient of the radiation temperature plus ${\cal F}_{\textrm{\tiny eq}}$, so that when $\theta_{\textrm{\tiny max}}$ occurs ${\cal F} = {\cal F}_{\textrm{\tiny eq}}$.
An important element in the argument by Drake and McClarren was that, when the angular dependence of the radiation is accounted for, at the location where $\theta_{\textrm{\tiny max}}$ occurs ${\cal F} \neq {\cal F}_{\textrm{\tiny eq}}$.
When the angular dependence of the radiation intensity is retained, the radiation flux (\ref{eq:nondimensional_radiation_flux}) is the negative gradient of the radiation pressure, ${\cal P} = f {\cal E}$, plus ${\cal F}_{\textrm{\tiny eq}}$, so the diffusion model requires making the VEF spatially constant.
The top and bottom plots of Figures \ref{fig:M2p7_Sn_zoomed}, \ref{fig:M3_Sn_zoomed} and \ref{fig:M3p3_Sn_zoomed} show $\theta$ and ${\cal F}$, zoomed-in around the location where $\theta_{\textrm{\tiny max}}$ occurs and it is seen that ${\cal F} \neq {\cal F}_{\textrm{\tiny eq}}$ in this region.

In a different set of work, Drake \cite{Drake2007a, Drake2007b} used an energy-balance analysis to conjecture that for a sufficiently strong shock the radiant heat flux produced under the Zel'dovich spike would raise the values of $T_{\textrm{\tiny p}}$ and $\theta_{\textrm{\tiny ps}}$, sufficiently, that it was possible to have $\theta_{\textrm{\tiny ps}} > T_1$ and $T_{\textrm{\tiny p}} > T_1$.
Drake further claimed that this would result in an adaptation zone near the embedded hydrodynamic shock, wherein the extra radiant heat is deposited over a short distance.
This is confirmed in the top plot of Figure \ref{fig:M3p3_Sn_zoomed}, where it is seen that the nonequilibrium-diffusion temperatures, $T_{\textrm{\tiny diff}}$ and $\theta_{\textrm{\tiny diff}}$, in the precursor region, approach the embedded hydrodynamic shock along almost straight lines, while the S$_{\textrm{n}}$-transport temperatures, $T_{\textrm{\tiny S}_{\textrm{\tiny n}}}$ and $\theta_{\textrm{\tiny S}_{\textrm{\tiny n}}}$, appear to curve upward as they approach the embedded hydrodynamic shock.

The adaptation zone can be understood by reviewing the analytic solution for the radiation intensity (\ref{eq:analytic_radiation_intensity}), and specifically looking at the integrand, $Q( x, \, \mu ) \exp[ - \tau( z, \, x ) ] \, / \mu$, and the definition of $\tau( z, \, x )$ (\ref{eq:tau_z_x}).
It is worth quickly noting that $Q \approx \left( \sigma_{\textrm{a}} \, T^4 + \sigma_{\textrm{s}} \, {\cal E} \right) / 4 \, \pi$, when dropping all terms of ${\cal O} \left( \beta \right)$ or higher, and for convenience below we assume that the material is purely absorbing, so that $\sigma_{\textrm{s}} = 0$.
As $\left| \mu \right| \rightarrow 0$, for fixed values of $z$ and $x$, the value of $\left| Q / \mu \right|$ increases, while the argument of the exponent becomes increasingly negative causing the exponent to rapidly decay.
However, the argument of the exponent is modified by the choice of $z$, such that when $z \approx x$ then $\tau( z, \, x ) \approx 0$, regardless of the value of $\mu$, and the product $Q \exp \left[ - \tau \left( z, x \right) \right] / \mu$ is then dominated by $Q / \mu$.
This is why, in the bottom plots of Figures \ref{fig:M2p7_with_intensities_zoomed} and \ref{fig:M5_with_intensities_zoomed}, the radiation intensities corresponding to $\left| \mu \right| \sim 0$ rise quickly under the Zel'dovich spike, over a distance which is dominated by the $Q / \mu$ term, but then decay rapidly as $\tau( z, \, x )$ increases.
This also explains the source of the oblique radiation shown in the polar plots in Figures \ref{fig:M2p7_polarplots} and \ref{fig:M5_polarplots}.

Similarly, in Figures \ref{fig:M2p7_with_intensities_zoomed} and \ref{fig:M5_with_intensities_zoomed}, the dominant contribution to the intensity temperatures $T_{I}(\left| \mu_{\textrm{\tiny m}} \right| \sim 0)$, at locations away from the Zel'dovich spike, comes from the $Q / \mu$ term over sufficiently short distances, where $z \approx x$, so that the exponential term is almost one.
Since $Q \approx \sigma_{\textrm{a}} T^4 /4 \pi$, the integral in (\ref{eq:analytic_radiation_intensity}) can be rewritten slightly:
\begin{align}
\nonumber
    I( x, \mu ) \approx \int_{x_{\textrm{\tiny eq}}}^x \frac{\sigma_{\textrm{a}} T^4}{4 \pi \mu} \, dz
& = \int_{\tau_{\textrm{\tiny eq}}}^\tau \frac{T^4}{4 \pi} \, d \left( \frac{\sigma_{\textrm{a}} z}{\mu} \right) \\
& = \int_{\tau_{\textrm{\tiny eq}}}^\tau \frac{T^4}{4 \pi} \, d \tau^{\prime} \, ,
\end{align}
where $\tau^{\prime} \equiv \sigma_{\textrm{a}} z / \mu$ is the modified optical depth.
For a given value of $dz$, as $\left| \mu \right| \rightarrow 0$ the modified optical depth increases implying that the system should be near thermal equilibrium, and appropriately, $I(\left| \mu \right| \sim 0) \approx T^4 / 4 \pi$, which is seen in Figures \ref{fig:M2p7_with_intensities_zoomed} and \ref{fig:M5_with_intensities_zoomed}, where the intensity temperature curves, $T_{I}(\left| \mu_{\textrm{\tiny m}} \right| \sim 0)$ are closest to the curve for $T$.

Before closing this subsection, we point out that all of the intensity temperature curves take their maximum values very near the location where $T_{\textrm{\tiny I}}$ crosses the curve for $T$.
This can be understood by reviewing equation (\ref{eq:dIdx}), and solving for $I$ when $dI / dx = 0$.
For the case of a purely absorbing material, and recognizing that the maximum value of $\beta$ is ${\cal M}_0 / {\cal C}_0 \approx 2 \times 10^{-3}$ which is negligibly small compared to 1, then $T_{I, \textrm{\tiny max}}(\mu) = T$.
This can be seen in the bottom plots in Figures \ref{fig:M2p7_with_intensities_zoomed} and \ref{fig:M5_with_intensities_zoomed}.

\section{Summary}
\label{sec:summary}
In this paper, we have presented new semi-analytic radiative-shock solutions where the radiation is modeled with grey S$_{\textrm{n}}$-transport, as originally recommended in the paper by Sen and Guess \cite{SG1957}.
We compared our solutions to the nonequilibrium-diffusion radiative-shock solutions presented by Lowrie and Edwards \cite{LE2008}.
{\color{black} It is our experience that the local Mach number is monotonic when producing nonequilibrium-diffusion solutions, but that this monotonicity may disappear in the precursor region when producing S$_{\text{n}}$-transport solutions.}
When the VEF deviates from one-third, significant quantitative and qualitative differences exist between these solutions.
We showed evidence for the conjectures made by Drake regarding an adaptation zone, and for the prediction by McClarren and Drake that anti-diffusive radiation flow exists under the Zel'dovich spike.
Subsequently, it is possible for the radiation temperature to be nonmonotonic and to have a local maximum under the Zel'dovich spike and for the radiation flux not to take its equilibrium value at this location, and for the material and radiation temperature values at the downstream precursor state, $\theta_{\textrm{\tiny ps}}$ and $T_{\textrm{\tiny p}}$, to both be greater than the downstream equilibrium temperature, $T_1$.
We showed important distinctions between the radiation flow solutions for the nonequilibrium-diffusion and S$_{\textrm{n}}$-transport radiation models by looking at polar plots of their radiation intensities at specific locations along the shock structure.
We analyzed and explained the structure of the S$_{16}$-transport radiation intensities across the shock structure by using the analytic solution to the time-independent radiation-transport equation.
Both, the S$_{16}$-transport intensity temperature solutions and the results shown in the polar plots showed that it is possible for the temperature-intensity solutions near a Zel'dovich spike, and in the adjacent relaxation region, to have values that are greater than the local material temperature.

Future work should seek semi-analytic solutions of the fully-relativistic hydrodynamic equations coupled to the fully-relativistic radiation-transport equation.
Similar work to this has already been performed by Farris \cite{Farris2008} in the equilibrium-diffusion approximation, but the same should be done for the nonequilibrium-diffusion and S$_{\textrm{n}}$-transport models of radiation.
From this, frequency-dependent solutions should also be considered which could investigate how edge effects and lines in the frequency domain affect the shock structure.
The frequency-dependent solutions for an equilibrium matter-radiation system have already been derived, and are being prepared for publication.
A more physically relevant EOS could also be implemented and investigated so as to more accurately describe flows found in laboratory settings.
Other work could determine under what conditions the local Mach number is monotonic, and the same for the other RH variables. \\

\noindent
\textbf{Acknowledgements}
One of us (JMF) would like to thank Don Shirk and Bob Singleton for many helpful comments, as well as Scott Doebling for continued support.
The authors would also like to thank the anonymous reviewers for their comments which helped to clarify key points in the paper.
This work was performed under the auspices of the US Department of Energy under contract DE-AC52-06NA25396 as LA-UR-16-28784.

\appendix
\section{Nondimensionalization of the ideal-gas equation-of-state}
\label{app:app5B}
The derivations presented in this appendix are intended to explain the nondimensionalization presented in Section \ref{sec:governing_equations} for the material, therefore, no consideration is given of radiation effects.
We assume an ideal-gas $\gamma$-law EOS for the material system:
\begin{gather}
\label{app5B:eq:ideal_gas_SM}
  p = \rho \, R_{\text{\tiny s}} \, T = \left( \gamma - 1 \right) \rho \, e \, ,
\end{gather}
where $R_{\text{\tiny s}}$ is the specific gas constant, and $e = c_V \, T$ is the specific internal-energy.
For an ideal-gas, $R_{\text{\tiny s}}$ is the difference between the specific heats at constant pressure and constant volume, $R_s = c_p - c_V$, and the ratio of these specific heats is the adiabatic index, $c_p / c_V \equiv \gamma$.
In a simplified model, sound waves are isentropic propagation waves of a small disturbance about the material's equilibrium state such that the squared sound speed is:
\begin{gather}
\label{app5B:eq:speed_of_sound_wave}
  a^2 = \frac{\gamma \, p_{\textrm{\tiny eq}}}{\rho_{\textrm{\tiny eq}}} = \gamma \left( \gamma - 1 \right) e_{\textrm{\tiny eq}} \, .
\end{gather}
It is apparent that the material's specific internal-energy has the dimensions of speed squared.

We now present the nondimensionalization of the material variables.
Dimensional variables are decomposed into their dimensional quantities which carry a tilde-$\tilde{\,}$ over them, and their nondimensional values, e.g., $\tilde{x} = x \tilde{L}$.
We use reference variables for the sound speed, $\tilde{a}_0$, and the temperature, $\tilde{T}_0$, such that they are nondimensionalized as $\tilde{a} = a \, \tilde{a}_0$ and $\tilde{T} = T \, \tilde{T}_0$, and using equation (\ref{app5B:eq:speed_of_sound_wave}) the material's specific internal-energy is nondimensionalized as $\tilde{e} = e \, \tilde{a}_0^2$.
Requiring consistency between these reference quantities produces:
\begin{gather}
\label{app5B:eq:sound_speed_dimension}
  \tilde{a}_0^2 = \gamma \left( \gamma - 1 \right) \tilde{e}_0 = \gamma \, \frac{\tilde{R}_{\text{\tiny s}}}{\tilde{c}_V} \, \tilde{e}_0 = \gamma \, \tilde{R}_{\text{\tiny s}} \, \tilde{T}_0 \, .
\end{gather}
The physical sound speed can now be written in two slightly different but equivalent forms:
\begin{subequations}
  \begin{gather}
    \tilde{a}^2 = a^2 \, \tilde{a}_0^2 = a^2 \, \gamma \, \tilde{R}_{\text{\tiny s}} \, \tilde{T}_0 \, ,
  \end{gather}
and
  \begin{gather}
    \tilde{a}^2 = \gamma \, \tilde{R}_{\text{\tiny s}} \, \tilde{T} = T \, \gamma \, \tilde{R}_{\text{\tiny s}} \, \tilde{T}_0 \, .
  \end{gather}
\end{subequations}
Comparison of the right-hand sides of these two expressions shows that $a^2 = T$.
Similarly, the specific internal-energy can be written in two slightly different but equivalent forms:
\begin{subequations}
  \begin{align}
  \nonumber
    \tilde{e}
      = e \, \tilde{a}_0^2 
    & = c_V \, T \, \gamma \, \tilde{R}_{\text{\tiny s}} \, \tilde{T}_0 \\
    & = c_V \, T \, \gamma \left( \gamma - 1 \right) \tilde{c}_V \, \tilde{T}_0 \, ,
  \end{align}
and
  \begin{gather}
    \tilde{e} = \tilde{c}_V \, \tilde{T} = T \, \tilde{c}_V \, \tilde{T}_0 \, .
  \end{gather}
\end{subequations}
Comparison of the right-hand sides of these two expressions shows that $c_V = \left[ \gamma \left( \gamma - 1 \right) \right]^{-1}$.
Having established the definitions of the nondimensional variables we now write the following nondimensional expressions that are used in the main body of the text:
\begin{subequations}
  \begin{gather}
    e = \frac{T}{\gamma \left( \gamma - 1 \right)} \, , \\
    p = \frac{\rho \, T}{\gamma} \, , \\
    {\cal M} \equiv \frac{\tilde{u}}{\tilde{a}} = \frac{u \, \tilde{a}_0}{a \, \tilde{a}_0} = \frac{u}{\sqrt{T}} \, .
  \end{gather}
\end{subequations}

\section{The solution procedure}
\label{app:the_solution_procedure}
In this appendix we describe the global solution procedure.
A similar solution procedure is given in Section 5 of the paper by Lowrie and Edwards \cite{LE2008}.

We seek to solve the RH ODEs (\ref{eqs:dxdM_dPdM}) as a two-point boundary-value problem, with a consistent VEF determined by solving each of the RT ODEs (\ref{eq:Sn_transport_broken}) as an initial-value problem.
We do this by following the summarized procedure below:
\begin{enumerate}
  \item[] \underline{The RH solve}
  \item Start with an initial equilibrium state which consists of values for the constants $\tilde{\rho}_0$, $\tilde{T}_0$ and ${\cal M}_0$, as well as the EOS and the cross sections as functions of the pair $(\rho, T)$, and an initial guess for the VEF.
To start the solution process we typically assume that the VEF is strictly one-third, however it is possible, in some cases, to initialize the problem using a VEF from a solution that is sufficiently similar to the desired solution.
  \item Determine the final equilibrium state via the Rankine-Hugoniot conditions, as described in Subsection \ref{subsec:Rankine_Hugoniot_jump_conditions}.
  \item Move the RH solution away from the initial and final equilibrium states.
For the first iteration we use a linearization procedure, as described in Subsection \ref{subsec:linearization_away_from_equilibrium}, in order to determine the values of $({\cal M}_{0 \epsilon}, {\cal P}_{0 \epsilon})$ and $({\cal M}_{1 \epsilon}, {\cal P}_{1 \epsilon})$.
For higher iterations we use the values of ${\cal M}_{0 \epsilon}$ and ${\cal M}_{1 \epsilon}$ to determine new values for ${\cal P}_{0 \epsilon}$ and ${\cal P}_{1 \epsilon}$ by linearly interpolating the most recent Mach solution from the RH solve against the most recent radiation pressure solution (\ref{eq:quadrature_P}) from the RT solve.
As a reminder, the initial values of $x_{0 \epsilon}$ and $x_{1 \epsilon}$ are arbitrary because the RH ODEs (\ref{eqs:dxdM_dPdM}) are shift invariant.
  \item As described in Subsection \ref{subsec:integrating_the_RH_ODEs}, integrate the RH ODEs (\ref{eqs:dxdM_dPdM}) in Mach-space from state $({\cal M}_{0 \epsilon}, {\cal P}_{0 \epsilon})$ to ${\cal M}_{\textrm{\tiny L}} = 1 + \epsilon_{\textrm{\tiny ASP}}$ to construct the integrated curve in the precursor region.
Similarly, integrate equations (\ref{eqs:dxdM_dPdM}) in Mach-space from state $({\cal M}_{1 \epsilon}, {\cal P}_{1 \epsilon})$ to ${\cal M}_{\textrm{\tiny R}} = 1 - \epsilon_{\textrm{\tiny ASP}}$ to construct the integration curve in the relaxation region.
The values of $\epsilon_{\textrm{\tiny ASP}}$ for the precursor and relaxation regions may be different.
During the integration procedure the value of the VEF is typically one-third for the first iteration.
For higher iterations we linearly interpolate the current local value of the radiation pressure against the most recent RT solutions of the radiation pressure (\ref{eq:quadrature_P}) and the VEF, i.e., given ${\cal P}_{\textrm{RT}}$ and $f$ from the most recent RT solve along with the current local value of ${\cal P}$, we linearly interpolate to obtain the current local value of $f$.
  \item {\color{black} Test whether $({\cal P}_{\textrm{\tiny R}} - {\cal P}_{\textrm{\tiny L}}) / {\cal P}_{\text{\tiny R}} < \epsilon_{\text{\tiny tol}}$, where the value of $\epsilon_{\text{\tiny tol}}$ is of the same order as $\epsilon_{\text{\tiny ASP}}$.}
If so, the solution is continuous in all variables. Otherwise, test whether ${\cal P}_{\textrm{\tiny L}} > {\cal P}_{\textrm{\tiny R}}$, in which case there is an embedded hydrodynamic shock.
Shift the $x$-values accordingly.
See Subsection \ref{subsec:continuity_conditions}.
  \item The constructed triplet $({\cal M}, {\cal P}, x)$ from the precursor and relaxation regions constitutes the RH solution.
All other variables can be constructed from the pair $({\cal P}, {\cal M})$.
Only the variables that are on the right-hand side of equation (\ref{eq:Sn_transport_broken}) are strictly needed. \\
  \item[] \underline{The RT solve}
  \item The regions of the RH solution between states $({\cal M}_0, {\cal P}_0)$ and $({\cal M}_{0 \epsilon}, {\cal P}_{0 \epsilon})$ in the precursor region, and between states $({\cal M}_1, {\cal P}_1)$ and $({\cal M}_{1 \epsilon}, {\cal P}_{1 \epsilon})$ in the relaxation region are linear.
However, the solutions of the S$_{\textrm{n}}$-transport radiation intensities and the VEF are linear over a smaller spatial domain.
Therefore, before we begin the RT solve we add points to the $x$-variable in these regions, as well as the nearby adjacent space, and we linearly interpolate to determine the values of all other variables from the RH solution at the points that we have just added.
  \item Compute the initial values of the n separate S$_{\textrm{n}}$-transport equations (\ref{eq:Sn_transport_broken}) from their equilibrium expressions (\ref{eqs:equilibrium_EFPf}), as described in Subsection \ref{subsec:RT_initial_values_and_linearization}.
  \item Move the n RT solutions from their initial values, $(x_{\textrm{\tiny eq}}, I_{\textrm{\tiny eq}})$, by using the linearization procedure described in Subsection \ref{subsec:RT_initial_values_and_linearization}, to the state at $(x_{\epsilon}, I_{\textrm{\tiny m}, \epsilon})$.
  \item Integrate the n RT ODEs (\ref{eq:Sn_transport_broken}) from the state at $(x_{\epsilon}, I_{\textrm{\tiny m}, \epsilon})$ to their boundary conditions at the opposite equilibrium state.
Since each RT ODE (\ref{eq:Sn_transport_broken}) represents an initial-value problem the integrated ODE naturally arrives at the other equilibrium state.
  \item Given the n radiation intensity solutions use quadrature integration to construct the radiation energy density, radiation flux, radiation pressure, and the VEF, as described in Subsection \ref{subsec:construct_RT_solutions}.
  \item Test whether the solution procedure should continue.  If so, repeat steps 3-11 until the solution procedure should stop.
\end{enumerate}

\section*{References}
\bibliographystyle{elsarticle-num}
\bibliography{References}

\begin{thebibliography}{10}
\expandafter\ifx\csname url\endcsname\relax
  \def\url#1{\texttt{#1}}\fi
\expandafter\ifx\csname urlprefix\endcsname\relax\def\urlprefix{URL }\fi
\expandafter\ifx\csname href\endcsname\relax
  \def\href#1#2{#2} \def\path#1{#1}\fi

\bibitem{LE2008}
R.~B. {Lowrie}, J.~D. {Edwards}, Radiative shock solutions with grey
  nonequilibrium diffusion, Shock Waves 18 (2008) 129--143.

\bibitem{Drake2007a}
R.~P. {Drake}, Energy balance and structural regimes of radiative shocks in
  optically thick media, Plasma Science, IEEE Transactions 35~(2) (2007)
  171--180.

\bibitem{Drake2007b}
R.~P. {Drake}, Theory of radiative shocks in optically thick media, Physics of
  Plasmas 14~(4) (2007) 043301.

\bibitem{McD2010}
R.~G. {McClarren}, R.~P. {Drake}, Anti-diffusive radiation flow in the cooling
  layer of a radiating shock, Journal of Quantitative Spectroscopy and
  Radiative Transfer 111~(14) (2010) 2095 -- 2105.

\bibitem{SG1957}
H.~K. {Sen}, A.~W. {Guess}, Radiation effects in shock-wave structure, Physical
  Review 108 (1957) 560--564.

\bibitem{LR2007}
R.~B. {Lowrie}, R.~M. {Rauenzahn}, Radiative shock solutions in the equilibrium
  diffusion limit, Shock Waves 16 (2007) 445--453.

\bibitem{HB1963}
M.~A. {Heaslet}, B.~S. {Baldwin}, Predictions of the structure of
  radiation-resisted shock waves, Physics of Fluids 6~(6) (1963) 781--791.

\bibitem{Chandrasekhar1960}
S.~{Chandrasekhar}, Radiative Transfer, Dover Books on Intermediate and
  Advanced Mathematics, Dover Publications, Mineola, N.Y., 1960.

\bibitem{ZR2002}
Y.~B. {Zel`dovich}, Y.~P. {Raizer}, Physics of Shock Waves and High-Temperature
  Hydrodynamic Phenomena, Dover Books on Physics Series, Dover Publications,
  Mineola, N.Y., 2002.

\bibitem{MM1999}
D.~{Mihalas}, B.~W. {Mihalas}, Foundations of Radiation Hydrodynamics, Dover
  Books on Physics Series, Dover, Mineola, N.Y., 1999.

\bibitem{JSD2014}
Y.~F. {Jiang}, J.~M. {Stone}, S.~W. {Davis}, {An algorithm for radiation
  magnetohydrodynamics based on solving the time-dependent transfer equation},
  The Astrophysical Journal Supplement Series 213~(1) (2014) 7.

\bibitem{McDMH2010}
{\color{blue} {McClarren}, R.~G. and {Drake}, R.~P. and {Morel}, J.~E. and
  {Holloway}, J.~P.}, {\color{blue} Theory of radiative shocks in the mixed,
  optically thick-thin case}, {\color{blue} Physics of Plasmas} {\color{blue}
  17}~({\color{blue} 9}) ({\color{blue} 2010}) {\color{blue} 093301}.

\bibitem{LMc2013}
{\color{blue} {Lane}, T.~K. and {McClarren}, R.~G.}, {{\color{blue} New
  self-similar radiation-hydrodynamics solutions in the high-energy density,
  equilibrium diffusion limit}}, {{\color{blue} New Journal of Physics}}
  {{\color{blue} 15}} ({\color{blue} 2013}) {{\color{blue} 095013--095029}},
  {{\color{blue} Focus on High-Energy-Density-Physics}}.

\bibitem{CA1986}
{\color{blue} {Coggeshall}, S.~V. and {Axford}, R.~A.}, {{\color{blue} Lie
  group invariance properties of radiation hydrodynamics equations and their
  associated similarity solutions}}, {{\color{blue} Physics of Fluids}}
  {{\color{blue} 29}}~({{\color{blue} 8}}) ({\color{blue} 1986}) {{\color{blue}
  2398}}.

\bibitem{FMB2011}
{\color{blue} {Falize}, E. and {Michaut}, C. and {Bouquet}, S.}, {{\color{blue}
  Similarity properties and scaling laws of radiation hydrodynamic flows in
  laboratory astrophysics}}, {{\color{blue} The Astrophysical Journal}}
  {{\color{blue} 730}}~({{\color{blue} 2}}) ({\color{blue} 2011})
  {{\color{blue} 96}}.

\bibitem{Barenblatt1996}
{\color{blue} Barenblatt, G.I.}, {\color{blue} Scaling, Self-similarity, and
  Intermediate Asymptotics: Dimensional Analysis and Intermediate Asymptotics},
  {\color{blue} Cambridge Texts in Applied Mathematics}, {\color{blue}
  Cambridge University Press}, {\color{blue} 1996}.

\bibitem{LK2000}
{\color{blue} {Liang}, E. and {Keilty}, K.}, {{\color{blue} An analytic
  approximation to radiative blast wave evolution}}, {{\color{blue} The
  Astrophysical Jouranl}} {{\color{blue} 533}}~({{\color{blue} 8}})
  ({\color{blue} 2000}) {{\color{blue} 890}}.

\bibitem{MWL2011}
{\color{blue} {Masser}, T.~O. and {Wohlbier}, J.~G. and {Lowrie}, R.~B.},
  {\color{blue} Shock wave structure for a fully ionized plasma}, {\color{blue}
  Shock Waves} {\color{blue} 21} ({\color{blue} 2011}) {\color{blue} 367--381}.

\bibitem{HFM2015}
{\color{blue} {Holgado}, A.~M. and {Ferguson}, J.~M. and {McClarren}, R.~G.},
  {\color{blue} Anti-diffusive-like-behavior in semi-analytic radiative shocks
  via multigroup Sn transport with constant cross sections}, {\color{blue} High
  Energy Density Physics} {\color{blue} 17, Part A} ({\color{blue} 2015})
  {\color{blue} 114 -- 118}, {\color{blue} {S}pecial {I}ssue: 10th
  International Conference on High Energy Density Laboratory Astrophysics}.

\bibitem{T1965}
S.~C. {Traugott}, Shock structure in a radiating, heat conducting, and viscous
  gas, Physics of Fluids 8~(5) (1965) 834--849.

\bibitem{Lamb1945}
H.~{Lamb}, Hydrodynamics, Dover Books on Physics Series, Dover, Mineola, N.Y.,
  1945.

\bibitem{Zeldovich1957}
I.~B. {Zel'dovich}, {Shock Waves of Large Amplitude in Air}, Soviet Journal of
  Experimental and Theoretical Physics 5~(5) (1957) 919.

\bibitem{Raizer1957}
I.~P. {Raizer}, {On the Structure of the Front of Strong Shock Waves in Gases},
  Soviet Journal of Experimental and Theoretical Physics 5~(6) (1957) 1242.

\bibitem{Farris2008}
B.~D. {Farris}, T.~K. {Li}, Y.~T. {Liu}, S.~L. {Shapiro}, {Relativistic
  radiation magnetohydrodynamics in dynamical spacetimes: Numerical methods and
  tests}, Physical Review D 78~(2) (2008) 024023.

\end{thebibliography}










\end{document}